\documentclass[12pt,twoside, a4paper]{article}
\def\pd{\partial}
\def\mc{\mathcal}
\def\ul{\underline}

\usepackage[dvips]{graphicx}
\usepackage{amssymb}
\usepackage{amssymb,amsmath}
\usepackage{graphicx}
\usepackage{dsfont}
\usepackage{caption}
\usepackage{subcaption}
\usepackage{mathtools}
\usepackage{verbatim}
\usepackage{graphicx}
\usepackage{multirow}
\usepackage[outline]{contour}
\usepackage{xcolor,colortbl}
\usepackage{pdflscape}
\usepackage{float}
\input{epsf.sty}  
\begin{document}
\begin{center}
\LARGE{\textbf{Supersymmetric $AdS_4$ black holes from matter-coupled $N=3,4$ gauged supergravities}}
\end{center}
\begin{center}
\large{\textbf{Parinya Karndumri}}
\end{center}
\begin{center}
String Theory and Supergravity Group, Department
of Physics, Faculty of Science, Chulalongkorn University, 254 Phayathai Road, Pathumwan, Bangkok 10330, Thailand
\end{center}
E-mail: parinya.ka@hotmail.com \vspace{1 cm}\\
\begin{abstract}
We study supersymmetric $AdS_4$ black holes in matter-coupled $N=3$ and $N=4$ gauged supergravities in four dimensions. In $N=3$ theory, we consider $N=3$ gauged supergravity coupled to three vector multiplets and $SO(3)\times SO(3)$ gauge group. The resulting gauged supergravity admits two $N=3$ supersymmetric $AdS_4$ vacua with $SO(3)\times SO(3)$ and $SO(3)$ symmetries. We find an $AdS_2\times H^2$ solution with $SO(2)\times SO(2)$ symmetry and an analytic solution interpolating between this geometry and the $SO(3)\times SO(3)$ symmetric $AdS_4$ vacuum. For $N=4$ gauged supergravity coupled to six vector multiplets with $SO(4)\times SO(4)$ gauge group, there exist four supersymmetric $AdS_4$ vacua with $SO(4)\times SO(4)$, $SO(4)\times SO(3)$, $SO(3)\times SO(4)$ and $SO(3)\times SO(3)$ symmetries. We find a number of $AdS_2\times S^2$ and $AdS_2\times H^2$ geometries together with the solutions interpolating between these geometries and all, but the $SO(3)\times SO(3)$, $AdS_4$ vacua. These solutions provide a new class of $AdS_4$ black holes with spherical and hyperbolic horizons dual to holographic RG flows across dimensions from $N=3,4$ SCFTs in three dimensions to superconformal quantum mechanics within the framework of four-dimensional gauged supergravity.
\end{abstract}
\newpage

\section{Introduction}
String/M-theory has provided a number of insights to various aspects of quantum gravity for many decades. In particular, a resolution for a long-standing problem of black hole entropy has been proposed in \cite{stronginger_vafa}. After this pioneering work, many other papers followed and clarified the issues of microscopic entropy of asymptotically flat black holes. For asymptotically $AdS_4$ black holes, a concrete result on the corresponding microscopic entropy, using AdS/CFT correspondence \cite{maldacena,Gubser_AdS_CFT,Witten_AdS_CFT}, has appeared recently in \cite{Zaffaroni_BH1,Zaffaroni_BH2,Zaffaroni_BH3}, see also \cite{twisted_index1,twisted_index2,twisted_index3,twisted_index4,twisted_index5}. 
\\
\indent On the gravity side, an important ingredient along this line is $AdS_4$ black hole solutions interpolating between asymptotic $AdS_4$ and $AdS_2\times \Sigma^2$ spaces with $\Sigma^2$ being a Riemann surface. The latter describes the geometry of the black hole horizon with the values of scalars determined by the attractor mechanism. These solutions holographically describe RG flows across dimensions from three-dimensional SCFTs, dual to the $AdS_4$ vacua, to superconformal quantum mechanics, dual to the $AdS_2$ factor of the horizons. The latter is obtained from twisted compactifications of the former which play an important role in computing Bekenstein-Hawking entropy of the black holes via twisted indices.         
\\
\indent In this paper, we are interested in supersymmetric $AdS_4$ black holes with the horizon geometry $AdS_2\times S^2$ and $AdS_2\times H^2$ with $S^2$ and $H^2$ being a two-sphere and a two-dimensional hyperbolic space, respectively. We will work in matter-coupled $N=3$ and $N=4$ gauged supergravities. This type of solutions has been extensively studied in $N=2$ gauged supergravity for a long time \cite{AdS4_BH1,AdS4_BH2,AdS4_BH3,AdS4_BH4,AdS4_BH5,Guarino_AdS2_1,Guarino_AdS2_2}, see also \cite{Kim_AdS2} for some results in $N=8$ gauged supergravity. Similar studies in other gauged supergravities have appeared only recently in \cite{N3_SU2_SU3,Trisasakian_AdS2,N5_flow,N6_flow}. In particular, a study of $AdS_2\times \Sigma^2$ solutions in $N=3$ with only magnetic charges has been initiated in \cite{N3_SU2_SU3}. We will extend this result by performing a more systematic analysis and including a possible dyonic generalization. We will consider a particular case of $N=3$ gauged supergravity coupled to three vector multiplets with a compact $SO(3)\times SO(3)$ gauge group. We will see that only one magnetic $AdS_2\times H^2$ solution with $SO(2)\times SO(2)$ symmetry exists. This is very similar to solutions in $N=5$ and $N=6$ gauged supergravities given in \cite{N5_flow} and \cite{N6_flow}.  
\\
\indent For $N=4$ case, we will consider $N=4$ gauged supergravity coupled to six vector multiplets with $SO(4)\times SO(4)$ gauge group. Unlike the $N=3$ theory with a purely electric gauging, any $N=4$ supergravity that admits supersymmetric $AdS_4$ vacua must be dyonically gauged \cite{AdS4_N4_Jan}. In this case, apart from an $AdS_2\times H^2$ solution similar to $N=3,5,6$ gauged supergravities, there exist a number of supersymmetric $AdS_2\times S^2$ and $AdS_2\times H^2$ solutions. It should also be pointed out that some $AdS_2\times \Sigma^2$ solutions in $N=4$ gauged supergravity obtained from a truncation of eleven-dimensional supergravity have also been found in \cite{Trisasakian_AdS2}. However, in that case, the gauge group is of non-semisimple form, and the resulting BPS equations are highly complicated. In the present work, we provide a number of much simpler examples of supersymmetric $AdS_4$ black holes in $N=4$ gauged supergravity. In particular, the two-form fields required by the consistency of incorporating magnetic gauge fields can be truncated out in the present case.
\\
\indent The paper is organized as follows. In section \ref{N3_SUGRA}, we will review the structure of $N=3$ gauged supergravity after translating the original construction in group manifold approach to the usual formulae in space-time. This is followed by a general analysis of relevant BPS equations for finding supersymmetric $AdS_4$ black hole solutions. An $AdS_2\times H^2$ solution with $SO(2)\times SO(2)$ symmetry together with the full flow solution interpolating between this fixed point and the supersymmetric $AdS_4$ vacuum with $SO(3)\times SO(3)$ symmetry are also given. Similar analysis is then performed in section \ref{N4_SUGRA} in which we will find a number of $AdS_2\times S^2$ and $AdS_2\times H^2$ fixed points and solutions interpolating between them and supersymmetric $AdS_4$ vacua with various unbroken symmetries in $N=4$ gauged supergravity. We end the paper by giving conclusions and comments on the results in section \ref{conclusion}. 

\section{$AdS_4$ black holes from $N=3$ gauged supergravity}\label{N3_SUGRA}
In this section, we consider matter-coupled $N=3$ gauged supergravity and possible supersymmetric $AdS_4$ black holes. We begin with a review of $N=3$ gauged supergravity and the analysis of relevant BPS equations. These are followed by the explicit solutions at the end of the section.

\subsection{Matter-coupled $N=3$ gauged supergravity} 
We now give a description of $N=3$ gauged supergravity coupled to $n$ vector
multiplets. This has been constructed by the geometric group manifold approach in \cite{N3_Ferrara}, see also \cite{N3_Ferrara2,Castellani_book}. However, the final form of the space-time Lagrangian has not been given, and the supersymmetry transformations of fermions have been given in a rather implicit form. We will first collect all these ingredients and specify to the case of $n=3$ vector multiplets later on. The interested reader can find a more detailed construction and some discussions on the structure of the scalar manifold and electric-magnetic duality in \cite{N3_Ferrara}. We will mostly follow the notations of \cite{N3_Ferrara} but in a mostly plus signature for the space-time metric and a slightly different convention for the gauge fields.
\\
\indent For $N=3$ supersymmetry in four dimensions, there are two types of supermultiplets, the gravity and vector multiplets. The former consists of the following component fields
\begin{displaymath}
(e^a_\mu, \psi_{\mu A}, A^A_{\mu }, \chi).
\end{displaymath}
$e^a_\mu$ is the graviton, and $\psi_{\mu A}$ are three
gravitini. Space-time and tangent space indices will be denoted by $\mu,\nu,\ldots$ and
$a,b,\ldots$, respectively. The gravity multiplet also contains three vector fields
$A^A_{\mu }$ with indices $A,B,\ldots=1,2,3$ denoting the fundamental representation of
the $SU(3)_R$ part of the full $SU(3)_R\times U(1)_R$ R-symmetry. There is also an $SU(3)_R$ singlet spinor field $\chi$.
\\
\indent $N=3$ supersymmetry allows the gravity multiplet to couple to an arbitrary number of vector multiplets, the only matter fields in this case. The component fields in a vector multiplet are given by the following field content
\begin{displaymath}
(A_\mu, \lambda_A, \lambda, z_A)
\end{displaymath}
consisting of a vector field $A_\mu$, four spinor
fields $\lambda$ and $\lambda_A$ which are respectively singlet and
triplet of $SU(3)_R$, and three complex scalars $z_A$ in the
fundamental of $SU(3)_R$. We will use indices $i,j,\ldots =1,\ldots, n$ to label each vector multiplet. 
\\
\indent The fermionic fields are subject to the chirality projection conditions
\begin{equation}
\psi_{\mu A}=\gamma_5\psi_{\mu A},\qquad \chi=\gamma_5\chi,\qquad
\lambda_A=\gamma_5\lambda_A,\qquad \lambda=-\gamma_5\lambda\, .
\end{equation}
These also imply $\psi_\mu^A=-\gamma_5\psi_\mu^A$ and
$\lambda^A=-\gamma_5\lambda^A$ for the corresponding conjugate spinors. 
\\
\indent In the matter-coupled supergravity with $n$ vector multiplets, there are $3n$ complex scalar
fields $z_A^{\phantom{A}i}$ parametrizing the coset space
$SU(3,n)/SU(3)\times SU(n)\times U(1)$. These scalars are
conveniently described by the coset representative
$L_\Lambda^{\phantom{\Lambda}\ul{\Sigma}}$. The
coset representative transforms under the global $G=SU(3,n)$ and the
local $H=SU(3)\times SU(n)\times U(1)$ symmetries by left and
right multiplications, respectively. The $SU(3,n)$ indices $\Lambda, \Sigma, \ldots$ will take values $1,\ldots, n+3$. On the other hand, it is convenient to split the $SU(3)\times SU(n)\times U(1)$ indices $\ul{\Lambda},\ul{\Sigma},\ldots$ as $(A,i)$. We can then write the coset representative as
\begin{equation}
L_\Lambda^{\phantom{\Lambda}\ul{\Sigma}}=(L_\Lambda^{\phantom{\Lambda}A},L_\Lambda^{\phantom{\Lambda}i}).
\end{equation}
\indent The $n+3$ vector fields from both the gravity and vector multiplets are combined into $A^\Lambda_\mu=(A^A_\mu,A^i_\mu)$. These are called electric vector fields that appear in the Lagrangian with the usual Yang-Mills (YM) kinetic terms. Accompanied by the corresponding magnetic dual $A_{\Lambda \mu}$, these vector fields transform in the fundamental representation $\mathbf{n+3}$ of the global symmetry group $SU(3,n)$, also called the duality group. 
\\
\indent For the gaugings of the matter-coupled $N=3$ supergravity, we will follow the original result of \cite{N3_Ferrara} since the complete modern approach using the embedding tensor has not been worked out so far. For general gaugings obtained from the embedding tensor formalism, both electric and magnetic gauge fields can participate in the gaugings. The construction of \cite{N3_Ferrara}, called electric gaugings, with only electric vector fields becoming the gauge fields results in gauge groups that only account for a smaller class of all possible gaugings. All gauge groups considered in \cite{N3_Ferrara} are subgroups of $SO(3,n)$ which is the electric subgroup of the full global symmetry $SU(3,n)$. 
\\
\indent After gauging a particular subgroup $G_0$ of $SO(3,n)\subset SU(3,n)$, the
corresponding non-abelian gauge field strengths are given by
\begin{equation}
F^\Lambda=dA^\Lambda+{f_{\Sigma\Gamma}}^{\Lambda}A^\Sigma\wedge A^{\Gamma}
\end{equation}
where ${f_{\Lambda\Sigma}}^\Gamma$ denote the structure constants of the gauge group. The gauge generators
$T_\Lambda$ satisfy
\begin{equation}
\left[T_\Lambda,T_\Sigma\right]=f_{\Lambda\Sigma}^{\phantom{\Lambda\Sigma}\Gamma}T_\Gamma\,
.
\end{equation}
Indices $\Lambda,\Sigma,\ldots$ can be raised and lowered by the $SU(3,n)$ invariant tensor
\begin{equation}
J_{\Lambda\Sigma}=J^{\Lambda\Sigma}=(\delta_{AB},-\delta_{ij})
\end{equation}
which will become the Killing form of the gauge group $G_0$. In order for the gaugings to be consistent with
supersymmetry, the structure constants $f_{\Lambda\Sigma\Gamma}$ need to satisfy the following constraint
\begin{equation}
f_{\Lambda\Sigma\Gamma}=f_{\Lambda\Sigma}^{\phantom{\Lambda\Sigma}\Delta}J_{\Delta\Gamma}=f_{[\Lambda\Sigma\Gamma]}
\end{equation}
which is equivalent to the linear constraint in the embedding tensor formalism. Some examples of possible gauge groups are $SO(3)\times H_n$, $SO(3,1)\times H_{n-3}$ and $SO(2,2)\times H_{n-3}$ with $H_n$ being an $n$-dimensional compact subgroup of $SO(n)\subset SU(n)$. These gaugings together with possible supersymmetric $AdS_4$ vacua and domain walls have already been studied in \cite{N3_4D_gauging}. 
\\
\indent With the fermion mass terms and the scalar potential included as required by supersymmetry, the bosonic Lagrangian of the $N=3$ gauged supergravity can be written as
\begin{eqnarray}
e^{-1}\mc{L}&=&\frac{1}{4}R-\frac{1}{4}P_\mu^{iA}P^\mu_{Ai}-a_{\Lambda\Sigma}F^{+\Lambda}_{\mu\nu}F^{+\Sigma\mu\nu}
-\bar{a}_{\Lambda\Sigma}F^{-\Lambda}_{\mu\nu}F^{-\Sigma\mu\nu}-V\, .\label{N3_Lar}
\end{eqnarray}
This Lagrangian is obtained from translating the first-order Lagrangian in the geometric group manifold approach given in \cite{N3_Ferrara} to the usual space-time Lagrangian. We have also multiplied the whole Lagrangian by a factor of $3$ resulting in a factor of $3$ in the scalar potential given below as compared to that given in \cite{N3_Ferrara}.
\\
\indent The self-dual and antiself-dual field strengths are defined by
\begin{equation}
F^{\pm\Lambda}_{ab}=\frac{1}{2}\left(F_{ab}^\Lambda \mp\frac{i}{2}\epsilon_{abcd}
F^{\Lambda cd}\right)
\end{equation}
which satisfy the following relations
\begin{equation}
\frac{1}{2}\epsilon_{abcd}F^{\pm \Lambda cd}=\pm i
F_{ ab}^{\pm \Lambda}\qquad \textrm{and}\qquad  F_{ab}^{\pm \Lambda}=(F_{ab}^{\mp \Lambda})^*.
\end{equation}
To write down the explicit form of the scalar matrix $a_{\Lambda\Sigma}$ in terms of the coset representative, we first identify various components of the coset representative as 
\begin{equation}
{L_\Lambda}^{\ul{\Sigma}}=\left(
                                         \begin{array}{cc}
                                           {L_A}^B& {L_A}^i\\
                                           {L_j}^B & {L_j}^i \\
                                         \end{array}
                                       \right).
\end{equation}
The symmetric matrix $a_{\Lambda \Sigma}$ can be written as
\begin{equation}
a_{\Lambda \Sigma}=({\mathbf{f}^\dagger}^{-1}
\mathbf{h}^\dagger)_{\Lambda \Sigma}
\end{equation}
in which the matrices ${\mathbf{f}_\Lambda}^{\ul{\Sigma}}=({L_\Lambda}^A,({L_\Lambda}^i)^*)$ and $\mathbf{h}_{\Lambda\ul{\Sigma}}=-(J\mathbf{f}J)_{\Lambda\ul{\Sigma}}$ are given explicitly by
\begin{equation}
{\mathbf{f}_\Lambda}^{\ul{\Sigma}}=\left(
                                         \begin{array}{cc}
                                           {L_A}^B& ({L_A}^i)^*\\
                                           {L_j}^B & ({L_j}^i)^* \\
                                         \end{array}
                                       \right)\qquad \textrm{and}\qquad 
\mathbf{h}_{\Lambda\ul{\Sigma}}=\left(
                                         \begin{array}{cc}
                                           {L_A}^B& -({L_A}^i)^*\\
                                           -{L_j}^B & ({L_j}^i)^* \\
                                         \end{array}
                                       \right).                                        
\end{equation}
\indent The scalar kinetic terms are written in terms of the vielbein on the $SU(3,n)/SU(3)\times SU(n)\times U(1)$ obtained from the Maurer-Cartan one-form
\begin{equation}
\Omega_{\ul{\Lambda}}^{\phantom{\Lambda}\ul{\Pi}}=(L^{-1})_{\ul{\Lambda}}^{\phantom{\Lambda}\Sigma}
dL_\Sigma^{\phantom{\Sigma}\ul{\Pi}}
+(L^{-1})_{\ul{\Lambda}}^{\phantom{\Lambda}\Sigma}{f_{\Sigma\Omega}}^\Gamma A^\Omega
L_\Gamma^{\phantom{\Gamma}\ul{\Pi}}
\end{equation} 
via the components 
\begin{equation}
P_i^{\phantom{i}A}=\Omega_i^{\phantom{i}A}=(\Omega_A^{\phantom{A}i})^*\, .
\end{equation}
We also note that the upper and lower indices of $SU(3)$ and $SU(n)$ are related by complex conjugation. Since ${L_\Lambda}^{\ul{\Sigma}}$ is an element of $SU(3,n)$, the inverse ${(L^{-1})_{\ul{\Lambda}}}^{\Sigma}$ satisfies the following relation
\begin{equation}
(L^{-1})_{\ul{\Lambda}}^{\phantom{\Lambda}\Sigma}=J_{\ul{\Lambda}\ul{\Pi}}J^{\Sigma\Delta}(L_\Delta^{\phantom{\Delta}\ul{\Pi}})^*\,
.
\end{equation}
\indent The composite connections $Q_A^{\phantom{A}B}$, $Q_i^{\phantom{i}j}$ and $Q$ for the $SU(3)\times SU(n)\times U(1)$ local symmetry are given by
\begin{equation}
\Omega_A^{\phantom{A}B}=Q_A^{\phantom{A}B}-n\delta^B_AQ\qquad \textrm{and}\qquad 
\Omega_i^{\phantom{i}j}= Q_i^{\phantom{i}j}+3\delta^j_iQ
\end{equation}
with $Q_A^{\phantom{A}A}=Q_i^{\phantom{i}i}=0$.
\\
\indent The scalar potential is given by
\begin{eqnarray}
V&=&-2S_{AB}S^{AB}+\frac{2}{3}\mc{U}_A\mc{U}^A+\frac{1}{6}\mc{N}_{iA}\mc{N}^{iA}
+\frac{1}{6}\mc{M}^{iB}_{\phantom{iB}A}\mc{M}_{iB}^{\phantom{iB}A}\nonumber \\
&=&\frac{1}{8}|C_{iA}^{\phantom{iA}B}|^2+\frac{1}{8}|C_i^{\phantom{A}PQ}|^2-\frac{1}{4}
\left(|C_A^{\phantom{A}PQ}|^2-|C_P|^2\right)
\end{eqnarray}
with $C_P=-C_{PM}^{\phantom{PM}M}$. Various components of the fermion-shift matrices are defined in terms of the ``boosted'' structure constants
\begin{equation}
{C^{\ul{\Lambda}}}_{\ul{\Pi}\ul{\Gamma}}={L_{\Lambda}}^{\ul{\Lambda}}
{(L^{-1})_{\ul{\Pi}}}^{\Pi}{(L^{-1})_{\ul{\Gamma}}}^{\Gamma}
{f_{\Pi\Gamma}}^{\Lambda}\qquad
\textrm{and}\qquad
{C_{\ul{\Lambda}}}^{\ul{\Pi}\ul{\Gamma}}=J_{\ul{\Lambda}\ul{\Lambda}'}J^{\ul{\Pi}\ul{\Pi}'}J^{\ul{\Gamma}\ul{\Gamma}'}
{(C^{\ul{\Lambda}'}}_{\ul{\Pi}'\ul{\Gamma}'})^*
\end{equation}
as
\begin{eqnarray}
S_{AB}&=&\frac{1}{4}\left(\epsilon_{BPQ}C_A^{\phantom{A}PQ}+\epsilon_{ABC}C_M^{\phantom{M}MC}\right)\nonumber \\
&=&\frac{1}{8}\left(C_A^{\phantom{A}PQ}\epsilon_{BPQ}+C_B^{\phantom{A}PQ}\epsilon_{APQ}\right),\nonumber \\
\mc{U}^A&=&-\frac{1}{4}C_M^{\phantom{A}MA},\qquad \mc{N}_{iA}=-\frac{1}{2}\epsilon_{APQ}C_i^{\phantom{A}PQ},\nonumber \\
\mc{M}_{iA}^{\phantom{iA}B}&=&\frac{1}{2}(\delta_A^BC_{iM}^{\phantom{iM}M}-2C_{iA}^{\phantom{iA}B}).
\end{eqnarray}
\indent Finally, the fermionic supersymmetry transformations obtained from the rheonomic parametrization of the
fermionic curvatures are given by\footnote{We also note an additional factor of $\frac{1}{2}$ in the gauge field strengths due to different conventions for differential forms, namely $F^\Lambda_{\textrm{here}}=\frac{1}{2}F^\Lambda_{\mu\nu}dx^\mu \wedge dx^\nu$ while $F^\Lambda_{\textrm{\cite{N3_Ferrara}}}=F^\Lambda_{\mu\nu}dx^\mu \wedge dx^\nu$.}
\begin{eqnarray}
\delta \psi_{\mu A}&=&D_\mu \epsilon_A-\epsilon_{ABC}G^B_{\mu\nu}\gamma^\nu\epsilon^C+S_{AB}\gamma_\mu \epsilon^B,\\
\delta \chi &=&-\frac{1}{4}G^A_{\mu\nu}\gamma^{\mu\nu}\epsilon_A+\mc{U}^A\epsilon_A,\\
\delta \lambda_i&=&-P_{i\mu}^{\phantom{i}A}\gamma^\mu \epsilon_A+\mc{N}_{iA}\epsilon^A,\\
\delta
\lambda_{iA}&=&-P_{i\mu}^{\phantom{i}B}\gamma^\mu\epsilon_{ABC}\epsilon^C-\frac{1}{2}G_{i\mu\nu}\gamma^{\mu\nu}\epsilon_A
+\mc{M}_{iA}^{\phantom{iA}B}\epsilon_B\, .
\end{eqnarray}
The covariant derivative for $\epsilon_A$ is defined by
\begin{equation}
D
\epsilon_A=d\epsilon_A+\frac{1}{4}\omega^{ab}\gamma_{ab}\epsilon_A+Q_A^{\phantom{A}B}\epsilon_B+\frac{1}{2}nQ\epsilon_A\,
.\label{D_epsilon}
\end{equation}
The field strengths appearing in the supersymmetry transformations are given by
\begin{eqnarray}
G^A_{\mu\nu}&=&\textrm{Re}\, a_{\Lambda\Sigma}{L_\Sigma}^AF^{+\Lambda}_{\mu\nu}=M^{AB}(L^{-1})_B^{\phantom{B}\Lambda}F^+_{\Lambda\mu\nu},\\
G^i_{\mu\nu}&=&\textrm{Re}\, a_{\Lambda\Sigma}({L_\Sigma}^i)^*F^{+\Lambda}_{\mu\nu}=-M^{ij}(L^{-1})_j^{\phantom{j}\Lambda}F^-_{\Lambda\mu\nu}
\end{eqnarray}
where $M^{ij}$ and $M^{AB}$ are respectively inverse matrices of
\begin{equation}
M_{ij}=(L^{-1})_i^{\phantom{j}\Lambda}(L^{-1})_j^{\phantom{j}\Pi}J_{\Lambda\Pi}\qquad
\textrm{and}\qquad
M_{AB}=(L^{-1})_A^{\phantom{A}\Lambda}(L^{-1})_B^{\phantom{B}\Pi}J_{\Lambda\Pi}\, .
\end{equation}

\subsection{BPS equations for supersymmetric $AdS_4$ black holes}
We now look at the BPS equations for supersymmetric $AdS_4$ black holes with the near horizon geometry given by $AdS_2\times \Sigma^2$. The metric ansatz is taken to be
\begin{equation}
ds^2=-e^{2f(r)}dt^2+dr^2+e^{2h(r)}(d\theta^2+F(\theta)^2d\phi^2)\label{metric_ansatz}
 \end{equation}  
with $F(\theta)$ defined by
\begin{equation}
F(\theta)=\sin\theta\qquad \textrm{and}\qquad F(\theta)=\sinh\theta
\end{equation}
for $\Sigma^2=S^2$ and $\Sigma^2=H^2$, respectively. The functions $f(r)$ and $h(r)$ together with all other non-vanishing fields only depend on the radial coordinate $r$. With the following choice of vielbein
\begin{equation}
e^{\hat{t}}=e^{f}dt,\qquad e^{\hat{r}}=dr,\qquad e^{\hat{\theta}}=e^{h}d\theta,\qquad e^{\hat{\phi}}=e^hF(\theta)d\phi,
\end{equation}
it is straightforward to compute non-vanishing components of the spin connection
\begin{eqnarray}
\omega^{\hat{t}\hat{r}}&=&f'e^{\hat{t}},\qquad \omega^{\hat{\theta}\hat{r}}=h'e^{\hat{\theta}},\nonumber \\
\omega^{\hat{\phi}\hat{r}}&=&h'e^{\hat{\phi}},\qquad \omega^{\hat{\theta}\hat{\phi}}=\frac{F'(\theta)}{F(\theta)}e^{-h}e^{\hat{\phi}}\, .
\end{eqnarray}
For clarity, we have used the values of flat indices as $a,b,\ldots, =(\hat{t},\hat{r},\hat{\theta},\hat{\phi})$.
\\
\indent In the present paper, we are interested in a simple $N=3$ gauged supergravity coupled to $n=3$ vector multiplets with a compact gauge group $SO(3)\times SO(3)$. The non-vanishing components of $f_{\Lambda\Sigma\Gamma}$ are given by
\begin{equation}
f_{\Lambda\Sigma}^{\phantom{\Lambda\Sigma}\Gamma}=(g_1\epsilon_{ABC},g_2\epsilon_{ijk}). \label{N3_structure_cons}
\end{equation}
We also recall that the $SO(3)\times SO(3)$ gauge group is electrically gauged with the corresponding gauge fields being the vector fields appearing in the ungauged Lagrangian with YM kinetic terms. To avoid confusion, we will call the first $SO(3)$ factor $SO(3)_R$ since this factor is embedded in $SU(3)_R$ R-symmetry. 
\\
\indent To preserve some amount of supersymmetry, we implement a topological twist by turning an $SO(2)\sim U(1)\subset SO(3)_R\subset SU(3)_R$ gauge field along $\Sigma^2$. In addition, we can also turn on an $SO(2)\subset SO(3)$ gauge field from the second $SO(3)$ factor. We will choose these gauge fields to be $A^3_\mu$ and $A^6_\mu$ with the following ansatz
\begin{eqnarray}
A^\Lambda=\tilde{q}^\Lambda(r)dt-p^\Lambda(r)F'(\theta)d\phi,\qquad \Lambda=3,6,\label{N3_A_ansatz}
\end{eqnarray}
for $F'(\theta)=\frac{dF(\theta)}{d\theta}$. The corresponding field strengths are given by
\begin{equation}
F^\Lambda=dA^\Lambda=\tilde{q}^{\Lambda'}dr\wedge dt -p^{\Lambda'} F'(\theta)dr\wedge d\phi+\kappa p^\Lambda F(\theta)d\theta\wedge d\phi\, . 
\end{equation}
 Throughout the paper, we will use $'$ to denote a derivative with respect to the radial coordinate $r$ with an exception for $F'(\theta)=\frac{dF(\theta)}{d\theta}$. In this equation, we have also introduced a parameter $\kappa$ via the relation $F''(\theta)=-\kappa F(\theta)$ with $\kappa=1$ and $\kappa=-1$ for $\Sigma^2=S^2$ and $\Sigma^2=H^2$, respectively. Imposing the Bianchi's identity $DF^\Lambda=0$ implies $p^{\Lambda'}=0$, so $p^\Lambda$ are constant and will be identified with magnetic charges.
 \\
 \indent It is useful to recall the definition of electric and magnetic charges given by
 \begin{equation}
 q_\Lambda=\frac{1}{4\pi}\int_{\Sigma^2}G_\Lambda\qquad \textrm{and}\qquad p^\Lambda =\frac{1}{4\pi}\int_{\Sigma^2}F^\Lambda\label{q_p_def}
   \end{equation}  
with $G_\Lambda=\frac{\delta S}{\delta F^\Lambda}$. 
To further fix the ansatz for the gauge fields, we consider the Lagrangian for the gauge fields
\begin{equation}
\mc{L}_{\textrm{gauge}}=-\frac{1}{2}R_{\Lambda\Sigma}*F^\Lambda \wedge F^\Sigma+\frac{1}{2}I_{\Lambda\Sigma}F^\Lambda\wedge F^\Sigma
\end{equation}
in which we have rewritten the relevant terms in the Lagrangian \eqref{N3_Lar} in differential form language. We have also used the following definition 
\begin{equation}
R_{\Lambda\Sigma}=\textrm{Re}\, a_{\Lambda\Sigma} \qquad \textrm{and}\qquad I_{\Lambda\Sigma}=\textrm{Im}\, a_{\Lambda\Sigma}\, .
\end{equation}
From the above Lagrangian, we find 
\begin{equation}
G_\Lambda=-R_{\Lambda\Sigma}*F^\Sigma+I_{\Lambda\Sigma}F^\Sigma
\end{equation}
which, together with the above definition of $(q_\Lambda,p^\Lambda)$ and $F^\Lambda_{\theta\phi}=\kappa p^\Lambda F(\theta)$, leads to 
\begin{equation}
F^\Lambda_{\hat{t}\hat{r}}=-e^{-2h}R^{\Lambda\Sigma}(\kappa I_{\Sigma\Gamma}p^\Gamma+q_\Sigma).\label{Ftr}
\end{equation}
We have written the inverse of $R_{\Lambda\Sigma}$ as $R^{\Lambda\Sigma}$. For later convenience, we also note the Maxwell equations obtained from the Lagrangian \eqref{N3_Lar} 
\begin{equation}
D_\nu(R_{\Lambda\Sigma}F^{\Sigma \mu\nu}+\frac{1}{2}I_{\Lambda\Sigma}e^{-1}\epsilon^{\mu\nu\rho\sigma}F^\Sigma_{\rho\sigma})={{P^\mu}_A}^i{(L^{-1})_i}^\Sigma {f_{\Lambda\Sigma}}^\Gamma {L_\Gamma}^A\, . \label{N3_YM}
\end{equation}
This can be rewritten in form language as 
\begin{equation}
*DG_\Lambda=P^{Ai}{(L^{-1})_i}^\Sigma {f_{\Lambda\Sigma}}^\Gamma {L_\Gamma}^A
\end{equation}
with $P^{Ai}=P^{Ai}_\mu dx^\mu$. It should also be emphasized that the left-hand side is related to a radial derivative of electric charges via the definition in \eqref{q_p_def}. Therefore, in general, electric charges are not conserved if the YM currents are non-vanishing as also pointed out in \cite{AdS4_BH5}. 

\indent We are now in a position to perform the analysis of BPS equations. The analysis is closely parallel to that in $N=2$ gauged supergravity given in \cite{AdS4_BH2} and \cite{AdS4_BH5}. We will work in Majorana representation with all $\gamma^a$ real but $\gamma_5=i\gamma^{\hat{t}}\gamma^{\hat{r}}\gamma^{\hat{\theta}}\gamma^{\hat{\phi}}$ purely imaginary. In this representation, the two chiral components $\epsilon_A$ and $\epsilon^A$ of the Killing spinors are related to each other by complex conjugation. In addition, in all of the solutions considered in this work, we assume that the Killing spinors depend only on the radial coordinate $r$. We are only interested in solutions with $SO(2)\times SO(2)$ and $SO(2)_{\textrm{diag}}$ symmetries, but in this section, we will consider the general structure of the BPS equations.  
\\
\indent We begin with the BPS equation from the variation $\delta \psi_{\hat{\phi}A}$ given by
\begin{equation}
0=\frac{1}{2}h'\gamma_{\hat{\phi}\hat{r}}\epsilon_A+\frac{1}{2}e^{-h}\frac{F'}{F}\gamma_{\hat{\phi}\hat{\theta}}\epsilon_A+{\Omega_{\hat{\phi}A}}^B\epsilon_B+S_{AB}\epsilon^B-\epsilon_{ABC}
G^B_{\hat{\phi}\hat{\theta}}\gamma^{\hat{\theta}}\epsilon^C\, .\label{dPsi_phi}
\end{equation}
The matrix $S_{AB}$ is symmetric and can be diagonalized. The corresponding eigenvalues will lead to the superpotential $\mc{W}$ in terms of which the scalar potential can be written. We then write, without summation on $A$,
\begin{equation}
S_{AB}=-\frac{1}{2}\mc{W}_A\delta_{AB}
\end{equation}
in which $\mc{W}_A$ denote eigenvalues of $S_{AB}$. It is also useful to define the central charge matrix as
\begin{equation}
\mc{Z}_{AB}=-2\epsilon_{ABC}G^C_{\hat{\theta}\hat{\phi}}=-2\epsilon_{ABC}M^{CD}{(L^{-1})_D}^\Lambda (\kappa p^\Lambda e^{-2h}-i\tilde{q}^{\Lambda'}e^{-f}).
\end{equation}
\\
\indent We now impose the following projector
\begin{equation}
\gamma^{\hat{r}}\epsilon^A=e^{-i\Lambda}\delta^{AB}\epsilon_B\qquad \textrm{or}\qquad \gamma^{\hat{r}}\epsilon_A=e^{i\Lambda}\delta_{AB}\epsilon^B \label{gamma_r_proj}
\end{equation}
and rewrite equation \eqref{dPsi_phi} as
\begin{equation}
0=\gamma_{\hat{\phi}}\left[h'e^{i\Lambda}\delta_{AB}-\mc{W}_A\delta_{AB}-Z_{AB}\gamma_{\hat{\phi}\hat{\theta}}\right]\epsilon^B+\gamma_{\hat{\phi}}\left[\frac{F'(\theta)}{F(\theta)}e^{-h}\gamma_{\hat{\phi}\hat{\theta}}\epsilon_A+2g_1{\epsilon_{AC}}^BA^C_{\hat{\phi}}\epsilon_B\right].\label{dPsi_phi_re}
\end{equation}
We have used the explicit form of ${\Omega_{\hat{\phi}_A}}^B=g_1{\epsilon_{ACD}}\delta^{DB}A^C_{\hat{\phi}}\epsilon_B$ which is valid for both cases we are interested in. We now notice that only the terms in the second bracket of equation \eqref{dPsi_phi_re} depend on $\theta$. Therefore, these terms must cancel against each other, and using the gauge field ansatz \eqref{N3_A_ansatz}, we find that
\begin{equation}
\gamma_{\hat{\phi}\hat{\theta}}\epsilon_A=2g_1{\epsilon_{AC}}^Bp^C\epsilon_B\, .
\end{equation}
Since only $p^3$ is non-vanishing, we find that the supersymmetry corresponding to $\epsilon^3$ must be broken. Imposing the twist condition
\begin{equation}
2g_1p^3=1
\end{equation}
and writting $\epsilon_{AB3}=\epsilon_{\hat{A}\hat{B}}$ for $\hat{A},\hat{B}=1,2$ and $\epsilon_{12}=1$, we obtain the following projector
\begin{equation}
\gamma_{\hat{\theta}\hat{\phi}}\epsilon_{\hat{A}}={\epsilon_{\hat{A}}}^{\hat{B}}\epsilon_{\hat{B}}\, .\label{gamma_th_ph_proj}
\end{equation}
In this analysis, we have written $\epsilon^A=(\epsilon^{\hat{A}},\epsilon^3)$. We also remark that indices $\hat{A},\hat{B},\ldots$ of $\epsilon_{\hat{A}\hat{B}}$ and $\epsilon^{\hat{A}\hat{B}}$ are simply raised and lowered by the Kronecker delta $\delta_{\hat{A}\hat{B}}$ and $\delta^{\hat{A}\hat{B}}$.
\\
\indent Using the projector \eqref{gamma_th_ph_proj} in the first bracket of \eqref{dPsi_phi_re} with $\epsilon^3=0$, we find the BPS condition
\begin{equation}
(h'e^{i\Lambda}-\mc{W}_{\hat{A}})\delta_{\hat{A}\hat{B}}-\mc{Z}_{\hat{A}\hat{C}}{\epsilon^{\hat{C}}}_{\hat{B}}=0\, . \label{N3_h_eq}
 \end{equation} 
In general, $\mc{W}_{\hat{A}}$ for a particular value of $\hat{A}$ gives the superpotential corresponding to the eigenvalue of $S_{\hat{A}\hat{B}}$ along the directions of the Killing spinors $\epsilon^{\hat{A}}$. We will simply denote this eigenvalue by $\mc{W}$. Moreover, it turns out that in the cases we will consider, only $G^3_{\mu\nu}$ is non-vanishing. We then find that 
\begin{equation}
\mc{Z}_{\hat{A}\hat{C}}{\epsilon^{\hat{C}}}_{\hat{B}}=\mc{Z}\delta_{\hat{A}\hat{B}}
\end{equation}
in which we have defined a complex number $\mc{Z}$ sometimes called the ``central charge'' as
\begin{equation}
\mc{Z}=2M^{3A}{(L^{-1})_A}^{\Lambda}(\kappa p^\Lambda e^{-2h}-i\tilde{q}^{\Lambda'}e^{-f}).
\end{equation}
With all these, we finally obtain the BPS equation from $\delta \psi_{\hat{\phi}A}$
\begin{equation}
h'e^{i\Lambda}=\mc{W}+\mc{Z}
\end{equation}
which implies
\begin{equation}
h'=\pm|\mc{W}+\mc{Z}|\qquad \textrm{and}\qquad e^{i\Lambda}=\pm\frac{\mc{W}+\mc{Z}}{|\mc{W}+\mc{Z}|}\, .\label{h_Lambda_eq}
\end{equation}
Using all of the results previously obtained, we can perform a similar analysis for $\delta\psi_{\hat{\theta}A}$. This results, as expected, in the same BPS equations given in \eqref{h_Lambda_eq}.
\\
\indent We now move to the variation $\delta \psi_{\hat{t}\hat{A}}$ of the form
\begin{eqnarray} 
0=\frac{1}{2}f'\gamma_{\hat{t}}\gamma_{\hat{r}}\epsilon_{\hat{A}}+{\mc{A}_{\hat{t}\hat{A}}}^{\hat{B}}\epsilon_{\hat{B}}+
\epsilon_{\hat{A}\hat{C}}G^3_{\hat{t}\hat{r}}
\gamma^{\hat{r}}\epsilon^{\hat{C}}+S_{\hat{A}\hat{B}}\gamma_{\hat{t}}\epsilon^{\hat{B}}\label{dPsi0}
\end{eqnarray}
with 
\begin{equation}
{\mc{A}_{\hat{t}A}}^B=-g_1\epsilon_{\hat{A}\hat{B}}A^3_{\hat{t}}\, .
\end{equation}
We then impose another projector
\begin{equation}
  \gamma^{\hat{t}}\epsilon^{\hat{A}}=ie^{-i\Lambda}\epsilon^{\hat{A}\hat{B}}\epsilon_{\hat{B}}\, .
\end{equation}
It should be noted that this is not an independent projector since it is implied by the $\gamma_{\hat{r}}$ and $\gamma_{\hat{\theta}\hat{\phi}}$ projectors given in \eqref{gamma_r_proj} and \eqref{gamma_th_ph_proj} by the relation $\gamma_5\epsilon^{\hat{A}}=-\epsilon^{\hat{A}}$.
\\
\indent We note here that the central charge matrix can also be written as 
\begin{equation}
\mc{Z}_{AB}=-2i\epsilon_{ABC}G^C_{\hat{t}\hat{r}}=2i\epsilon_{ABC}M^{CD}{(L^{-1})_D}^\Lambda (\tilde{q}^{\Lambda'}e^{-f}+i\kappa p^\Lambda e^{-2h}).
\end{equation}
With all the previous results, we can write equation \eqref{dPsi0} as
\begin{equation}
\left[f'-e^{-i\Lambda}(\mc{W}+\mc{Z})\right]\epsilon_{\hat{A}\hat{B}}-2ig_1\epsilon_{\hat{A}\hat{B}}A^3_{\hat{t}}=0
\end{equation}
which implies
\begin{eqnarray}
f'&=&\textrm{Re}\, \left[e^{-i\Lambda} (\mc{W}-\mc{Z})\right],\\ 
\textrm{and}\qquad g_1A^3_{\hat{t}}&=&-\frac{1}{2}\textrm{Im}\, \left[e^{-i\Lambda} (\mc{W}-\mc{Z})\right].
\end{eqnarray}
The second equation fixes the form of $A^3_t$.
\\
\indent Finally, we consider the variation $\delta \psi_{\hat{r}A}$ which gives
\begin{equation}
\epsilon'_{\hat{A}}-\frac{1}{2}e^{-i\Lambda}(\mc{W}-\mc{Z})\epsilon_{\hat{A}}+\frac{3}{2}Q_r\epsilon_{\hat{A}}+{Q_{r\hat{A}}}^{\hat{B}}\epsilon_{\hat{B}}=0\, .
\end{equation}
In all the cases we will consider, it turns out that $Q_r=0$ and ${Q_{r\hat{A}}}^{\hat{B}}=0$. Using $\delta \psi_{\hat{t}\hat{A}}=0$ equation, we can rewrite this equation as
\begin{equation}
\epsilon'_{\hat{A}}=\frac{1}{2}(f'-2ig_1A^3_{\hat{t}})\epsilon_{\hat{A}}
 \end{equation} 
which gives 
\begin{equation}
\epsilon_{\hat{A}}= e^{\frac{f}{2}-i\int g_1A^3_{\hat{t}}dr}\epsilon^{(0)}_{\hat{A}} 
\end{equation}
with $\epsilon^{(0)}_{\hat{A}}$ being $r$-independent spinors subject to the projectors
\begin{equation}
\gamma_{\hat{r}}\epsilon^{(0)}_{\hat{A}}=\delta_{\hat{A}\hat{B}}\epsilon^{(0)\hat{B}}\qquad \textrm{and}\qquad \gamma_{\hat{\theta}\hat{\phi}}\epsilon^{(0)}_{\hat{A}}={\epsilon_{\hat{A}}}^{\hat{B}}\epsilon^{(0)}_{\hat{B}}\, .
\end{equation}
Consistency with the projector \eqref{gamma_r_proj} leads to a flow equation for the phase $\Lambda$
\begin{equation}
\Lambda'+2ig_1A^3_{\hat{t}}=0\, .
\end{equation}
\indent Since all scalars depend only on the radial coordinate $r$, the BPS equations obtained from $\delta\chi$, $\delta\lambda_i$ and $\delta \lambda^A_i$ only involve $\gamma_{\hat{r}}$. By using the projector \eqref{gamma_r_proj} and phase factor in \eqref{h_Lambda_eq} in these variations, we eventually obtain flow equations for scalars. Before giving the solutions, we end this section with the conditions for the near horizon geometry $AdS_2\times \Sigma^2$
\begin{equation}
f'=\frac{1}{L_{AdS_2}},\qquad h'=0,\qquad ({z_i}^A)'=0
\end{equation}
meaning that the function $h$ and all scalars are constant, and $f$ is linear in $r$ in this limit. We will also choose an upper sign choice in \eqref{h_Lambda_eq} for definiteness.

\subsection{Solutions with $SO(2)\times SO(2)$ symmetry}
We now consider supersymmetric solutions to the BPS equations with the general structure given in the previous section. We begin with explicit parametrization of the $SU(3,3)/SU(3)\times SU(3)\times U(1)$ coset manifold. It is convenient to introduce a basis for $GL(6,\mathbb{R})$ matrices
\begin{equation}
(e_{\Lambda\Sigma})_{\Pi\Gamma}=\delta_{\Lambda\Pi}\delta_{\Sigma\Gamma}\,
.
\end{equation}
With the structure constants given in \eqref{N3_structure_cons}, the $SO(3)_R\times SO(3)$ gauge generators are given by
\begin{equation}
(T^{(1)}_A)_{\Pi}^{\phantom{\Pi}\Gamma}=f_{A\Pi}^{\phantom{\Lambda\Pi}\Gamma}\qquad \textrm{and} \qquad
(T^{(2)}_i)_{\Pi}^{\phantom{\Pi}\Gamma}=f_{i+3,\Pi}^{\phantom{i+3,\Pi}\Gamma}\, .
\end{equation}
\indent The residual $SO(2)\times SO(2)$ symmetry is generated by $T^{(1)}_3$ and $T^{(2)}_3$. There are two singlet scalars corresponding to the following $SU(3,3)$ non-compact generators
\begin{equation}
\hat{Y}_1=e_{36}+e_{63}\qquad \textrm{and}\qquad \hat{Y}_2=ie_{63}-ie_{36}\, .
\end{equation}
The coset representative can be written as
\begin{equation}
L=e^{\phi_1 \hat{Y}_1}e^{\phi_2 \hat{Y}_2}\, .\label{N3_SO2SO2_L}
\end{equation}
In this case, the YM currents vanish, so the electric charges are constant. The scalar potential is given by
\begin{equation}
V=-\frac{1}{2}g_1^2e^{-2\phi_1}\left[e^{2\phi_1}+\cosh2\phi_2 (1+e^{4\phi_1})\right].\label{Poten1_N3}
\end{equation}
This potential admits a unique $N=3$ supersymmetric vacuum at $\phi_1=\phi_2=0$ with the cosmological constant $V_0=-\frac{3}{2}g_1^2$. The $AdS_4$ radius is given by the relation
\begin{equation}
L_{AdS_4}=\sqrt{-\frac{3}{2V_0}}=\frac{1}{g_1}
\end{equation}
in which we have taken $g_1>0$ for convenience. We also note that truncating all vector multiplets out gives rise to pure $N=3$ gauged supergravity with $SO(3)_R$ gauge group and cosmological constant $-\frac{3}{2}g_1^2$ constructed in \cite{pureN3_1} and \cite{pureN3_2}, see also a more recent result \cite{Varela_pureN3} in which pure $N=3$ gauged supergravity is embedded in massive type IIA theory. 
\\
\indent The matrix $S_{AB}$ is given by
\begin{equation}
S_{AB}=-\frac{1}{2}\textrm{diag}(\mc{W}_1,\mc{W}_1,\mc{W}_2)
\end{equation}
in which $\mc{W}_1$ and $\mc{W}_2$ are given by
\begin{eqnarray}
\mc{W}_1&=&g_1\cosh\phi_1\cosh\phi_2,\nonumber \\
\mc{W}_2&=&g_1(\cosh\phi_1\cosh\phi_2+i\sinh\phi_1\sinh\phi_2).
\end{eqnarray}
It turns out that only $\mc{W}_2$ gives the superpotential in terms of which the scalar potential \eqref{Poten1_N3} can be written as, see more detail in \cite{N3_4D_gauging},
\begin{equation}
V=-\frac{1}{2\cosh^22\phi_2}\left(\frac{\pd W_2}{\pd\phi_1}\right)^2-\frac{1}{2}\left(\frac{\pd W_2}{\pd\phi_2}\right)^2-\frac{3}{2}W^2_2
\end{equation}
with $W_2=|\mc{W}_2|$. In this case, the supersymmetry associated with $\epsilon^{1,2}$, which are relevant to the present work, is broken. For $\phi_2=0$, $\mc{W}_1$ can give rise to the superpotential leading to unbroken supersymmetry along $\epsilon^{1,2}$, and in this case, $\mc{W}_1$ and $\mc{W}_2$ are equal. We then set $\phi_2=0$ in the following analysis. We will also write $\mc{W}=\mc{W}_1=\mc{W}_2$ and $\phi=\phi_1$. In addition, it is worth noting that setting pseudo-scalars, corresponding to imaginary parts of the complex scalars ${z_A}^i$, to zero always gives $I_{\Lambda\Sigma}=0$. This implies that the components $F^\Lambda_{\hat{t}\hat{r}}$ are given only in terms of electric charges and vanish for purely magnetic solutions.
\\ 
\indent With $\epsilon^3=0$, we find that $\delta\chi=0$ and $\delta\lambda_i=0$ identically. By using the coset representative \eqref{N3_SO2SO2_L} with $\phi_2=0$, we find a consistent BPS equation for $\phi$ from $\delta\lambda^A_i$ provided that one of these two conditions is satified
\begin{equation}
q_3=q_6=0\qquad \textrm{or}\qquad p^6=q_6=0\, .
 \end{equation} 
The first one corresponds to a purely magnetic case while the second one is a dyonic case with only $q_3$ and $p^3$ non-vanishing. 
\\
\indent Setting $q_3=q_6=0$ and using the BPS equations given in the previous section, we find the following set of BPS equations 
\begin{eqnarray}
f'&=&\frac{1}{2}e^{-\phi-2h}\left[g_1e^{2h}-g_1e^{2\phi}(p^3-p^6)\kappa +g_1e^{2(\phi+h)}-\kappa (p^3+p^6)\right],\\
h'&=&|\mc{W}+\mc{Z}|\nonumber \\
&=&\frac{1}{2}e^{-\phi-2h}\left[g_1e^{2h}+g_1e^{2\phi}(p^3-p^6)\kappa +g_1e^{2(\phi+h)}+\kappa (p^3+p^6)\right],\\
\phi'&=&-\frac{\pd |\mc{W}+\mc{Z}|}{\pd \phi}\nonumber \\
&=&\frac{1}{2}e^{-\phi-2h}\left[g_1e^{2h}-g_1e^{2\phi}(p^3-p^6)\kappa -g_1e^{2(\phi+h)}+\kappa (p^3+p^6)\right].
\end{eqnarray}  
We note that both $\mc{W}$ and $\mc{Z}$ are real giving rise to $e^{i\Lambda}= \pm 1$. The existence of $AdS_2\times \Sigma^2$ fixed points requires $p^6=0$. In this case, we find the fixed point given by
\begin{equation}
\phi=\phi_0,\qquad h=\frac{1}{2}\ln\left[-\frac{\kappa p^3}{g_1}\right],\qquad f'=\frac{1}{L_{AdS_2}}=2g_1\cosh\phi_0\label{N3_AdS2_1}
  \end{equation}  
for a constant $\phi_0$. For real $h$ and $2g_1p^3=1>0$, we need to take $\kappa=-1$, so this is an $AdS_2\times H^2$ fixed point.
\\
\indent For $p^6=q_6=0$, we find
\begin{equation}
\mc{W}+\mc{Z}=\frac{1}{2}e^{-\phi-2h}(1+e^{2\phi})(e^{2h}g_1+\kappa p^3+iq_3)
\end{equation}
leading to the BPS equations
\begin{eqnarray}
f'&=&\frac{e^{-\phi-2h}(1+e^{2\phi})(e^{4h}g_1^2-q_3^3-(p^3)^2)}{2\sqrt{(e^{2h}g_1+\kappa p^3)^2+q_3^2}},\\
h'&=&|\mc{W}+\mc{Z}|=\frac{1}{2}e^{-\phi-2h}(1+e^{2\phi})\sqrt{(\kappa p^3+g_1e^{2h})^2+q_3^2},\\
\phi'&=&-\frac{\pd|\mc{W}+\mc{Z}|}{\pd \phi}=-\frac{1}{2}e^{-\phi-2h}(e^{2\phi}-1)\sqrt{(\kappa p^3+g_1e^{2h})^2+q_3^2}
\end{eqnarray}
together with 
\begin{equation}
\tilde{q}^3=-\frac{q_3e^{-\phi+f}(1+e^{2\phi})}{2\sqrt{(e^{2h}g_1+\kappa p^3)^2+q_3^2}}
\end{equation}
which fixes the time component of the gauge field ansatz. We also note that upon using the BPS equations for $f'$, $h'$ and $\phi'$, we find
\begin{equation}
\tilde{q}^{3'}=\frac{1}{2}q_3e^{-2\phi+f-2h}(1+e^{4\phi})
\end{equation}
in agreement with the gauge field ansatz given in \eqref{Ftr}.
\\
\indent The existence of $AdS_2\times \Sigma^2$ fixed points requires $q_3=0$. This can be clearly seen from the condition $h'=0$. With $q_3=0$, the $AdS_2\times \Sigma^2$ fixed point is just the $AdS_2\times H^2$ vacuum given in \eqref{N3_AdS2_1}. We then find that all supersymmetric black hole solutions will be magnetically charged without any dyonic generalization. 
\\
\indent We now look for a solution interpolating between the supersymmetric $AdS_4$ vacuum and this $AdS_2\times H^2$ critical point. To find the relevant solution, we can further set $p^6=0$ and $q_3=0$ in the two sets of the BPS equations. In this case, the two sets lead to the same BPS equations which we repeat here for convenience
\begin{eqnarray}
f'&=&\frac{1}{2}e^{-\phi-2h}(1+e^{2\phi})(g_1e^{2h}-\kappa p^3),\\
h'&=&\frac{1}{2}e^{-\phi-2h}(1+e^{2\phi})(g_1e^{2h}+\kappa p^3),\\
\phi'&=&-\frac{1}{2}e^{-\phi-2h}(e^{2\phi}-1)(g_1e^{2h}+\kappa p^3).
\end{eqnarray}   
These equations are very similar to those given in $N=5$ and $N=6$ gauged supergravities studied in \cite{N5_flow} and \cite{N6_flow}. By a similar analysis, we can obtain an analytic solution
\begin{eqnarray}
h&=&\phi-\ln (1-e^{2\phi}),\\
f&=&\ln[\kappa p^3(1+e^{4\phi})+(g_1-2\kappa p^3)e^{2\phi}]-\ln (1-e^{2\phi})-\phi,\\
2g_1\rho&=&\ln[\kappa p^3(1+e^{4\phi})+(g_1-2\kappa p^3)e^{2\phi}]-2\ln(1-e^{2\phi})\nonumber \\
& &+2\sqrt{\frac{g_1}{4\kappa p^3-g_1}}\tan^{-1}\left[\frac{g_1+2\kappa p^3(e^{2\phi}-1)}{\sqrt{g_1(4\kappa p^3-g_1)}}\right]
\end{eqnarray}
with the new radial coordinate $\rho$ defined by $\frac{d\rho}{dr}=e^{-\phi}$.
\\
\indent As $\phi\sim 0$, we find that the solution becomes
\begin{equation}
f\sim h\sim g_1r\qquad \textrm{and}\qquad \phi\sim e^{-g_1\rho}\sim e^{-g_1r}\sim e^{-\frac{r}{L_{AdS_4}}}
\end{equation}
which is an asymptotically locally $AdS_4$ preserving the full $N=3$ supersymmetry. On the other hand, for $\phi\sim \phi_0$ with
\begin{equation}
\phi_0=\frac{1}{2}\ln\left[\frac{2\kappa p^3-g_1+\sqrt{g_1(g_1-4\kappa p^3)}}{2\kappa p^3}\right],
\end{equation}
the solution approaches the $AdS_2\times H^2$ fixed point with
\begin{equation}
h\sim \frac{1}{2}\ln \left[-\frac{\kappa p^3}{g_1}\right],\qquad \phi\sim e^{-\frac{r}{L_{AdS_2}}},\qquad f\sim \frac{r}{L_{AdS_2}}
\end{equation}
for $L_{AdS_2}=\frac{1}{g_1}\sqrt{\frac{\kappa p^3}{4\kappa p^3-g_1}}$.
\\
\indent We end this section by a comment on the solution of pure $N=3$ gauged supergravity in which we set $\phi=0$ for the entire solution. This simply gives the following solution
\begin{equation}
h=\frac{1}{2}\ln \left[\frac{e^{2g_1(r-r_0)}-\kappa p^3}{g_1}\right]\quad \textrm{and}\quad f=2g_1r-\frac{1}{2}\ln(e^{2g_1(r-r_0)}-\kappa p^3)
 \end{equation} 
for a constant $r_0$. This solution can be embedded in massive type IIA theory via $S^6$ truncation given in \cite{Varela_pureN3}. Alternatively, this solution can also be embedded in eleven dimensions using a consistent truncation on a trisasakian manifold given in \cite{N010_truncation_Cassani}.

\subsection{Solutions with $SO(2)_{\textrm{diag}}$ symmetry}
We now consider solutions with $SO(2)_{\textrm{diag}}$ symmetry generated by $T^{(1)}_3+T^{(2)}_3$. There are six singlet scalars corresponding to the following non-compact generators
\begin{eqnarray}
\bar{Y}_1&=&e_{36}+e_{63},\qquad
\bar{Y}_2=-ie_{36}+ie_{63},\nonumber \\
\bar{Y}_3&=&e_{25}+e_{52}+e_{14}+e_{41},\qquad
\bar{Y}_4=-ie_{25}+ie_{52}-ie_{14}+ie_{41},\nonumber \\
\bar{Y}_5&=&e_{15}+e_{51}-e_{24}-e_{42},\qquad
\bar{Y}_6=-ie_{15}+ie_{51}+ie_{24}-ie_{42}\, .
\end{eqnarray}
The coset representative is given by
\begin{equation}
L=e^{\phi_1\bar{Y}_1}e^{\phi_2\bar{Y}_2}e^{\phi_3\bar{Y}_3}e^{\phi_4\bar{Y}_4}e^{\phi_5\bar{Y}_5}e^{\phi_6\bar{Y}_6}\, .
\end{equation}
In this case, the scalar potential turns out to be highly complicated. We refrain from giving its explicit form here, but it is useful to note that there are two supersymmetric $AdS_4$ vacua, see more detail in \cite{N3_4D_gauging}. The first one is the $N=3$ supersymmetric $AdS_4$ vacuum with all scalars vanishing and the full $SO(3)\times SO(3)$ gauge group unbroken. This is the same as the $AdS_4$ critical point mentioned in the previous section. The second one is another $N=3$ $AdS_4$ critical point with $SO(3)_{\textrm{diag}}\subset SO(3)\times SO(3)$ symmetry given by 
\begin{equation}
\phi_1=\pm \phi_3=\frac{1}{2}\ln\left[\frac{g_2-g_1}{g_1+g_2}\right],\qquad V_0=-\frac{3}{2}\frac{g_1^2g_2^2}{g_2^2-g_1^2}
\end{equation}
with all other scalars vanishing. We now repeat the same analysis as in the previous case with an additional condition $g_2A^6=g_1 A^3$ implementing the $SO(2)_{\textrm{diag}}$ subgroup. This condition results in the same component ${Q_{\hat{\phi}A}}^B$ as in the $SO(2)\times SO(2)$ case, so the twist can be performed by the same procedure. We will not repeat all the details here to avoid repetition.
\\
\indent As in the previous case, it turns out that all pseudo-scalars must be truncated out in order to preserve supersymmetry along $\epsilon^1$ and $\epsilon^2$. Therefore, we need to set $\phi_2=\phi_4=\phi_6=0$. Consistency for the scalar equations also requires all electric charges to vanish resulting in a real phase $e^{i\Lambda}=\pm 1$. We will accordingly set $q_\Lambda=0$ and obtain the following BPS equations
\begin{eqnarray}
f'&=&\frac{1}{16g_2}e^{-\phi_1}\left[g_2e^{-2\phi_3-2\phi_5}(1+e^{4\phi_3})(1+e^{4\phi_5})[(g_1+g_2)e^{2\phi_1}+g_1-g_2] \right. \nonumber \\ 
& &\left. +4e^{-2h}[(g_1-g_2)(e^{2h}g_2+2\kappa p^3)e^{2\phi_1}+(g_1+g_2)(e^{2h}g_2-2\kappa p^3)]\right],\\
h'&=&\frac{1}{16g_2}e^{-\phi_1}\left[g_2e^{-2\phi_3-2\phi_5}(1+e^{4\phi_3})(1+e^{4\phi_5})[(g_1+g_2)e^{2\phi_1}+g_1-g_2] \right. \nonumber \\ 
& &\left. +4e^{-2h}[(g_1-g_2)(e^{2h}g_2-2\kappa p^3)e^{2\phi_1}+(g_1+g_2)(e^{2h}g_2+2\kappa p^3)]\right],\\
\phi_1'&=&-\frac{1}{16g_2}e^{-\phi_1}\left[g_2e^{-2\phi_3-2\phi_5}(1+e^{4\phi_3})(1+e^{4\phi_5})[(g_1+g_2)e^{2\phi_1}+g_2-g_1] \right.\nonumber \\
& &\left. -4e^{-2h}[(g_1+g_2)(e^{2h}g_2+2\kappa p^3)-(g_1-g_2)(e^{2h}g_2-2\kappa p^3)e^{2\phi_1}]\right],\\
\phi_3'&=&-\frac{1}{2}\textrm{sech}2\phi_5\sinh2\phi_3(g_1\cosh\phi_1+g_2\sinh\phi_1),\\
\phi_5'&=&-\frac{1}{2}\cosh2\phi_3\sinh2\phi_5(g_1\cosh\phi_1+g_2\sinh\phi_1).
\end{eqnarray}
We also note that these equations can be written more compactly as
\begin{eqnarray}
& &f'=|\mc{W}-\mc{Z}|,\qquad h'=|\mc{W}+\mc{Z}|,\qquad \phi_1'=-\frac{\pd |\mc{W}+\mc{Z}|}{\pd \phi_1},\nonumber \\
& &\phi_3'=-\frac{1}{2}\textrm{sech}^22\phi_5\frac{\pd |\mc{W}+\mc{Z}|}{\pd \phi_3},\qquad \phi_5'=-\frac{1}{2}\frac{\pd |\mc{W}+\mc{Z}|}{\pd \phi_5}\, . 
\end{eqnarray}
For $AdS_2\times \Sigma^2$ fixed points to exist, we immediately see from $\phi_3'$ and $\phi_5'$ equations that there are two possibilities; $\phi_3=\phi_5=0$ or $\phi_1=\frac{1}{2}\ln\left[\frac{g_2-g_1}{g_2+g_1}\right]$. However, both of these choices do not lead to any $AdS_2\times \Sigma^2$ fixed point, so there are no supersymmetric $AdS_4$ black holes with $SO(2)_{\textrm{diag}}$ symmetry. 
\\
\indent At this point, it should be noted that similar BPS equations have been considered in \cite{N3_SU2_SU3} with more vector multiplets ($n=8$), and a number of $AdS_2\times \Sigma^2$ fixed points have been given. A truncation of that results to three vector multiplets can be performed resulting in the BPS equations given above. It is worth pointing out here that there is a sign error in the BPS equations considered in \cite{N3_SU2_SU3} regarding to the contribution of the gauge fields to the supersymmetry transformations. The corresponding equations from the present analysis are correct and compatible with the second-order field equations. Therefore, the $AdS_2\times \Sigma^2$ fixed points with $SO(2)_{\textrm{diag}}\times SO(2)$ symmetry found in \cite{N3_SU2_SU3} do not exist.

\section{$AdS_4$ black holes from $N=4$ gauged supergravity}\label{N4_SUGRA}
In this section, we repeat the same analysis as in the previous section for matter-coupled $N=4$ gauged supergravity. Unlike the $N=3$ gauged supergravity considered in the previous section, gaugings of $N=4$ supergravity that can give rise to supersymmetric $AdS_4$ vacua need to be dyonic, involving both electric and magnetic vector fields. However, there always exists a symplectic frame in which the resulting gaugings are purely electric. As in the previous section, we will begin with a review of $N=4$ gauged supergravity coupled to $n$ vector multiplets.

\subsection{Matter-coupled $N=4$ gauged supergravity} 
Unlike the $N=3$ gauged supergravity, $N=4$ gauged supergravity has completely been constructed in the embedding tensor formalism in \cite{N4_gauged_SUGRA}. We will mainly follow the construction and notation used in \cite{N4_gauged_SUGRA}. 
\\
\indent Similar to the $N=3$ theory, $N=4$ supersymmetry in four
dimensions only allows for the graviton and vector multiplets. Unlike $N=3$ supersymmetry, the graviton multiplet in $N=4$ supersymmetry does contain scalars with the full field content given by
\begin{equation}
(e^{\hat{\mu}}_\mu,\psi^i_\mu,A_\mu^m,\chi^i,\tau).
\end{equation}
The component fields are given by the graviton
$e^{\hat{\mu}}_\mu$, four gravitini $\psi^i_\mu$, six vectors
$A_\mu^m$, four spin-$\frac{1}{2}$ fields $\chi^i$ and one complex
scalar $\tau$ parametrizing the $SL(2,\mathbb{R})/SO(2)$ coset. In this case, indices $m,n=1,\ldots, 6$ and
$i,j=1,2,3,4$ respectively describe the vector and chiral spinor
representations of the $SO(6)_R\sim SU(4)_R$ R-symmetry. The former is equivalent to a second-rank anti-symmetric tensor representation of $SU(4)_R$. Furthermore, in this section, we denote flat space-time indices by $\hat{\mu},\hat{\nu},\ldots$ to avoid confusion with indices labeling the vector multiplets to be introduced later.
\\
\indent As in the $N=3$ theory, the supergravity multiplet can couple to an arbitrary number $n$ of vector multiplets. Each vector multiplet will be labeled by indices $a,b=1,\ldots, n$ and contain the following field content
\begin{equation}
(A^a_\mu,\lambda^{ia},\phi^{ma})
\end{equation}
corresponding to vector fields $A^a_\mu$, gaugini $\lambda^{ia}$ and scalars $\phi^{ma}$. The $6n$ scalar fields can be described by $SO(6,n)/SO(6)\times SO(n)$ coset. We also note the well-known fact that the field contents of the vector multiplet in $N=3$ and $N=4$ supersymmetries are the same.
\\
\indent All fermionic fields and supersymmetry parameters that transform in the fundamental representation of $SU(4)_R$ R-symmetry are subject to the chirality projections
\begin{equation}
\gamma_5\psi^i_\mu=\psi^i_\mu,\qquad \gamma_5\chi^i=-\chi^i,\qquad \gamma_5\lambda^i=\lambda^i\, .
\end{equation}
Similarly, the conjugate fields transforming in the anti-fundamental
representation of $SU(4)_R$ satisfy
\begin{equation}
\gamma_5\psi_{\mu i}=-\psi_{\mu i},\qquad \gamma_5\chi_i=\chi_i,\qquad \gamma_5\lambda_i=-\lambda_i\, .
\end{equation}
\indent The most general gaugings of the matter-coupled $N=4$ supergravity
can be efficiently described by the embedding tensor $\Theta$. There are two components of the embedding tensor $\xi^{\alpha M}$ and $f_{\alpha MNP}$ with $\alpha=(+,-)$ and $M,N=(m,a)=1,\ldots,
n+6$ denoting respectively fundamental representations of $SL(2,\mathbb{R})\times SO(6,n)$ global symmetry. The electric vector fields $A^{M+}=(A^m_\mu,A^a_\mu)$ together with their magnetic dual $A^{M-}$, collectively denoted by $A^{M\alpha}$, form a doublet of $SL(2,\mathbb{R})$. The existence of $AdS_4$ vacua requires $\xi^{\alpha M}=0$ \cite{AdS4_N4_Jan}, so we will consider gaugings with only $f_{\alpha MNP}$ non-vanishing and set $\xi^{\alpha M}$ to zero from now on. 
\\
\indent The embedding tensor implements the minimal coupling to various fields via the covariant derivative
\begin{equation}
D_\mu=\nabla_\mu-gA_\mu^{M\alpha}f_{\alpha M}^{\phantom{\alpha
M}NP}t_{NP}
\end{equation}
where $\nabla_\mu$ is the space-time covariant derivative including (possibly) the spin connections. $t_{MN}$ denote $SO(6,n)$ generators which can be chosen as
\begin{equation}
(t_{MN})_P^{\phantom{P}Q}=2\delta^Q_{[M}\eta_{N]P},\label{SO6_n_generator}
\end{equation}
with $\eta_{MN}=\textrm{diag}(-1,-1,-1,-1,-,1-,1,1,1,\ldots,1)$ being the $SO(6,n)$ invariant tensor. The gauge
coupling constant $g$ can also be absorbed in the definition of the embedding tensor $f_{\alpha MNP}$. 
\\
\indent In addition to $\xi^{\alpha M}=0$, the existence of $AdS_4$ vacua requires the gaugings to be dyonic involving both electric and magnetic vector fields. In this case, both $A^{M+}$ and $A^{M-}$
enter the Lagrangian, and $f_{\alpha MNP}$ with $\alpha=\pm$ are non-vanishing. Consistency requires the presence of two-form fields when magnetic vector fields are included. In the case of $\xi^{\alpha M}=0$, the two-forms transform as an anti-symmetric
tensor under $SO(6,n)$ and will be denoted by $B^{MN}_{\mu\nu}=B^{[MN]}_{\mu\nu}$. The two-forms are also needed to define covariant gauge field strengths given by
\begin{equation}
\mc{H}^{M\pm}=dA^{M\pm}-\frac{1}{2}\eta^{MQ}f_{\alpha QNP}A^{N\alpha}\wedge A^{P\pm}\pm\frac{1}{2}\eta^{MQ}f_{\mp QNP}B^{NP}\, .
\end{equation}
In particular, for non-vanishing $f_{-MNP}$ the electric field strengths $\mc{H}^{M+}$ acquire a contribution from the two-form
fields.
\\
\indent The scalar coset manifold $SL(2,\mathbb{R})/SO(2)$ in the graviton multiplet can be described by a coset representative
\begin{equation}
\mc{V}_\alpha=\frac{1}{\sqrt{\textrm{Im} \tau}}\left(
                                         \begin{array}{c}
                                           \tau \\
                                           1 \\
                                         \end{array}
                                       \right)\label{Valpha}
\end{equation}
or equivalently by a symmetric matrix
\begin{equation}
M_{\alpha\beta}=\textrm{Re} (\mc{V}_\alpha\mc{V}^*_\beta)=\frac{1}{\textrm{Im}
\tau}\left(
                                    \begin{array}{cc}
                                      |\tau|^2 & \textrm{Re} \tau \\
                                      \textrm{Re} \tau & 1 \\
                                    \end{array}
                                  \right).
\end{equation}
We also note the relation $\textrm{Im}(\mc{V}_\alpha\mc{V}^*_\beta)=\epsilon_{\alpha\beta}$. The complex scalar $\tau$ can in turn be written in terms of the dilaton $\phi$ and the axion $\chi$ as
\begin{equation}
\tau=\chi+ie^\phi\, .
\end{equation}
\indent For the $SO(6,n)/SO(6)\times SO(n)$ coset from vector multiplets, we introduce the
coset representative $\mc{V}_M^{\phantom{M}A}$ transforming by left and right multiplications under $SO(6,n)$ and $SO(6)\times SO(n)$, respectively. The $SO(6)\times SO(n)$ index will be split as
$A=(m,a)$ according to which the coset representative can be written as
\begin{equation}
\mc{V}_M^{\phantom{M}A}=(\mc{V}_M^{\phantom{M}m},\mc{V}_M^{\phantom{M}a}).
\end{equation}
Being an element of $SO(6,n)$, the matrix $\mc{V}_M^{\phantom{M}A}$ satisfies the relation
\begin{equation}
\eta_{MN}=-\mc{V}_M^{\phantom{M}m}\mc{V}_N^{\phantom{M}m}+\mc{V}_M^{\phantom{M}a}\mc{V}_N^{\phantom{M}a}\,
.
\end{equation}
The $SO(6,n)/SO(6)\times SO(n)$ coset can also be parametrized in terms of a symmetric matrix
defined by
\begin{equation}
M_{MN}=\mc{V}_M^{\phantom{M}m}\mc{V}_N^{\phantom{M}m}+\mc{V}_M^{\phantom{M}a}\mc{V}_N^{\phantom{M}a}
\end{equation}
with a manifest $SO(6)\times SO(n)$ invariance.
\\
\indent The bosonic Lagrangian of the $N=4$ gauged supergravity for $\xi^{\alpha M}=0$ is given by
\begin{eqnarray}
e^{-1}\mc{L}&=&\frac{1}{2}R+\frac{1}{16}\mc{D}_\mu M_{MN}\mc{D}^\mu
M^{MN}-\frac{1}{4(\textrm{Im}\tau)^2}\pd_\mu \tau \pd^\mu \tau^*-V\nonumber \\
& &-\frac{1}{4}\textrm{Im}\,\tau M_{MN}\mc{H}^{M+}_{\mu\nu}\mc{H}^{N+\mu\nu}-\frac{1}{8}\textrm{Re}\,\tau e^{-1}\epsilon^{\mu\nu\rho\sigma}\eta_{MN}\mc{H}^{M+}_{\mu\nu}\mc{H}^{N+}_{\rho\sigma}\nonumber \\
& &-\frac{1}{2}\left[f_{-MNP}A^{M-}_\mu A^{N+}_\nu\pd_\rho A^{P-}_\sigma +\frac{1}{4}f_{\alpha MNR}f_{\beta PQS}\eta^{RS}A^{M\alpha}_\mu A^{N+}_\nu A^{P\beta}_\rho A^{Q-}_\sigma \right.\nonumber \\
& &-\frac{1}{4}f_{-MNP}B^{NP}_{\mu\nu}\left(2\pd_\rho A^{M-}_\sigma -\frac{1}{2}\eta^{MS}f_{\alpha SQR}A^{Q\alpha}_\rho A^{R-}_\sigma\right)\nonumber \\
& &\left.-\frac{1}{16}f_{+MNR}f_{-PQS}\eta^{RS}B^{MN}_{\mu\nu}B^{PQ}_{\rho\sigma}
\right]e^{-1}\epsilon^{\mu\nu\rho\sigma}
\end{eqnarray}
where $e$ is the vielbein determinant. 
\\
\indent The scalar potential is given by
\begin{eqnarray}
V&=&\frac{g^2}{16}\left[f_{\alpha MNP}f_{\beta
QRS}M^{\alpha\beta}\left[\frac{1}{3}M^{MQ}M^{NR}M^{PS}+\left(\frac{2}{3}\eta^{MQ}
-M^{MQ}\right)\eta^{NR}\eta^{PS}\right]\right.\nonumber \\
& &\left.-\frac{4}{9}f_{\alpha MNP}f_{\beta
QRS}\epsilon^{\alpha\beta}M^{MNPQRS}\right]
\end{eqnarray}
where $M^{MN}$ is the inverse of $M_{MN}$, and $M^{MNPQRS}$ is
defined by
\begin{equation}
M_{MNPQRS}=\epsilon_{mnpqrs}\mc{V}_{M}^{\phantom{M}m}\mc{V}_{N}^{\phantom{M}n}
\mc{V}_{P}^{\phantom{M}p}\mc{V}_{Q}^{\phantom{M}q}\mc{V}_{R}^{\phantom{M}r}\mc{V}_{S}^{\phantom{M}s}\label{M_6}
\end{equation}
with indices raised by $\eta^{MN}$. The covariant derivative of $M_{MN}$ is defined by
\begin{equation}
\mc{D}M_{MN}=dM_{MN}+2A^{P\alpha}\eta^{QR}f_{\alpha QP(M}M_{N)R}\, .
\end{equation}
\indent The magnetic vectors and two-form fields do not have kinetic terms. They are auxiliary fields and enter the Lagrangian through topological terms. The corresponding field equations give rise to the duality relation between
two-forms and scalars and the electric-magnetic duality between $A^{M+}$ and $A^{M-}$, respectively. The field equations resulting from varying the Lagrangian with respect to $A^{M\pm}_\mu$ and $B^{MN}_{\mu\nu}$ are given by
\begin{eqnarray}
\eta_{MN}*\mc{D}\mc{H}^{N-}&=&-\frac{1}{4}{f_{+MP}}^NM_{NQ}\mc{D}M^{QP},\label{YM1}\\
\eta_{MN}*\mc{D}\mc{H}^{N+}&=&\frac{1}{4}{f_{-MP}}^NM_{NQ}\mc{D}M^{QP},\label{YM2}\\
\mc{H}^{M-}&=&\textrm{Im}\, \tau M^{MN}\eta_{NP}*\mc{H}^{P+}-\textrm{Re}\, \tau\mc{H}^{M+}\label{YM3}
\end{eqnarray}
written in differential form language for computational convenience. By substituting $\mc{H}^{M-}$ from
\eqref{YM3} in \eqref{YM1}, we obtain the usual Yang-Mills equations
for $\mc{H}^{M+}$ while equation \eqref{YM2} simply gives the
relation between the Hodge dual of the three-form field strengths and
the scalars due to the usual Bianchi identity of the gauge field
strengths defined by
\begin{equation}
\mc{F}^{M\pm}=dA^{M\pm}-\frac{1}{2}\eta^{MQ}f_{\alpha QNP}A^{N\alpha}\wedge A^{P\pm}\, .
\end{equation}
\indent The supersymmetry transformations of
fermionic fields are given by
\begin{eqnarray}
\delta\psi^i_\mu &=&2D_\mu \epsilon^i-\frac{2}{3}gA^{ij}_1\gamma_\mu
\epsilon_j+\frac{i}{4}(\mc{V}_\alpha)^*{\mc{V}_M}^{ij}\mc{H}^{M\alpha}_{\nu\rho}\gamma^{\nu\rho}\gamma_\mu\epsilon_j,\\
\delta \chi^i &=&-\epsilon^{\alpha\beta}\mc{V}_\alpha D_\mu
\mc{V}_\beta\gamma^\mu \epsilon^i-\frac{4}{3}igA_2^{ij}\epsilon_j+\frac{1}{2}\mc{V}_\alpha{\mc{V}_M}^{ij}\mc{H}^{M\alpha}_{\mu\nu}\gamma^{\mu\nu}\epsilon_j,\\
\delta \lambda^i_a&=&2i\mc{V}_a^{\phantom{a}M}D_\mu
\mc{V}_M^{\phantom{M}ij}\gamma^\mu\epsilon_j-2igA_{2aj}^{\phantom{2aj}i}\epsilon^j
-\frac{1}{4}\mc{V}_\alpha\mc{V}_{Ma}\mc{H}^{M\alpha}_{\mu\nu}\gamma^{\mu\nu}\epsilon^i
\end{eqnarray}
with the fermion shift matrices defined by
\begin{eqnarray}
A_1^{ij}&=&\epsilon^{\alpha\beta}(\mc{V}_\alpha)^*\mc{V}_{kl}^{\phantom{kl}M}\mc{V}_N^{\phantom{N}ik}
\mc{V}_P^{\phantom{P}jl}f_{\beta M}^{\phantom{\beta M}NP},\nonumber
\\
A_2^{ij}&=&\epsilon^{\alpha\beta}\mc{V}_\alpha\mc{V}_{kl}^{\phantom{kl}M}\mc{V}_N^{\phantom{N}ik}
\mc{V}_P^{\phantom{P}jl}f_{\beta M}^{\phantom{\beta M}NP},\nonumber
\\
A_{2ai}^{\phantom{2ai}j}&=&\epsilon^{\alpha\beta}\mc{V}_\alpha
{\mc{V}_a}^M{\mc{V}_{ik}}^N\mc{V}_P^{\phantom{P}jk}f_{\beta
MN}^{\phantom{\beta MN}P}
\end{eqnarray}
where $\mc{V}_M^{\phantom{M}ij}$ is defined in terms of the `t Hooft
matrices $G^{ij}_m$ and $\mc{V}_M^{\phantom{M}m}$ as
\begin{equation}
\mc{V}_M^{\phantom{M}ij}=\frac{1}{2}\mc{V}_M^{\phantom{M}m}G^{ij}_m
\end{equation}
and similarly for its inverse
\begin{equation}
{\mc{V}_{ij}}^M=-\frac{1}{2}{\mc{V}_m}^M(G^{ij}_m)^*\,
.
\end{equation}
We note that $G^{ij}_m$ satisfy the relations
\begin{equation}
G_{mij}=(G^{ij}_m)^*=\frac{1}{2}\epsilon_{ijkl}G^{kl}_m\, .
\end{equation}
We will choose the explicit form of these matrices as follows
\begin{eqnarray}
G_1^{ij}&=&\left[
                     \begin{array}{cccc}
                       0 & 1 & 0 & 0 \\
                       -1 & 0 & 0 & 0 \\
                       0 & 0 & 0  & 1\\
                        0 & 0 & -1  & 0\\
                     \end{array}
                   \right],\qquad
G_2^{ij}=\left[
                     \begin{array}{cccc}
                       0 & 0 & 1 & 0\\
                       0 & 0 & 0 & -1\\
                       -1 & 0 & 0  & 0\\
                        0 & 1 & 0  & 0\\
                     \end{array}
                   \right],\nonumber \\
G_3^{ij}&=&\left[
                     \begin{array}{cccc}
                       0 & 0 & 0 & 1\\
                       0 & 0 & 1 & 0\\
                       0 & -1 & 0  & 0\\
                        -1 & 0 & 0  & 0\\
                     \end{array}
                   \right],\qquad
G_4^{ij}=\left[
                     \begin{array}{cccc}
                       0 & i & 0 & 0\\
                       -i & 0 & 0 & 0\\
                       0 & 0 & 0  & -i\\
                        0 & 0 & i  & 0\\
                     \end{array}
                   \right],\nonumber \\
G_5^{ij}&=&\left[
                     \begin{array}{cccc}
                       0 & 0 & i & 0\\
                       0 & 0 & 0 & i\\
                       -i & 0 & 0  & 0\\
                        0 & -i & 0  & 0\\
                     \end{array}
                   \right],\qquad
G_6^{ij}=\left[
                     \begin{array}{cccc}
                       0 & 0 & 0 & i\\
                       0 & 0 & -i & 0\\
                       0 & i & 0  & 0\\
                        -i & 0 & 0  & 0\\
                     \end{array}
                   \right].  
\end{eqnarray}
The covariant derivative of $\epsilon^i$ is given by
\begin{equation}
D_\mu \epsilon^i=\pd_\mu \epsilon^i+\frac{1}{4}{\omega_\mu}^{\hat{\mu}\hat{\nu}}\gamma_{\hat{\mu}\hat{\nu}}\epsilon^i+{Q_{\mu j}}^i\epsilon^j\, .
\end{equation}
\indent Finally, it should be noted that the scalar potential can be written in terms of $A_1$ and $A_2$ tensors as
\begin{equation}
V=-\frac{1}{3}A^{ij}_1A_{1ij}+\frac{1}{9}A^{ij}_2A_{2ij}+\frac{1}{2}A_{2ai}^{\phantom{2ai}j}
A_{2a\phantom{i}j}^{\phantom{2a}i}
\end{equation}
which is usually referred to as supersymmetric Ward's identity. We also recall that upper and lower $i,j,\ldots$ indices are related by complex conjugation.
\\
\indent We end this section by giving some relations which are very useful in deriving the BPS equations in subsequent analysis. With the explicit form of $\mc{V}_\alpha$ given in
\eqref{Valpha} and equation \eqref{YM3}, it is straightforward to
derive the following identities
\begin{eqnarray}
i\mc{V}_\alpha{\mc{V}_M}^{ij}\mc{H}^{M\alpha}_{\mu\nu}\gamma^{\mu\nu}&=&-(\mc{V}_-)^{-1}{\mc{V}_M}^{ij}\mc{H}^{M+}_{\mu\nu}\gamma^{\mu\nu}(1-\gamma_5),\\
i\mc{V}_\alpha{\mc{V}_M}^a\mc{H}^{M\alpha}_{\mu\nu}\gamma^{\mu\nu}&=&-(\mc{V}_-)^{-1}{\mc{V}_M}^a\mc{H}^{M+}_{\mu\nu}\gamma^{\mu\nu}(1+\gamma_5),\\
i(\mc{V}_\alpha)^*{\mc{V}_M}^{ij}\mc{H}^{M\alpha}_{\mu\nu}\gamma^{\mu\nu}\gamma_\rho
&=&(\mc{V}_-)^{-1}{\mc{V}_M}^{ij}\mc{H}^{M+}_{\mu\nu}\gamma^{\mu\nu}\gamma_\rho(1-\gamma_5)
\end{eqnarray}
in which we have used the following relations for
the $SO(6,n)$ coset representative \cite{Eric_N4_4D}
\begin{eqnarray}
\eta_{MN}&=&-\frac{1}{2}\epsilon_{ijkl}{\mc{V}_M}^{ij}{\mc{V}_N}^{kl}+{\mc{V}_M}^{a}{\mc{V}_N}^{a},\qquad {\mc{V}_M}^a{\mc{V}_{ij}}^M=0,\nonumber \\
{\mc{V}_M}^{ij}{\mc{V}_{kl}}^{M}&=&-\frac{1}{2}(\delta^i_k\delta^j_l-\delta^i_l\delta^j_k),\qquad {\mc{V}_M}^a{\mc{V}_b}^M=\delta^a_b\, .
\end{eqnarray}
It should be noted that these relations are slightly different from those given in
\cite{N4_gauged_SUGRA} due to a different convention on $\mc{V}_\alpha$ in terms of the scalar
$\tau$ namely $\mc{V}_\alpha$ used in this paper satisfies $\mc{V}_+/\mc{V}_-=\tau$ while that used in
\cite{N4_gauged_SUGRA} gives $\mc{V}_+/\mc{V}_-=\tau^*$. 

\subsection{Solutions with $SO(2)\times SO(2)\times SO(2)\times SO(2)$ symmetry}
In this paper, we are interested in $N=4$ gauged supergravity with $n=6$ vector multiplets and $SO(4)\times SO(4)\sim SO(3)\times SO(3)\times SO(3)\times SO(3)$ gauge group. The corresponding embedding tensor takes the following form \cite{dS_Roest}
\begin{eqnarray}
f_{+\hat{m}\hat{n}\hat{p}}&=&g_1\epsilon_{\hat{m}\hat{n}\hat{p}},\qquad f_{+\hat{a}\hat{b}\hat{c}}=\tilde{g}_1\epsilon_{\hat{a}\hat{b}\hat{c}},\nonumber \\
f_{-\tilde{m}\tilde{n}\tilde{p}}&=&g_2\epsilon_{\tilde{m}\tilde{n}\tilde{p}},\qquad f_{-\tilde{a}\tilde{b}\tilde{c}}=\tilde{g}_2\epsilon_{\tilde{a}\tilde{b}\tilde{c}}\, .\label{embedding_tensor}
\end{eqnarray}
We have used the convention on the $SO(6,6)$ index $M=(m,a)=(\hat{m},\tilde{m},\hat{a},\tilde{a})$ with $\hat{m}=1,2,3$, $\tilde{m}=4,5,6$, $\hat{a}=7,8,9$ and $\tilde{a}=10,11,12$. The two $SO(4)$ factors are electrically and magnetically embedded in $SO(6,6)$ and will be denoted by $SO(4)_+\times SO(4)_-$. In terms of the $SO(3)$ factors corresponding to the embedding tensor in \eqref{embedding_tensor}, we will write the gauge group as $SO(3)_+\times SO(3)_-\times SO(3)_+\times SO(3)_-$ with the first two factors embedded in the $SU(4)_R\sim SO(6)_R$.  
\\
\indent We now consider solutions preserving $SO(2)\times SO(2)\times SO(2)\times SO(2)$ symmetry. To proceed further, we first give an explicit parametrization of the $SO(6,6)/SO(6)\times SO(6)$ coset. The scalar sector of $SO(2)\times SO(2)\times SO(2)\times SO(2)$ singlets have already been studied recently in \cite{N4_Janus}. We will mostly take various results from \cite{N4_Janus} in which more details can be found. By using $SO(6,6)$ generators in the fundamental representation of the form given in \eqref{SO6_n_generator}, we can identify the $SO(6,6)$ non-compact generators as
\begin{equation}
Y_{ma}=t_{m,a+6}\, .
\end{equation}
There are four $SO(2)\times SO(2)\times SO(2)\times SO(2)$ singlet scalars from the $SO(6,6)/SO(6)\times SO(6)$ coset. With the $SO(2)\times SO(2)\times SO(2)\times SO(2)$ generators chosen to be $X_{+3}$, $X_{-6}$, $X_{+9}$ and $X_{-12}$, the non-compact generators corresponding to these singlets are given by $Y_{33}$, $Y_{36}$, $Y_{63}$ and $Y_{66}$ in terms of which the coset representative can be written as
\begin{equation}
\mc{V}=e^{\phi_1 Y_{33}}e^{\phi_2 Y_{36}}e^{\phi_3 Y_{63}}e^{\phi_4 Y_{66}}\, .\label{SO2_4_coset}
\end{equation}
Together with the dilaton and axion, there are six scalars in the $SO(2)\times SO(2)\times SO(2)\times SO(2)$ sector. The scalar potential for these singlet scalars is given by
\begin{equation}
V=-\frac{1}{2}e^{-\phi}(g_1^2+e^{2\phi}g_2^2+g_2^2\chi^2)-2g_1g_2\cosh\phi_1\cosh\phi_2\cosh\phi_3\cosh\phi_4\label{Potential}
\end{equation}
which admits a unique $AdS_4$ critical point at
\begin{equation}
\phi=\ln\left[\frac{g_1}{g_2}\right]\qquad \textrm{and}\qquad \phi_1=\phi_2=\phi_3=\phi_4=\chi=0
\end{equation}
with the cosmological constant and $AdS_4$ radius given by
\begin{equation}
V_0=-3g_1g_2\qquad \textrm{and}\qquad L=\sqrt{-\frac{3}{V_0}}=\frac{1}{\sqrt{g_1g_2}}\, .
\end{equation}
This $AdS_4$ vacuum preserves $N=4$ supersymmetry and the full $SO(4)\times SO(4)$ symmetry. We can also choose $g_2=g_1=g$, by shifting the dilaton, to make the dilaton vanish at this critical point. Holographic RG flows and Janus solutions in this sector have been extensively studied in \cite{N4_Janus}. In the present work, we look for supersymmetric $AdS_4$ black holes with the horizons of $AdS_2\times \Sigma^2$ geometry. The analysis is parallel to the $N=3$ case considered in the previous section with some modifications to incorporate the magnetic gauge fields. Similar analyses can be found in \cite{Guarino_AdS2_1,Guarino_AdS2_2,Klemm_symplectic} and \cite{Trisasakian_AdS2} in the contexts of $N=2$ and $N=4$ gauged supergravities, respectively. We will closely follow the procedure in \cite{Trisasakian_AdS2}.
\\
\indent We first consider the ansatz for $SO(2)\times SO(2)\times SO(2)\times SO(2)$ gauge fields of the form
\begin{eqnarray}
A^{M+}&=&A^M_tdt-p^MF'(\theta)d\phi,\qquad M=3,6,9,12\\
A^{M-}&=&\tilde{A}^M_tdt-e_MF'(\theta)d\phi\, .
\end{eqnarray}
We also note that the gauge fields participating in the $SO(4)\times SO(4)$ gauging are given by $A^{3+}$, $A^{6-}$, $A^{9+}$ and $A^{12-}$ while the above ansatz includes all of their electric-magnetic duals. The ansatz for relevant two-form fields is given by 
\begin{eqnarray}
B^{12}&=&b_3(r)F(\theta)d\theta \wedge d\phi,\qquad B^{45}=b_6(r)F(\theta)d\theta \wedge d\phi,\nonumber \\
B^{78}&=&b_9(r)F(\theta)d\theta \wedge d\phi,\qquad B^{10,11}=b_{12}(r)F(\theta)d\theta \wedge d\phi\, .
\end{eqnarray}
The metric ansatz is still given by \eqref{metric_ansatz}. In addition, to avoid some confusion and make various expressions less cumbersome, we will denote the magnetic charges with a subscript, $p^M=(p_3,p_6,p_9,p_{12})$. 
\\
\indent With the embedding tensor \eqref{embedding_tensor}, it is straightforward to compute the covariant gauge field strengths 
\begin{eqnarray}
\mc{H}^{3+}&=&A^{3'}_tdr\wedge dt+\kappa p_3 F(\theta)d\theta \wedge d\phi,\nonumber \\
\mc{H}^{3-}&=&\tilde{A}^{3'}_tdr\wedge dt+(\kappa e_3-g_1b_3) F(\theta)d\theta \wedge d\phi,\nonumber \\
\mc{H}^{6+}&=&A^{6'}_tdr\wedge dt+(\kappa p_6+\tilde{g}_1b_6) F(\theta)d\theta \wedge d\phi,\nonumber \\
\mc{H}^{6-}&=&\tilde{A}^{6'}_tdr\wedge dt+\kappa e_6 F(\theta)d\theta \wedge d\phi,\nonumber \\
\mc{H}^{9+}&=&A^{9'}_tdr\wedge dt+\kappa p_9 F(\theta)d\theta \wedge d\phi,\nonumber \\
\mc{H}^{9-}&=&\tilde{A}^{9'}_tdr\wedge dt+(\kappa e_9-g_2b_9) F(\theta)d\theta \wedge d\phi,\nonumber \\
\mc{H}^{12+}&=&A^{12'}_tdr\wedge dt+(\kappa p_{12}+\tilde{g}_2b_{12}) F(\theta)d\theta \wedge d\phi,\nonumber \\
\mc{H}^{12-}&=&\tilde{A}^{12'}_tdr\wedge dt+\kappa e_{12} F(\theta)d\theta \wedge d\phi\, .
\end{eqnarray}
In this $SO(2)\times SO(2)\times SO(2)\times SO(2)$ sector, it turns out that all components of YM current are zero
\begin{equation}
{f_{\pm MP}}^NM_{NQ}DM^{QP}=0\, .
\end{equation}
Equations \eqref{YM1} and \eqref{YM2} then imply that $D\mc{H}^{M\pm}=0$. Therefore, we find that all the fields $b_i(r)$ and electric charges $e_{i}$ are constant.
\\
\indent As pointed out in \cite{N4_Janus}, supersymmetric solutions with $SO(2)\times SO(2)\times SO(2)\times SO(2)$ symmetry can arise from two possibilities, $\chi=\phi_2=\phi_3=0$ or $\chi=\phi_1=\phi_4=0$. For definiteness, we will choose the first possibility. Choosing the second one results in relabeling the scalars. With $\textrm{Re}\, \tau=\chi=0$, equations \eqref{YM3} gives
\begin{eqnarray}
A^{3'}_t&=&e^{f-\phi-2h}\left[(\kappa e_3-g_1b_3)\cosh 2\phi_1+(\kappa e_9-g_2b_9)\sinh 2\phi_1\right],\nonumber\\
A^{6'}_t&=&\kappa e^{f-\phi-2h}(e_6\cosh 2\phi_4+e_{12}\sinh 2\phi_4),\nonumber \\
A^{9'}_t&=&-e^{f-\phi-2h}\left[(\kappa e_9-g_2b_9)\cosh 2\phi_1+(\kappa e_3-g_1b_3)\sinh 2\phi_1\right],\nonumber\\
A^{12'}_t&=&-\kappa e^{f-\phi-2h}(e_{12}\cosh 2\phi_4+e_{6}\sinh 2\phi_4),\nonumber \\
\tilde{A}^{3'}_t&=&-\kappa e^{f+\phi-2h}(p_3\cosh 2\phi_1+p_{9}\sinh 2\phi_1),\nonumber \\
\tilde{A}^{6'}_t&=&-e^{f+\phi-2h}\left[(\kappa p_6+\tilde{g}_1b_6)\cosh 2\phi_4+(\kappa p_{12}+\tilde{g}_2b_{12})\sinh 2\phi_4\right],\nonumber\\
\tilde{A}^{9'}_t&=&\kappa e^{f+\phi-2h}(p_9\cosh 2\phi_1+p_{3}\sinh 2\phi_1),\nonumber \\
\tilde{A}^{12'}_t&=&e^{f+\phi-2h}\left[(\kappa p_{12}+\tilde{g}_2b_{12})\cosh 2\phi_4+(\kappa p_6+\tilde{g}_1b_6)\sinh 2\phi_4\right].
\end{eqnarray}
All these relations fix the ansatz for the $H^{M\alpha}_{0r}$ components of the field strengths in terms of scalars and various charges.
\\
\indent We now consider topological twists along $\Sigma^2$. The scalar coset representative \eqref{SO2_4_coset} gives the composite connection of the form
\begin{equation}
{Q_{\mu i}}^j=\frac{1}{2}g_1A^{3+}_\mu{(i\sigma_2\otimes \sigma_1)_i}^j+\frac{1}{2}g_2A^{6-}_\mu {(\sigma_1\otimes i\sigma_2)_i}^j\label{N4_composite}
\end{equation}
with $\sigma_a$, $a=1,2,3$, are usual Pauli matrices. To perform a twist, we consider relevant terms in the variation $\delta\psi^i_{\hat{\phi}}$ of the form
\begin{equation}
\frac{1}{2}e^{-h}\frac{F'(\theta)}{F(\theta)}\left[\gamma_{\hat{\phi}\hat{\theta}}\delta_j^i-g_1p_3{(i\sigma_2\otimes \sigma_1)_j}^i-g_2e_6{(\sigma_1\otimes i\sigma_2)_j}^i\right]\epsilon^j=0\, .\label{twist_N4}
\end{equation}
There are a few possibilities to satisfy this condition. These are given by the following two main categories:
\begin{itemize}   
\item $N=4$ twists: By setting either $p_3=0$ or $e_6=0$, all four $\epsilon^i$ can be non-vanishing. These two choices lead to the following twist conditions and projectors
\begin{eqnarray}
e_6=0;& &\qquad g_1p_3=1,\qquad \gamma_{\hat{\theta}\hat{\phi}}\epsilon^i={(i\sigma_2\otimes \sigma_1)^i}_j\epsilon^j,\\
p_3=0;& &\qquad g_2e_6=1,\qquad \gamma_{\hat{\theta}\hat{\phi}}\epsilon^i={(\sigma_1\otimes \sigma_2)^i}_j\epsilon^j\, .  
\end{eqnarray}  
We will refer to these two cases as $N=4$ twists which have a similar structure to the $N=3$ theory.
\item $N=2$ twists: By using the relation
\begin{equation}
(\sigma_3\otimes \sigma_3)(\sigma_1\otimes i\sigma_2)=(\sigma_1\otimes i\sigma_2)(\sigma_3\otimes \sigma_3)=i\sigma_2\otimes \sigma_1,
\end{equation}
we can rewrite the condition \eqref{twist_N4} as
\begin{equation}
\gamma_{\hat{\phi\hat{\theta}}}\epsilon^i=\left[g_1p_3{(\sigma_3\otimes \sigma_3)_j}^k+g_2e_6\delta_j^k\right]{(\sigma_1\otimes i\sigma_2)_k}^i\epsilon^j\, .
\end{equation}
This can be solved by imposing the following conditions
\begin{equation}
g_1p_3+g_2e_6=1,\qquad \gamma_{\hat{\theta}\hat{\phi}}\epsilon^i={(\sigma_1\otimes i\sigma_2)^i}_j\epsilon^j, \qquad {(\sigma_3\otimes \sigma_3)^i}_j\epsilon^j=\epsilon^i\, .\label{twist_N2_N4}
\end{equation}
The last projector simply sets $\epsilon^2=\epsilon^3=0$ reducing half of the original supersymmetry. Accordingly, we will call this case $N=2$ twists.  
\end{itemize}
We also note that the situation is very similar to $AdS_5$ black strings in five-dimensional $N=4$ gauged supergravity considered in \cite{5Dtwist}. In addition, the two possibilities of $N=4$ twists correspond to the $H$-twist and $C$-twist of the dual $N=4$ SCFT in three dimensions considered in \cite{H_C_twist}.
\\
\indent By a similar analysis performed in the $N=3$ theory, we find a general structure of the BPS equations given by
\begin{eqnarray}
h'=|\mc{W}+\mc{Z}|\qquad\textrm{and}\qquad f'=\textrm{Re}\, [e^{-i\Lambda}(\mc{W}-\mc{Z})]
\end{eqnarray}
together with an algebraic constraint
\begin{equation}
 g_1A^3_t+g_2A^6_t=e^f\textrm{Im}\, [e^{-i\Lambda}(\mc{W}-\mc{Z})].\label{N4_constraint}
\end{equation}
In these equations, $\mc{W}$ is the superpotential obtained from the eigenvalue of the $A^{ij}_1$ tensor along the Killing spinors, and $\mc{Z}$ is the central charge as in the previous section. We have also imposed the following projector
\begin{equation}
\gamma_{\hat{r}}\epsilon_i=e^{i\Lambda} \delta_{ij}\epsilon^j\qquad \textrm{with}\qquad e^{i\Lambda}=\frac{\mc{W}+\mc{Z}}{|\mc{W}+\mc{Z}|}\, .
\end{equation}
Using this projector in the supersymmetry transformations $\delta\chi^i$ and $\delta\lambda^i_a$ leads to the BPS equations for scalars in the gravity and vector multiplets, respectively.

\subsubsection{Solutions with $N=4$ twists}
We begin with the case of $N=4$ twist by $A^{3+}$. In addition to setting $e_6=0$, unbroken $N=4$ supersymmetry also requires
\begin{equation}
b_6=b_{12}=e_{12}=p_6=p_{12}=0\, .
\end{equation}
Moreover, consistency of the scalar equations imposes further conditions of the form 
\begin{equation}
e_3=e_9=b_3=b_9=0\, .
\end{equation}
All these lead to the following set of consistent BPS equations
\begin{eqnarray}
f'&=&|\mc{W}-\mc{Z}|\nonumber \\
&=&\frac{1}{2}e^{-\frac{\phi}{2}}\left[g_2\cosh\phi_1+e^\phi g_2\cosh\phi_4-\kappa e^{\phi-2h}(p_3\cosh\phi_1+p_9\sinh\phi_1)\right],\qquad\\
h'&=&|\mc{W}+\mc{Z}|\nonumber \\
&=&\frac{1}{2}e^{-\frac{\phi}{2}}\left[g_2\cosh\phi_1+e^\phi g_2\cosh\phi_4+\kappa e^{\phi-2h}(p_3\cosh\phi_1+p_9\sinh\phi_1)\right],\qquad\\
\phi_1'&=&-2\frac{\pd |\mc{W}+\mc{Z}|}{\pd \phi_1}\nonumber \\
&=&-e^{-2h-\frac{\phi}{2}}\left[e^\phi \kappa(p_3\sinh\phi_1+p_9\cosh\phi_1)+e^{2h}g_1\sinh\phi_1\right],\\
\phi_4'&=&-2\frac{\pd |\mc{W}+\mc{Z}|}{\pd \phi_4}=-g_2e^{\frac{\phi}{2}}\sinh\phi_4,\\
\phi'&=&-4\frac{\pd |\mc{W}+\mc{Z}|}{\pd \phi}\nonumber \\
&=&e^{-\frac{\phi}{2}}\left[g_1\cosh\phi_1-e^{\phi-2h}(g_2e^{2h}\cosh\phi_4+\kappa p_3\cosh\phi_1+\kappa p_9\sinh\phi_1)\right].\qquad
\end{eqnarray}
However, there do not exist any $AdS_2\times \Sigma^2$ fixed points in these equations.
\\
\indent We then look at the case of $N=4$ twist by $A^{6-}$ in which consistency similarly requires the following conditions
\begin{equation}
b_3=b_9=e_3=e_9=p_9=b_6=b_{12}=p_6=p_{12}=0\, .
\end{equation}
The BPS equations are given by
\begin{eqnarray}
f'&=&\frac{1}{2}e^{-\frac{\phi}{2}}\left[g_1\cosh\phi_1+e^{\phi}g_2\cosh\phi_4-\kappa e^{-2h}(e_6\cosh\phi_4+ e_{12}\sinh\phi_4)\right],\qquad\\
h&=&\frac{1}{2}e^{-\frac{\phi}{2}}\left[g_1\cosh\phi_1+e^{\phi}g_2\cosh\phi_4+\kappa e^{-2h}(e_6\cosh\phi_4+ e_{12}\sinh\phi_4)\right],\qquad\\
\phi_1'&=&-g_1e^{-\frac{\phi}{2}}\sinh\phi_1,\\
\phi_4'&=&-e^{-2h-\frac{\phi}{2}}[(e^{2h+\phi}+\kappa e_6)\sinh\phi_4+\kappa e_{12}\cosh\phi_4],\\
\phi'&=&e^{-2h-\frac{\phi}{2}}\left[e^{2h}g_1\cosh\phi_1+(\kappa e_6-g_2e^{2h+\phi})\cosh\phi_4+\kappa e_{12}\sinh\phi_4\right]
\end{eqnarray}
which do not admit any $AdS_2\times \Sigma^2$ fixed points as in the case of $A^{3+}$ twist.

\subsubsection{Solutions with $N=2$ twists}
We now move to a more interesting and more complicated case of $N=2$ twists by both $A^{3+}$ and $A^{6-}$. The resulting BPS conditions are much more involved than those in the previous case. However, we are able to find a number of solutions for special values of electric and magnetic charges. 
\begin{itemize}
\item Solutions from pure $N=4$ gauged supergravity
\end{itemize}
We will begin with a simple case of pure $N=4$ gauged supergravity with $\phi_1=\phi_4=0$ and $A^{9+}=A^{12-}=0$. 
\\
\indent In this case, the constraint \eqref{N4_constraint} requires $e_3=p_6=0$, and we find
\begin{eqnarray}
\mc{W}&=&\frac{1}{2}e^{-\frac{\phi}{2}}[g_1+g_2e^\phi+ig_2\chi],\\
\mc{Z}&=&\frac{1}{2}e^{-\frac{\phi}{2}-2h}\kappa [e_6+p_3e^\phi+ip_3\chi].
\end{eqnarray}
We then find the following BPS equations
\begin{eqnarray}
\chi'&=&-4e^{2\phi}\frac{\pd |\mc{W}+\mc{Z}|}{\pd\chi}=-\frac{e^{-4h+\phi}(\kappa g_2e^{2h}+p_3)^2\chi}{|\mc{W}+\mc{Z}|},\\
\phi'&=&-4\frac{\pd |\mc{W}+\mc{Z}|}{\pd \phi}\nonumber \\
&=&\frac{e^{-4h-\phi}}{2|\mc{W}+\mc{Z}|}\left[(e_6+\kappa g_1e^{2h})^2-(\kappa g_2e^{2h}+p_3)^2(e^{2\phi}-\chi^2)\right],\\
h'&=&|\mc{W}+\mc{Z}|,\\
f'&=&\frac{e^{-4h-\phi}}{4|\mc{W}+\mc{Z}|}\left[e^{4h}(g_1+g_2e^\phi)^2-(e_6+p_3e^\phi)^2+(e^{4h}g_2^2-p_3^2)^2\chi^2\right]
\end{eqnarray}
with 
\begin{equation}
|\mc{W}+\mc{Z}|=\frac{1}{2}e^{-2h-\frac{\phi}{2}}\sqrt{[e^{2h}(g_1+e^{\phi}g_2)+\kappa e_6+\kappa p_3e^\phi]^2+(e^{2h}g_2+\kappa p_3)^2\chi^2}\, .
\end{equation}
\indent From these equations, we find an $AdS_2\times H^2$ fixed point given by
\begin{equation}
h=\frac{1}{2}\ln\left[-\frac{\kappa p_3}{g_2}\right],\qquad L_{AdS_2}=\frac{1}{g_1e^{-\frac{\phi_0}{2}}+g_2e^{\frac{\phi_0}{2}}}\label{AdS2_N4_1}
\end{equation}
for constants $\phi=\phi_0$ and $\chi=\chi_0$ provided that $g_2e_6=g_1p_3$. We note that for $\chi=0$, the above BPS equations and the $AdS_2\times H^2$ fixed point are the same as those considered in \cite{flow_acrossD_bobev} with an appropriate change of symplectic frame to purely electric $SO(4)$ gauge group. We have slightly generalized the equations in \cite{flow_acrossD_bobev} by including a non-vanishing axion. We now give the flow solutions interpolating between the $AdS_4$ vacuum and the $AdS_2\times H^2$ geometry. Before giving explicit solutions, we first simplify the expressions by setting $g_2=g_1$ according to which the twist condition gives $p_3=e_6=\frac{1}{2g_1}$. 
\\
\indent For $\chi=0$ and $\kappa=-1$, we find a much simpler set of BPS equations
\begin{eqnarray}
\phi'&=&-e^{-2h-\frac{\phi}{2}}(e^\phi-1)(e^{2h}g_1-p_3),\\
h'&=&\frac{1}{2}e^{-2h-\frac{\phi}{2}}(1+e^{\phi})(e^{2h}g_1-p_3),\\
f'&=&\frac{1}{2}e^{-2h-\frac{\phi}{2}}(1+e^{\phi})(e^{2h}g_1+p_3).
\end{eqnarray}
These equations take a very similar form to those of $N=5,6$ gauged supergravities and $N=3$ gauged supergravity given in the previous section. We then expect that the resulting solutions are related to each other by truncations of $N=6$ gauged supergravity to gauged supergravities with lower amounts of supersymmetry. The solution is given by
\begin{eqnarray}
g_1(r-r_0)&=&\tanh^{-1}\sqrt{\frac{1+\cosh\phi}{2}}\nonumber \\
& &-2\sqrt{\frac{p_3}{g_1+4p_3}}\tanh^{-1}\sqrt{\frac{2p_3(1+\cosh\phi)}{g_1+4p_3}},\\
h&=&\frac{\phi}{2}-\ln(1-e^\phi),\\
f&=&\ln\left[p_3(1+e^{2\phi})-(g_1+2p_3)e^\phi\right]-\ln (1-e^\phi)-\frac{\phi}{2}\, .
\end{eqnarray}
This solution flows to the $AdS_2\times H^2$ fixed point \eqref{AdS2_N4_1} for $\phi_0$ given by 
\begin{equation}
\phi_0=\ln\left[\sqrt{\frac{g_1(g_1+4p_3)}{4p_3^2}}+\frac{g_1}{2p_3}+1\right].
\end{equation}
\indent For $\chi\neq 0$, we have the BPS equations 
\begin{eqnarray}
f'&=&\frac{1}{2}e^{-2h-\frac{\phi}{2}}(e^{2h}g_1+p_3)\sqrt{(1+e^\phi)^2+\chi^2},\\
h'&=&\frac{1}{2}e^{-2h-\frac{\phi}{2}}(e^{2h}g_1-p_3)\sqrt{(1+e^\phi)^2+\chi^2},\\
\phi'&=&\frac{e^{-2h-\frac{\phi}{2}}(e^{2h}g_1-p_3)(1-e^{2\phi}+\chi^2)}{\sqrt{(1+e^\phi)^2+\chi^2}},\\
\chi'&=&-\frac{2e^{-2h+\frac{3\phi}{2}}(e^{2h}g_1-p_3)\chi}{\sqrt{(1+e^\phi)^2+\chi^2}}
\end{eqnarray}
with the solution given by
\begin{eqnarray}
\phi&=&\frac{1}{2}\ln(1-\chi^2+C_0\chi),\\
f&=&\ln(e^{2h}g_1-p_3)-h,\\
h&=&\frac{1}{8}\ln\left[\frac{1+C_0\chi -\chi^2}{\chi^4}\right]+\frac{1}{4}\ln\left[2+C_0\chi+2\sqrt{1+C_0\chi-\chi^2}\right]
\end{eqnarray}
for a constant $C_0$. However, we are not able to find an analytic solution for $\chi(r)$. The solution flows to the $AdS_2\times H^2$ fixed point if 
\begin{equation}
\chi_0=\frac{g_1\tilde{C}[\tilde{C}^2-2p_3^2+g_1C_0\tilde{C}]}{2(g_1^2\tilde{C}^2+p_3^4)}
\end{equation}
with $\tilde{C}=\sqrt{g_1^2(4+C_0^2)+4p_3^2}$.
\begin{itemize}
\item Solutions from matter-coupled $N=4$ gauged supergravity
\end{itemize}
We now consider solutions from matter-coupled $N=4$ gauged supergravity with $\phi_1,\phi_4\neq 0$. Consistency for setting $\phi_2=\phi_3=0$ in $\delta\lambda_a^i$ conditions also requires setting $A^{9+}=0$. The residual symmetry of the solutions in this case is then enhanced to $SO(2)\times SO(2)\times SO(3)\times SO(2)$. With all these, we find two sets of BPS equations consistent with the constraint \eqref{N4_constraint}. These are given by
\begin{eqnarray}
i&:&\qquad \chi=\phi_2=0,\qquad e_3=p_6=p_{12}=0,\\
ii&:&\qquad \chi=\phi_4=0,\qquad e_3=p_6=e_{12}=0\, .
\end{eqnarray} 
$\square$ Case $i$:
\\
\indent In this case, we find the following BPS equations
\begin{eqnarray}
f'&=&\frac{1}{4}e^{-2h-\frac{\phi}{2}-\phi_4}\left[e^\phi[e^{2h}g_2(1+e^{2\phi_4})-2\kappa p_3e^{\phi_4}]\right. \nonumber \\
& &\left.+2g_1e^{2h+\phi_4}+\kappa e_{12}(1-e^{2\phi_4})-\kappa e_6 (1+e^{2\phi_4}) \right],\\
h'&=&\frac{1}{4}e^{-2h-\frac{\phi}{2}-\phi_4}\left[e^\phi[e^{2h}g_2(1+e^{2\phi_4})+2\kappa p_3e^{\phi_4}]\right. \nonumber \\
& &\left.+2g_1e^{2h+\phi_4}-\kappa e_{12}(1-e^{2\phi_4})+\kappa e_6 (1+e^{2\phi_4}) \right],\\
\phi'&=&-\frac{1}{4}e^{-2h-\frac{\phi}{2}-\phi_4}\left[e^{2h+\phi}g_2-2g_1e^{2h+\phi_4}+e^{2h+\phi+2\phi_4}g_2 \right. \nonumber \\
& &\left.+\kappa (e_{12}-e_6)-\kappa (e_{12}+e_6)e^{2\phi_4}+2\kappa p_3 e^{\phi+\phi_4} \right],\\
\phi_4'&=&-\frac{1}{2}e^{-2h-\frac{\phi}{2}-\phi_4}\left[e^{2h+\phi}g_2(e^{2\phi_4}-1)+\kappa (e_{12}-e_6)+\kappa (e_{12}+e_6)e^{2\phi_4}\right].\quad
\end{eqnarray}
There is a family of $AdS_2\times \Sigma^2$ fixed points given by
\begin{eqnarray}
\phi&=&\ln\left[\frac{(1+e^{2\phi_4})[e_{12}(1+e^{2\phi_4})+e_6(e^{2\phi_4}-1)]}{2p_3(e^{2\phi_4}-1)}\right]-\phi_4,\nonumber \\
h&=&\frac{\phi_4}{2}-\frac{1}{2}\ln\left[-\frac{g_2(1+e^{2\phi_4})}{2\kappa p_3}\right],\nonumber \\
\phi_4&=&\frac{1}{2}\ln\left[\frac{2g_1p_3-e_6g_2+\sqrt{e_{12}^2g_2^2+4g_1p_3(g_1p_3-g_2e_6)}}{g_2(e_{12}+e_6)}\right].
\end{eqnarray}
It can be verified that for appropriate values of the parameters, this critical point is valid for both $\kappa=1$ and $\kappa=-1$ resulting in a class of $AdS_2\times S^2$ and $AdS_2\times H^2$ geometries. Since $p_{12}=0$ in this case, the solutions carry only electric charges of $A^{12-}$. 
\\
\indent Examples of solutions interpolating between $AdS_4$ and $AdS_2\times H^2$ vacua with
\begin{equation}
g_2=g_1=1,\qquad p_3=\frac{3}{2},\qquad \kappa=-1
\end{equation}
and $e_{12}=1,2,3$ are shown in figure \ref{Fig1}. We also note that the value of $e_6$ is fixed by the twist condition $g_1(p_3+e_6)=1$.
\\
\indent A number of interpolating solutions between $AdS_4$ and $AdS_2\times S^2$ critical points are shown in figure \ref{Fig2} with the following numerical values 
\begin{equation}
g_2=g_1=1,\qquad p_3=-2,\qquad \kappa=1
\end{equation}
and $e_{12}=4,6,8$.\\
$\square$ Case $ii$:

\begin{figure}
  \centering
  \begin{subfigure}[b]{0.45\linewidth}
    \includegraphics[width=\linewidth]{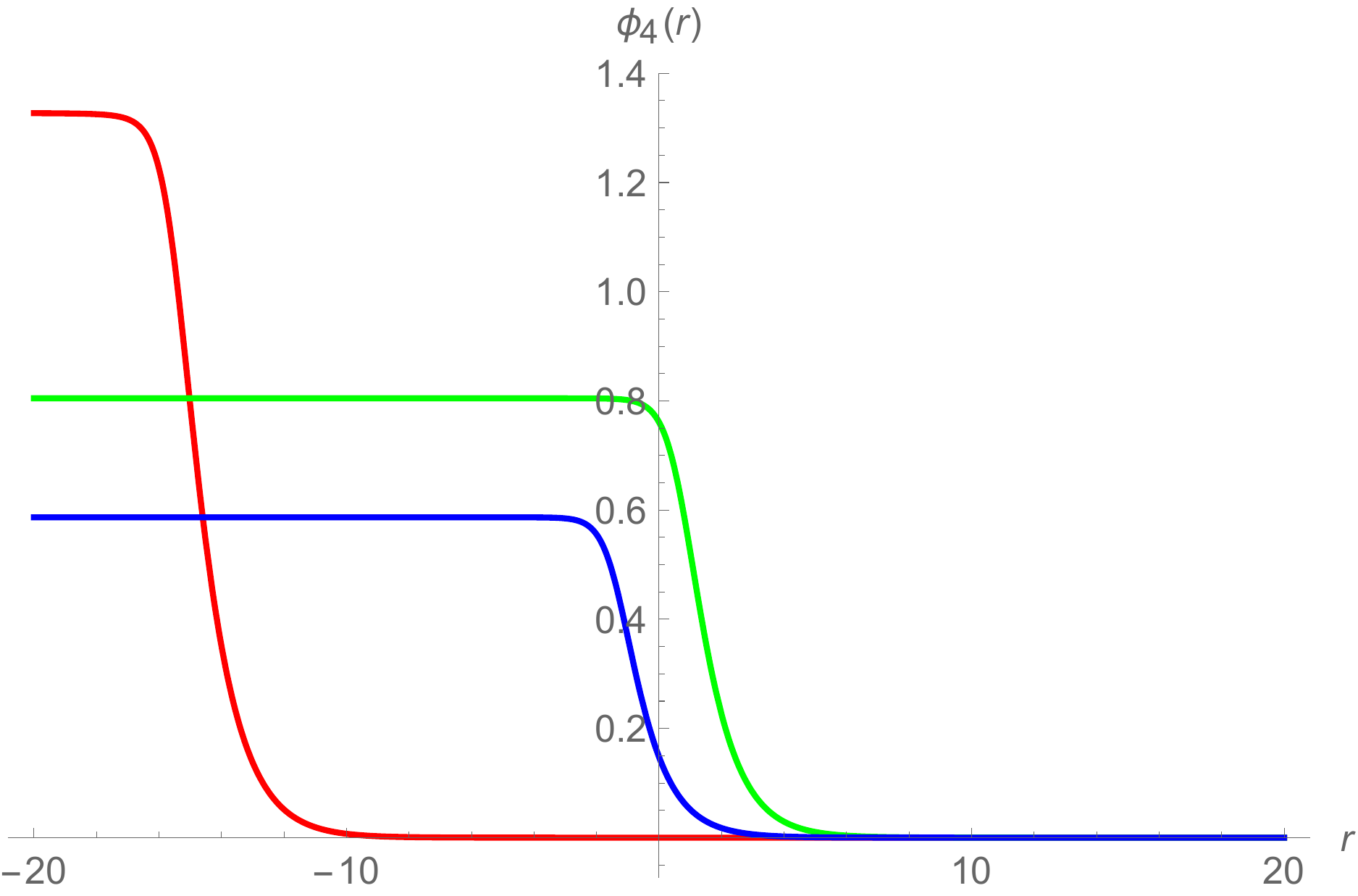}
  \caption{$\phi_4(r)$ solution}
  \end{subfigure}
  \begin{subfigure}[b]{0.45\linewidth}
    \includegraphics[width=\linewidth]{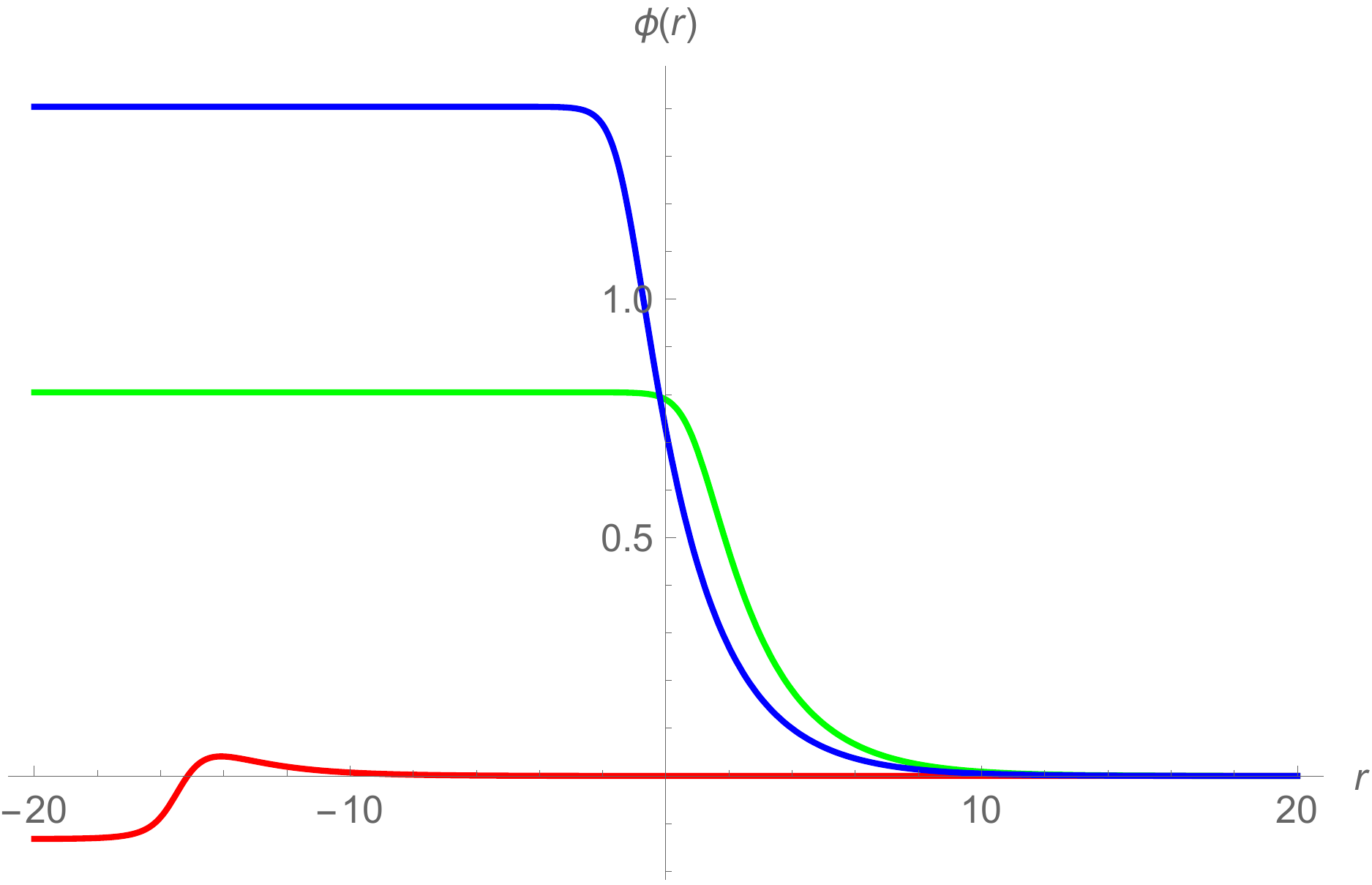}
  \caption{$\phi(r)$ solution}
  \end{subfigure}\\
   \begin{subfigure}[b]{0.45\linewidth}
    \includegraphics[width=\linewidth]{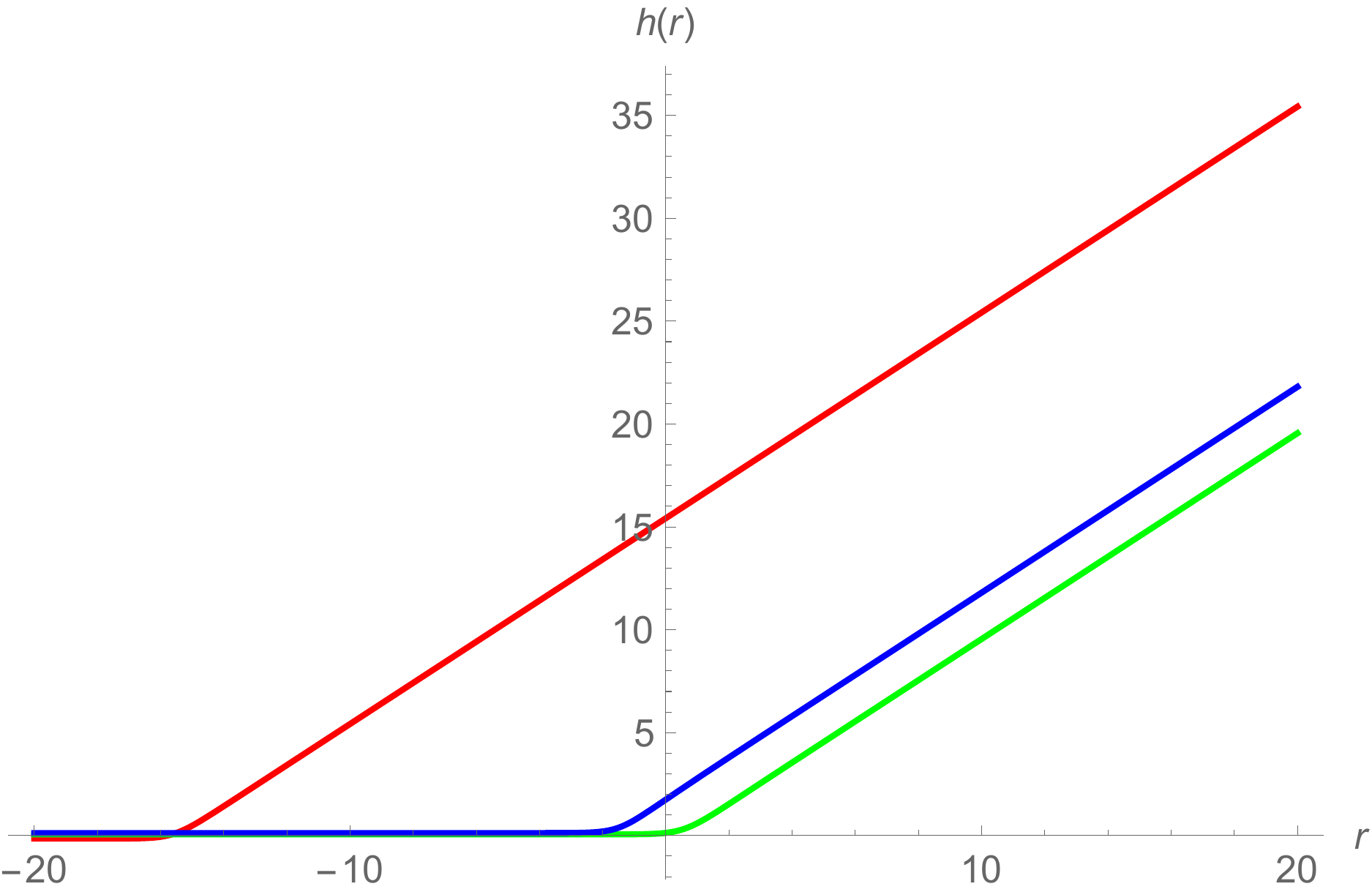}
  \caption{$h(r)$ solution}
   \end{subfigure} 
 \begin{subfigure}[b]{0.45\linewidth}
    \includegraphics[width=\linewidth]{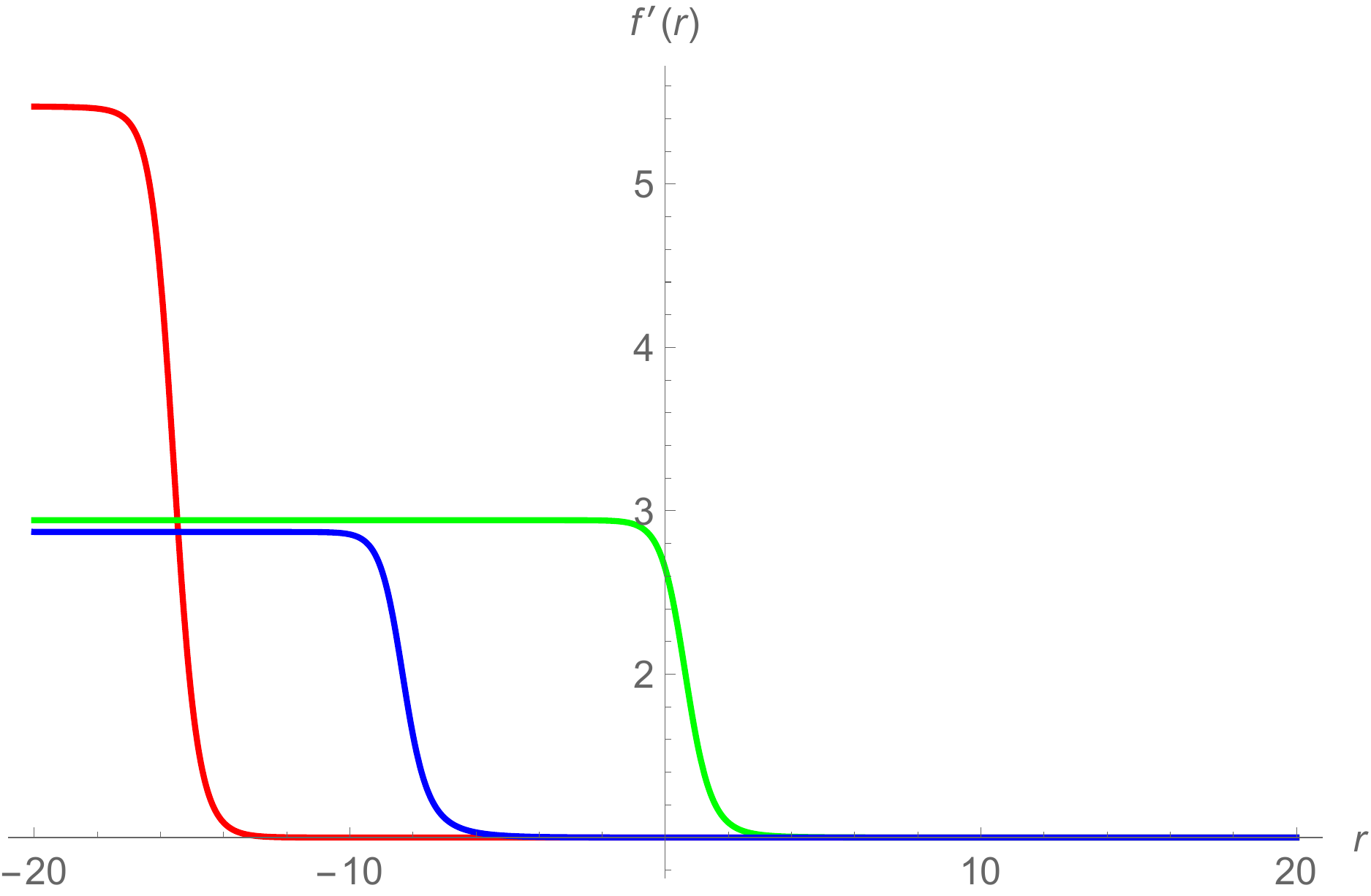}
  \caption{$f'(r)$ solution}
   \end{subfigure} 
  \caption{Supersymmetric $AdS_4$ black holes with $AdS_2\times H^2$ horizon for $g_2=g_1=1$, $p_3=\frac{3}{2}$, $\kappa=-1$ and $e_{12}=1 (\textrm{red}), 2 (\textrm{green}), 3 (\textrm{blue})$.}
  \label{Fig1}
\end{figure}

\begin{figure}
  \centering
  \begin{subfigure}[b]{0.45\linewidth}
    \includegraphics[width=\linewidth]{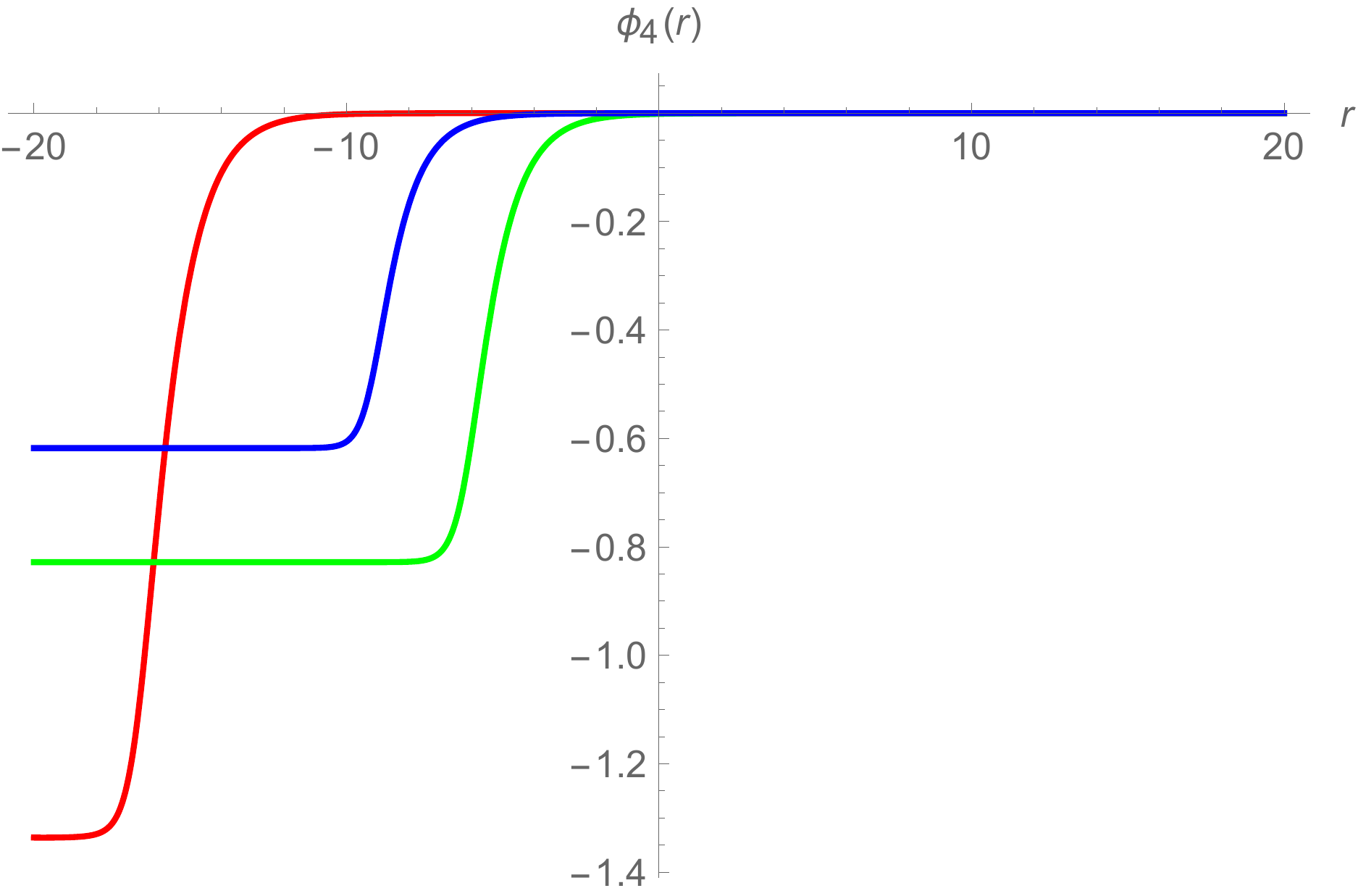}
  \caption{$\phi_4(r)$ solution}
  \end{subfigure}
  \begin{subfigure}[b]{0.45\linewidth}
    \includegraphics[width=\linewidth]{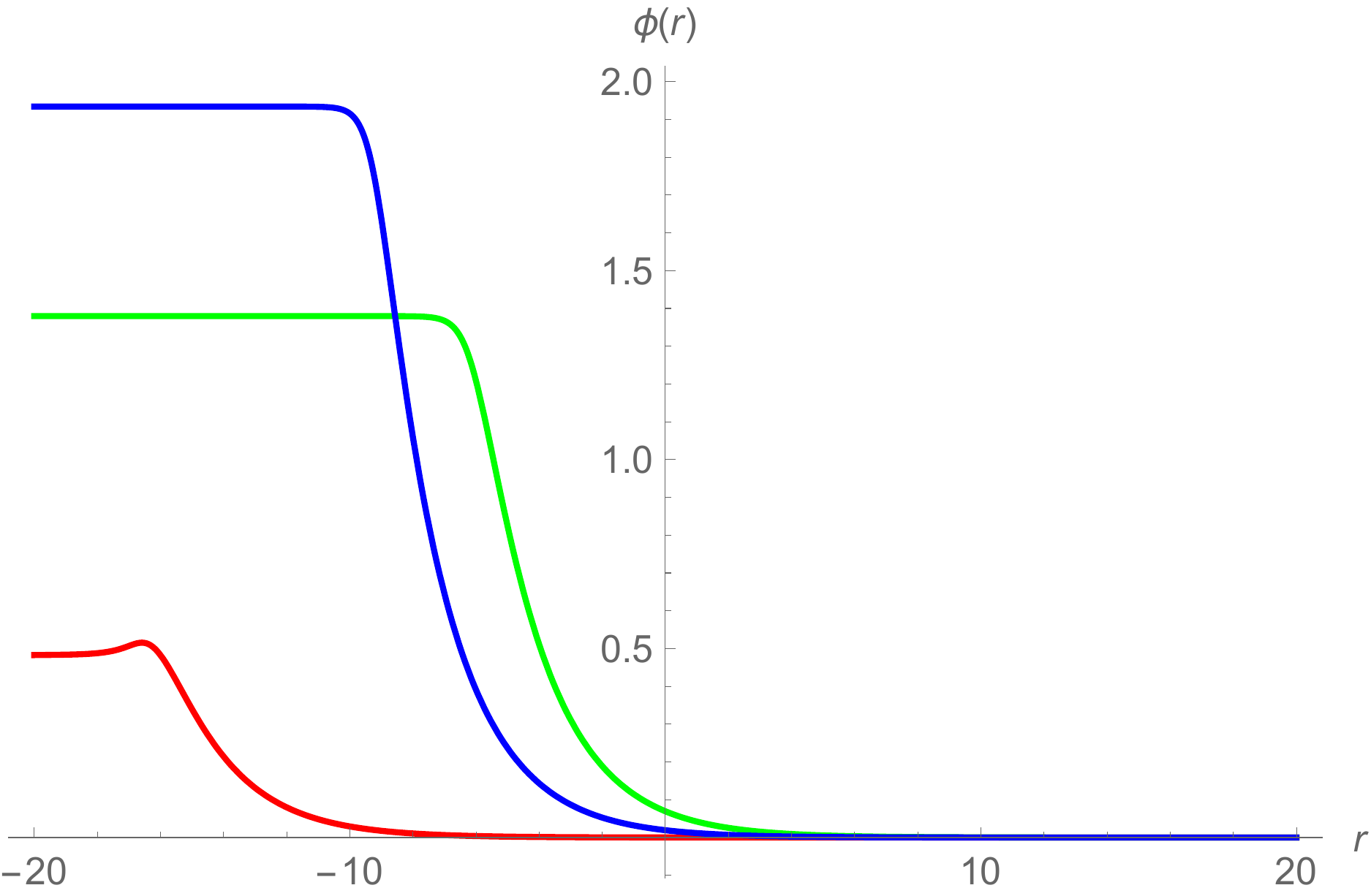}
  \caption{$\phi(r)$ solution}
  \end{subfigure}\\
   \begin{subfigure}[b]{0.45\linewidth}
    \includegraphics[width=\linewidth]{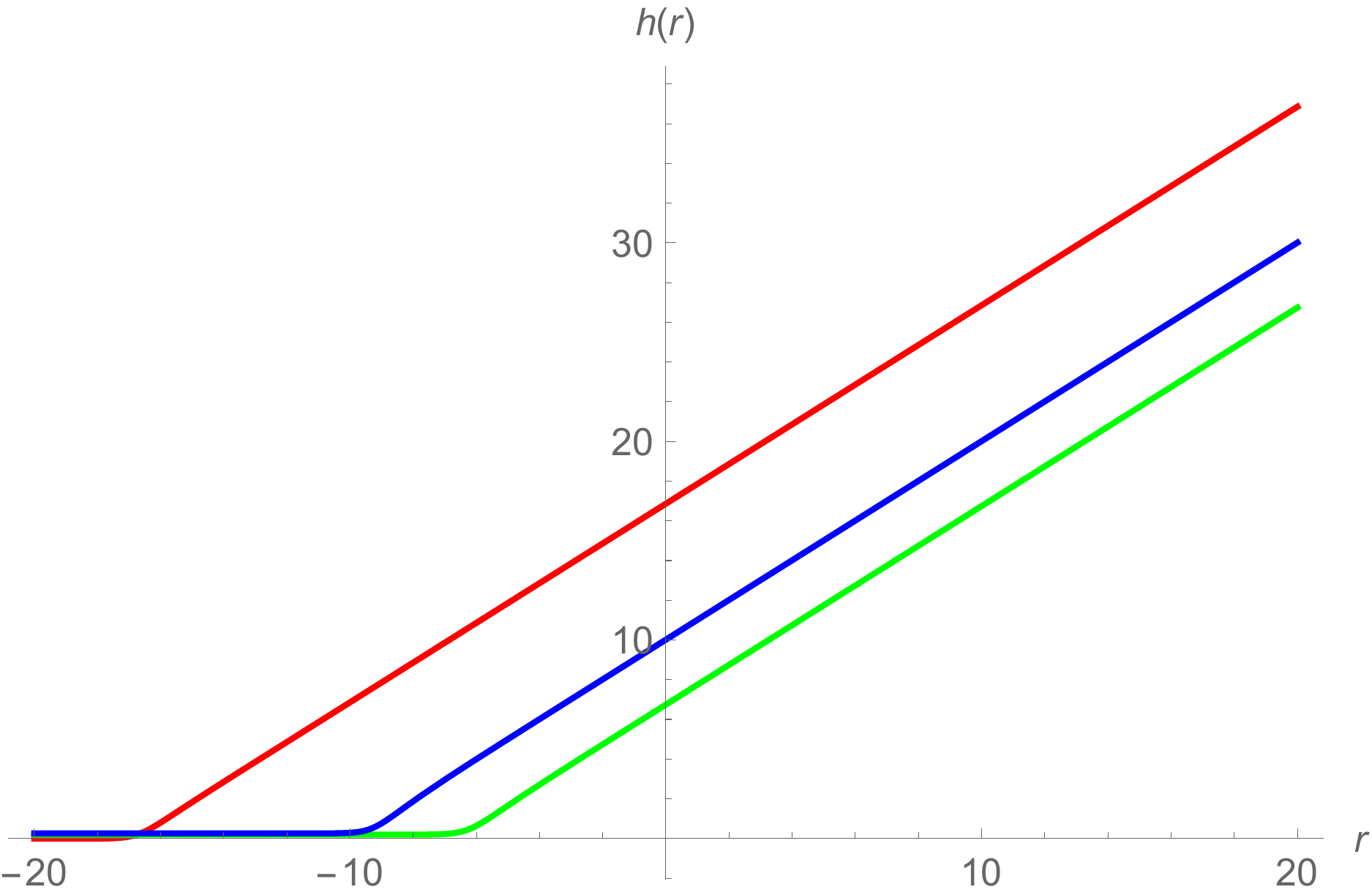}
  \caption{$h(r)$ solution}
   \end{subfigure} 
 \begin{subfigure}[b]{0.45\linewidth}
    \includegraphics[width=\linewidth]{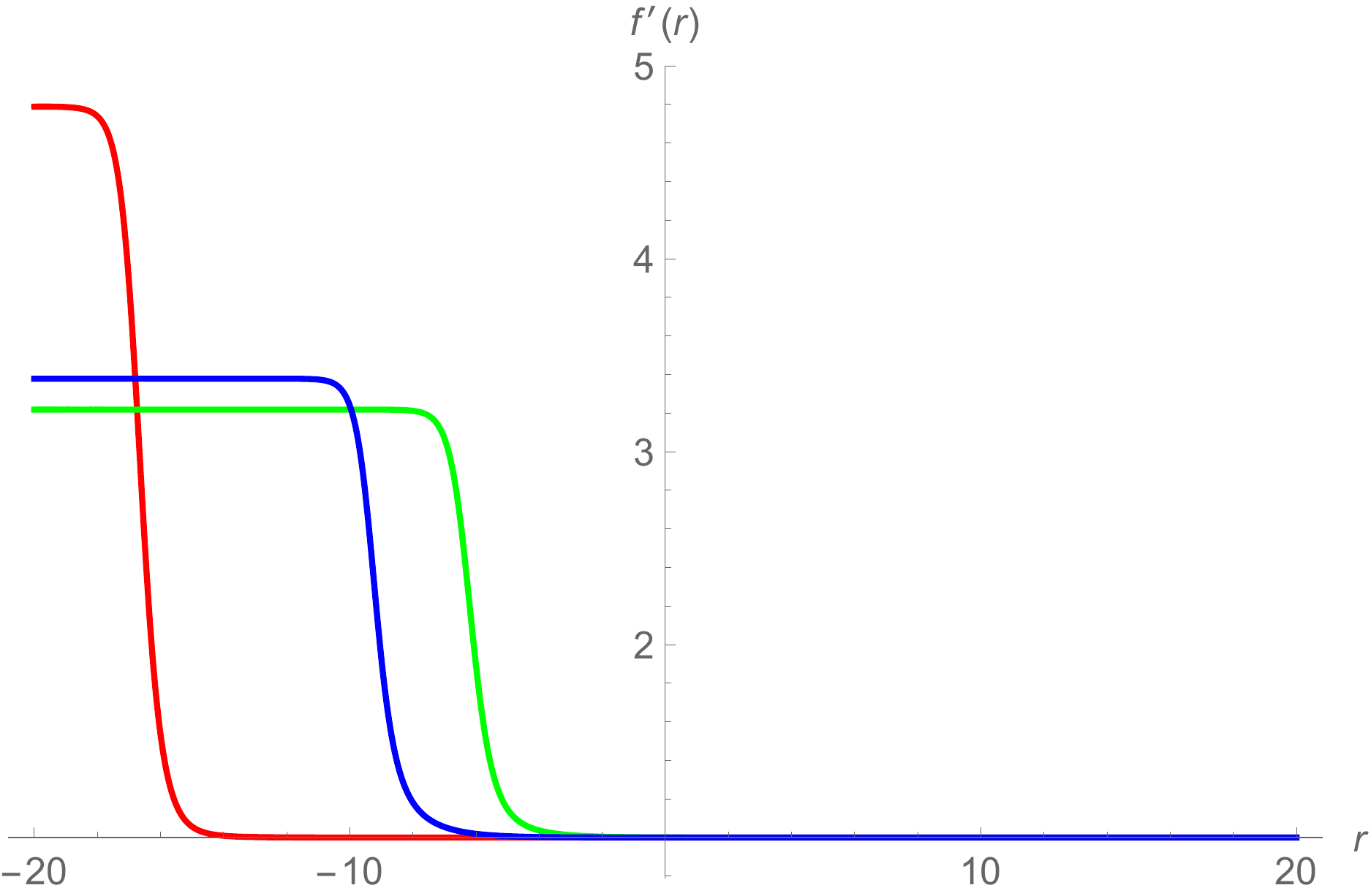}
  \caption{$f'(r)$ solution}
   \end{subfigure} 
  \caption{Supersymmetric $AdS_4$ black holes with $AdS_2\times S^2$ horizon for $g_2=g_1=1$, $p_3=-2$, $\kappa=1$ and $e_{12}=4 (\textrm{red}), 6 (\textrm{green}), 8 (\textrm{blue})$.}
  \label{Fig2}
\end{figure}

\begin{figure}[H]
  \centering
  \begin{subfigure}[b]{0.45\linewidth}
    \includegraphics[width=\linewidth]{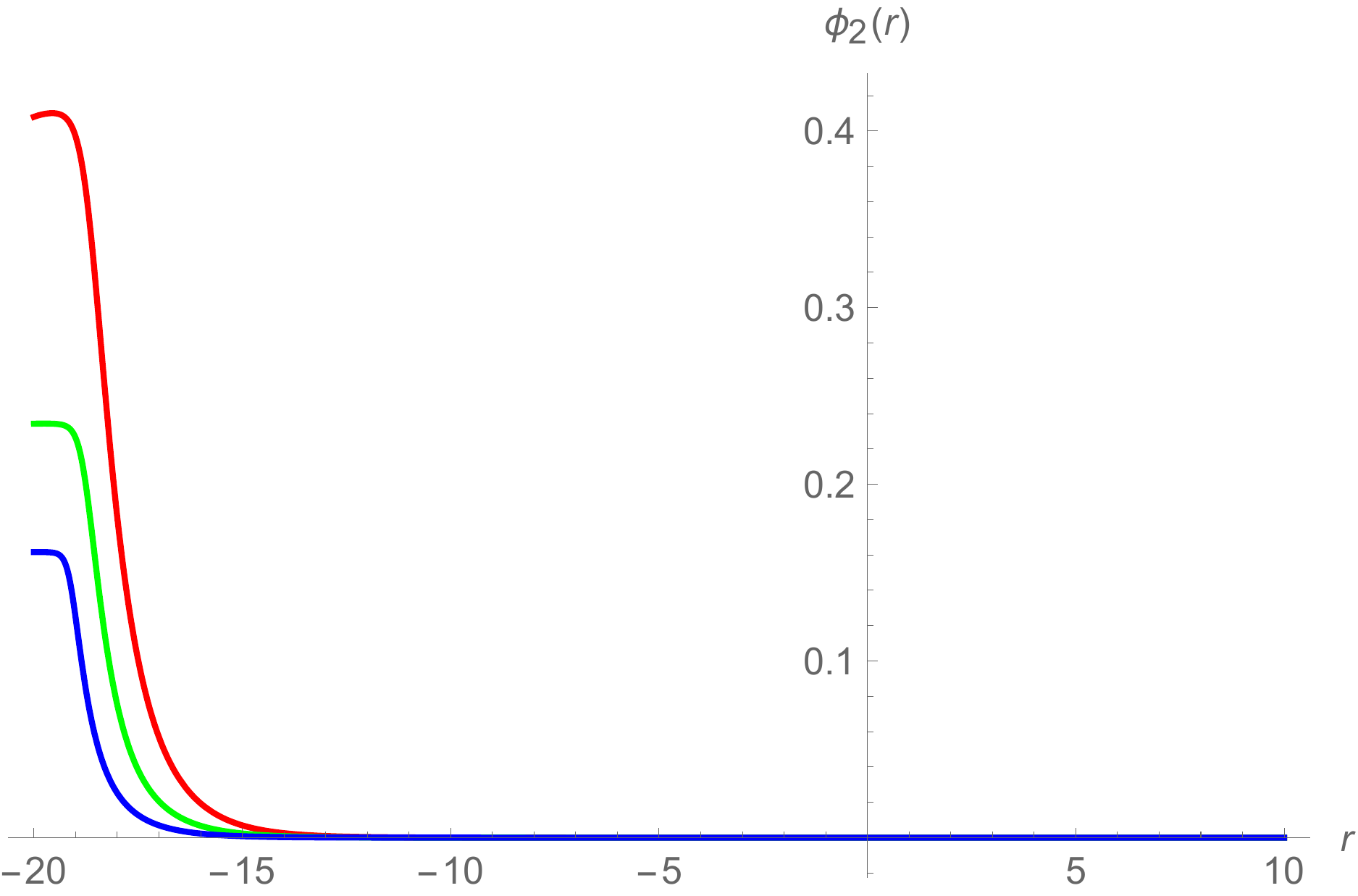}
  \caption{$\phi_2(r)$ solution}
  \end{subfigure}
  \begin{subfigure}[b]{0.45\linewidth}
    \includegraphics[width=\linewidth]{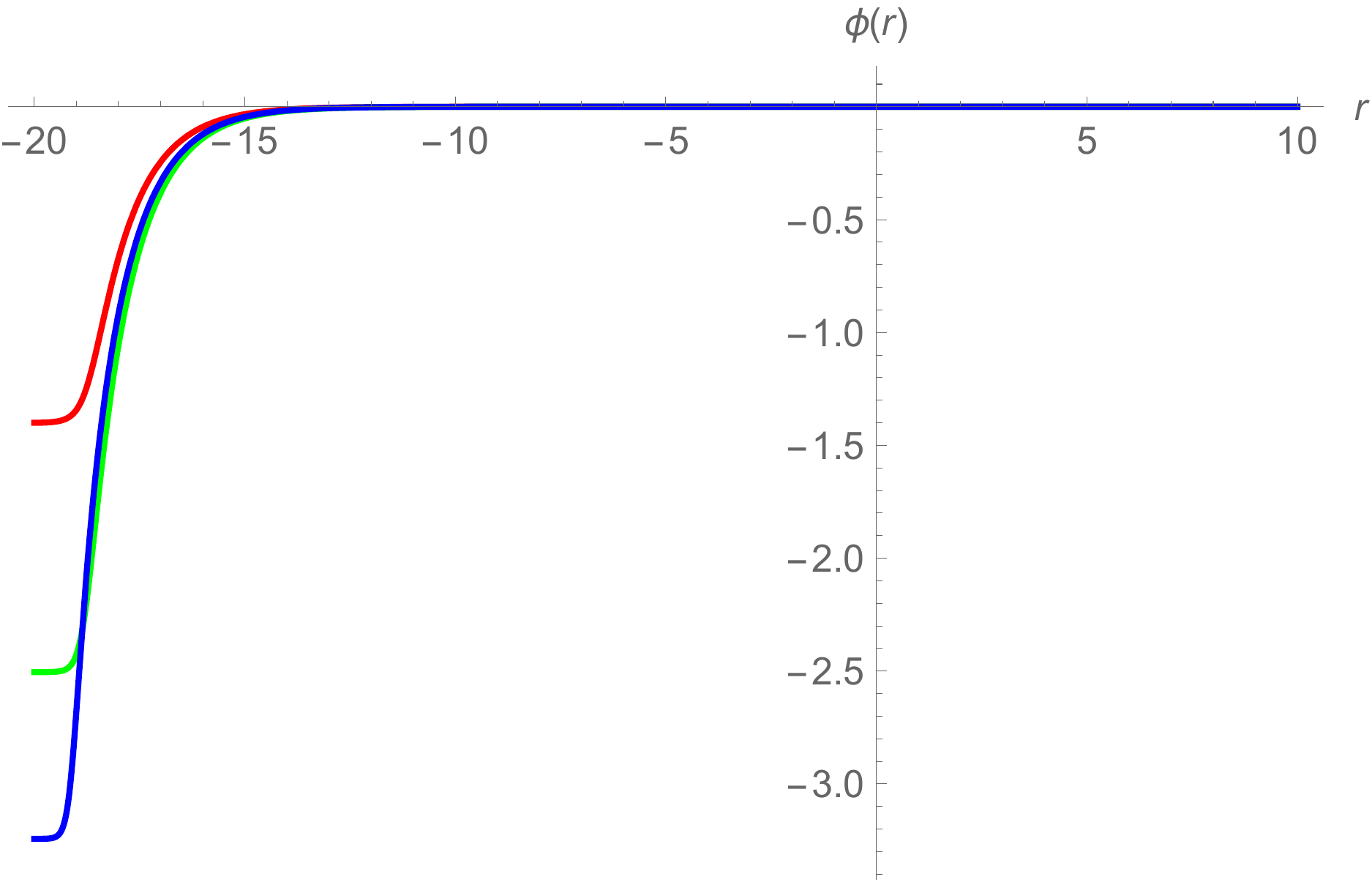}
  \caption{$\phi(r)$ solution}
  \end{subfigure}\\
   \begin{subfigure}[b]{0.45\linewidth}
    \includegraphics[width=\linewidth]{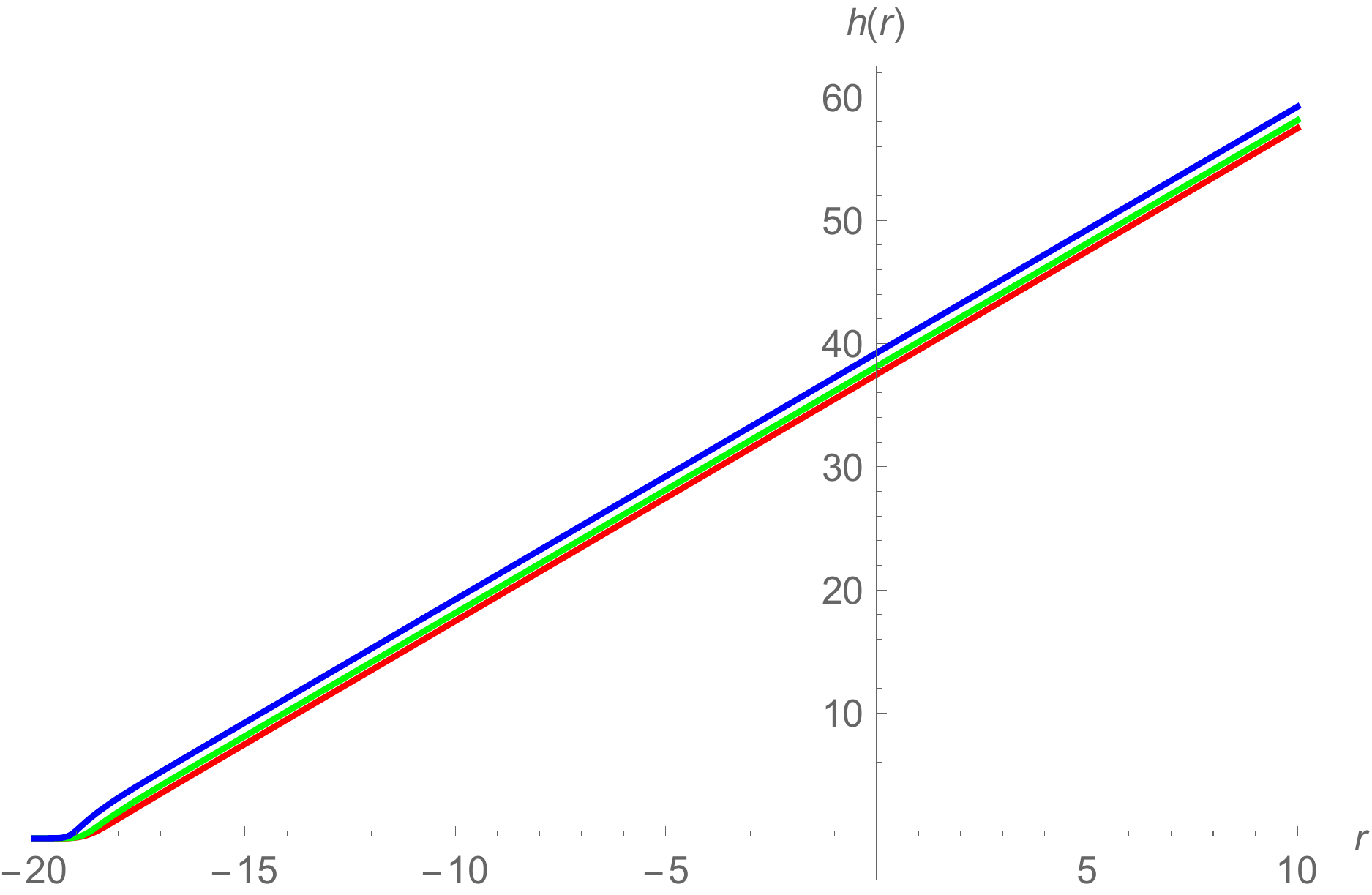}
  \caption{$h(r)$ solution}
   \end{subfigure} 
 \begin{subfigure}[b]{0.45\linewidth}
    \includegraphics[width=\linewidth]{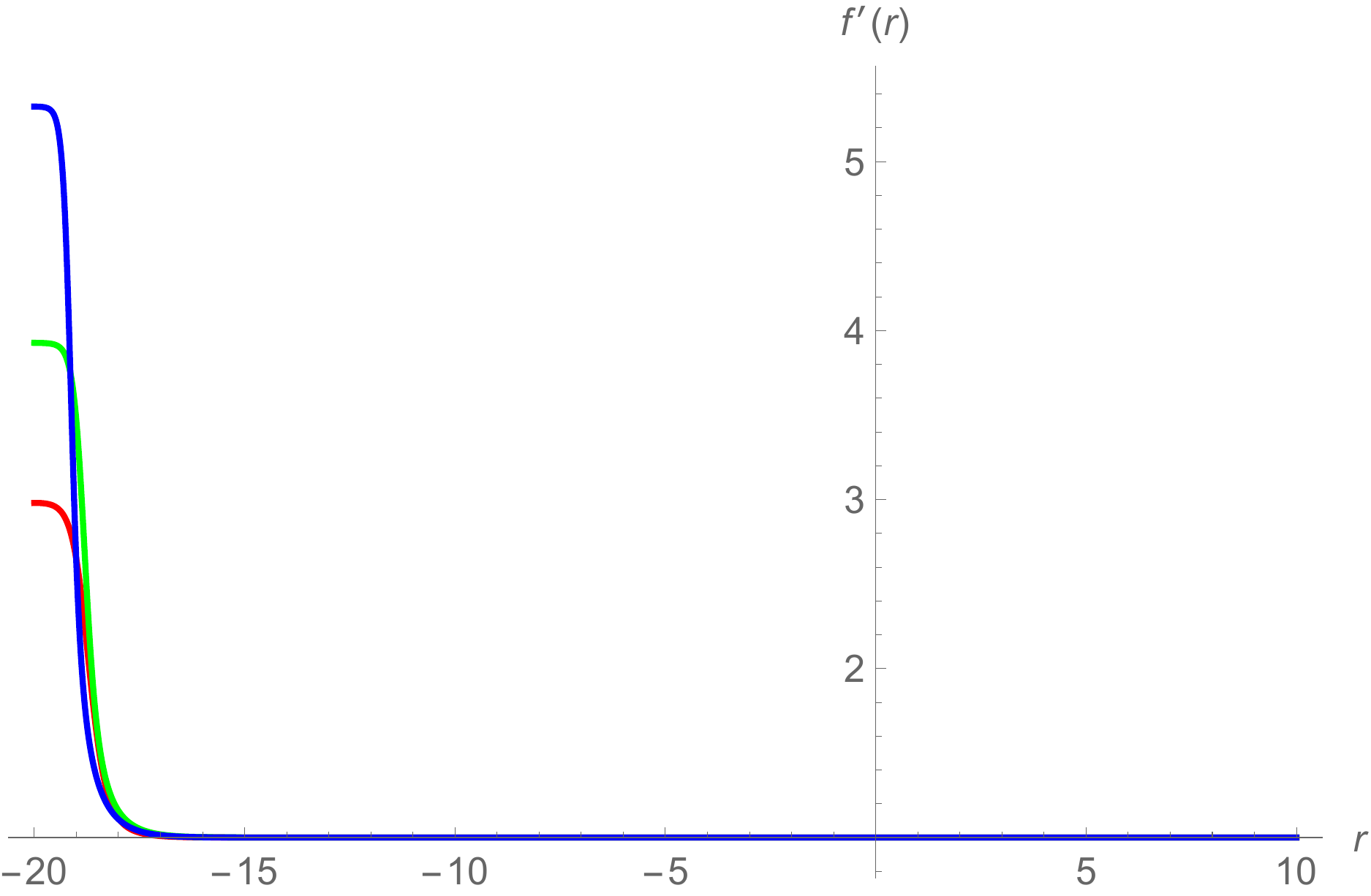}
  \caption{$f'(r)$ solution}
   \end{subfigure} 
  \caption{Supersymmetric $AdS_4$ black holes with $AdS_2\times H^2$ horizon for $g_2=g_1=1$, $p_3=\frac{1}{4}$, $\kappa=-1$ and $p_{12}=1 (\textrm{red}), 2 (\textrm{green}), 3 (\textrm{blue})$.}
  \label{Fig3}
\end{figure}

\begin{figure}[H]
  \centering
  \begin{subfigure}[b]{0.45\linewidth}
    \includegraphics[width=\linewidth]{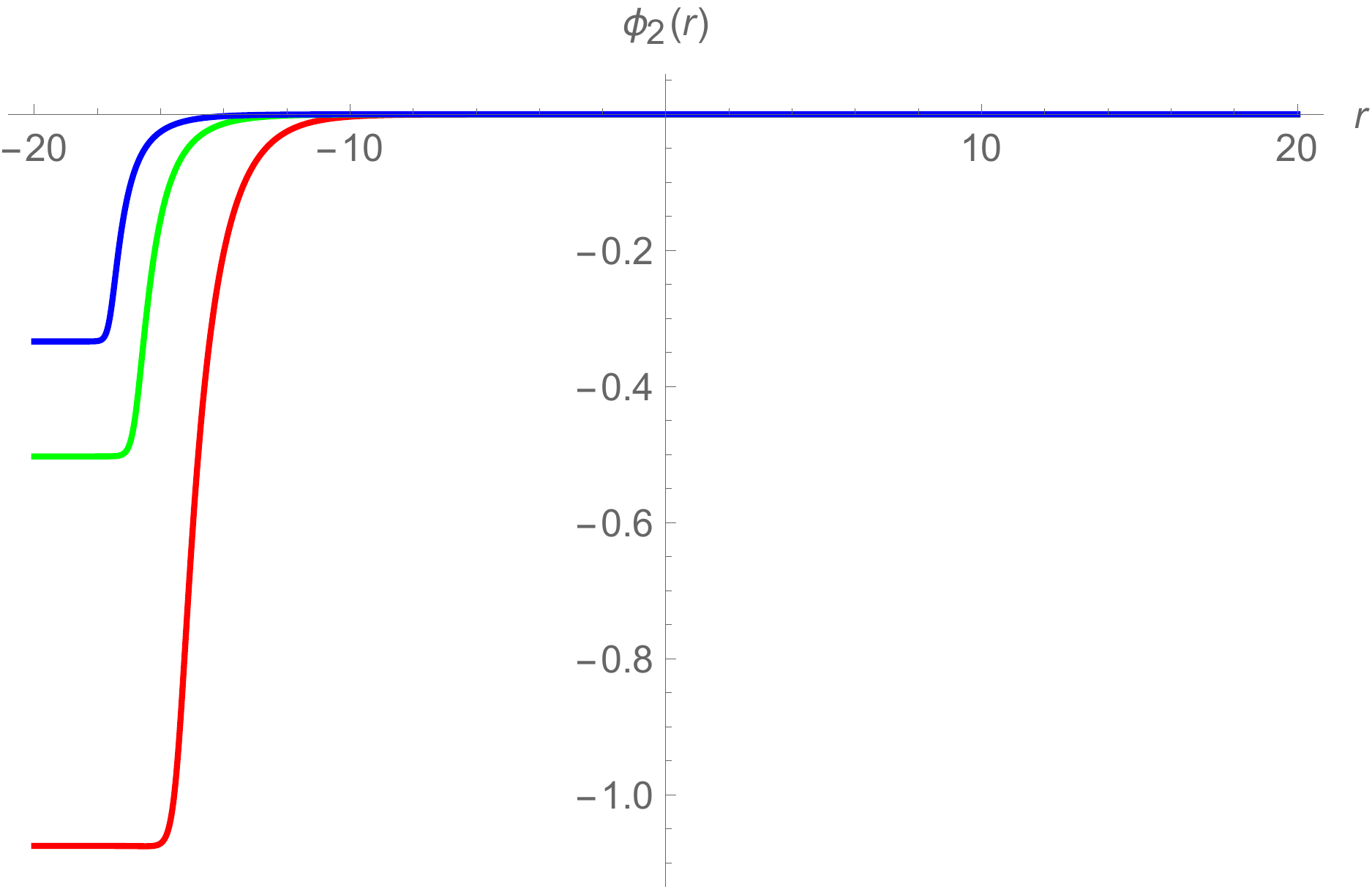}
  \caption{$\phi_2(r)$ solution}
  \end{subfigure}
  \begin{subfigure}[b]{0.45\linewidth}
    \includegraphics[width=\linewidth]{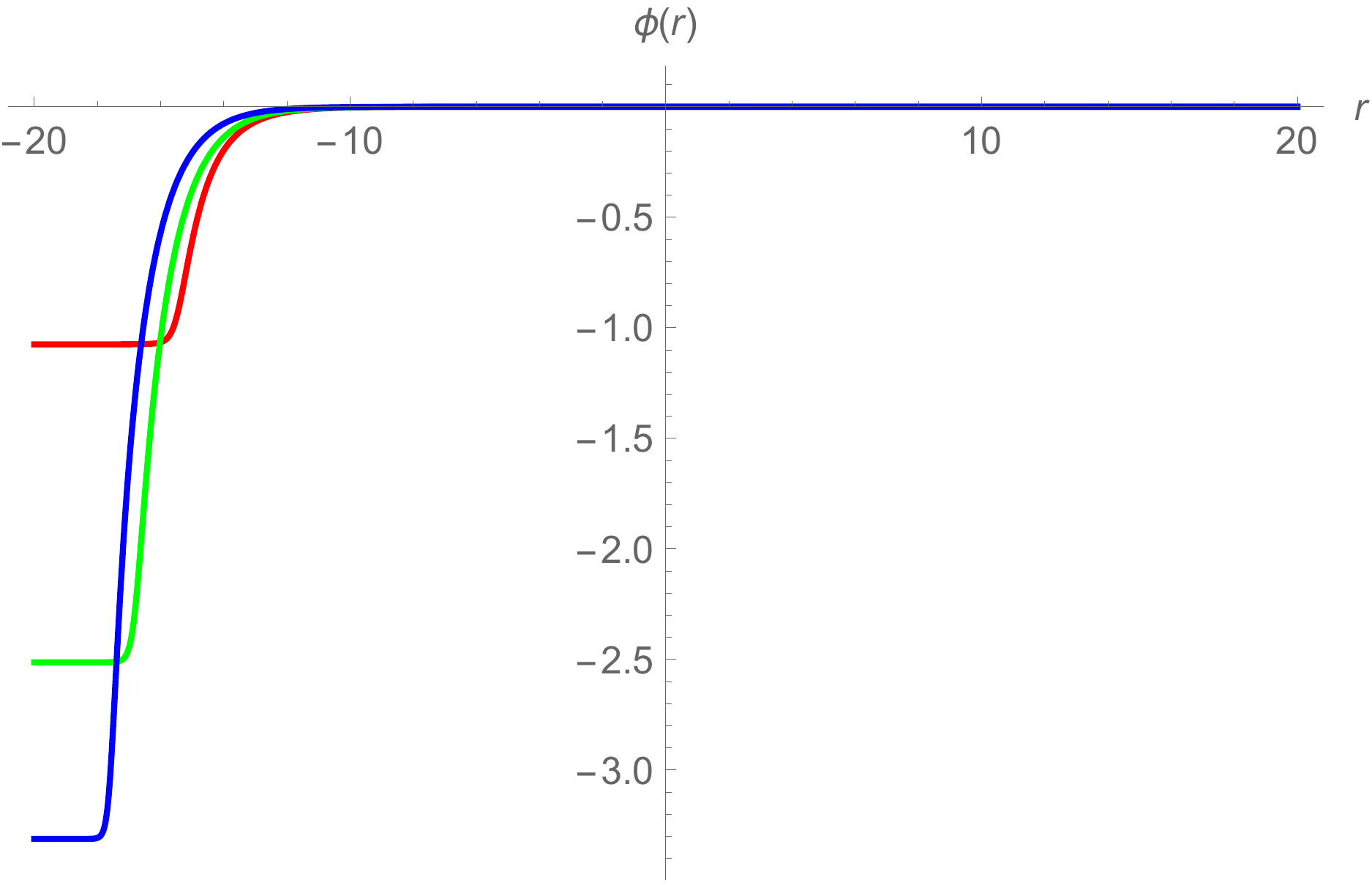}
  \caption{$\phi(r)$ solution}
  \end{subfigure}\\
   \begin{subfigure}[b]{0.45\linewidth}
    \includegraphics[width=\linewidth]{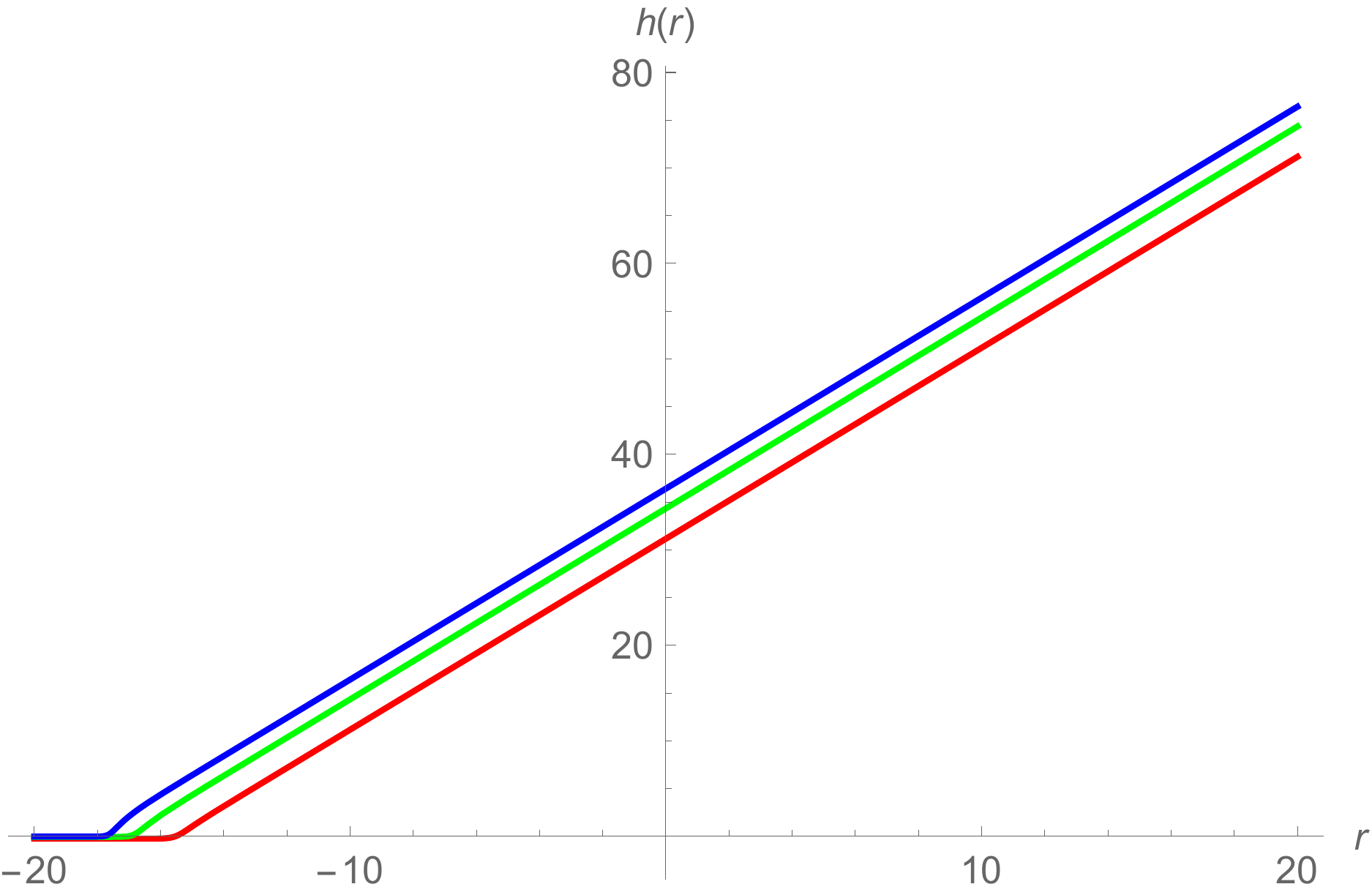}
  \caption{$h(r)$ solution}
   \end{subfigure} 
 \begin{subfigure}[b]{0.45\linewidth}
    \includegraphics[width=\linewidth]{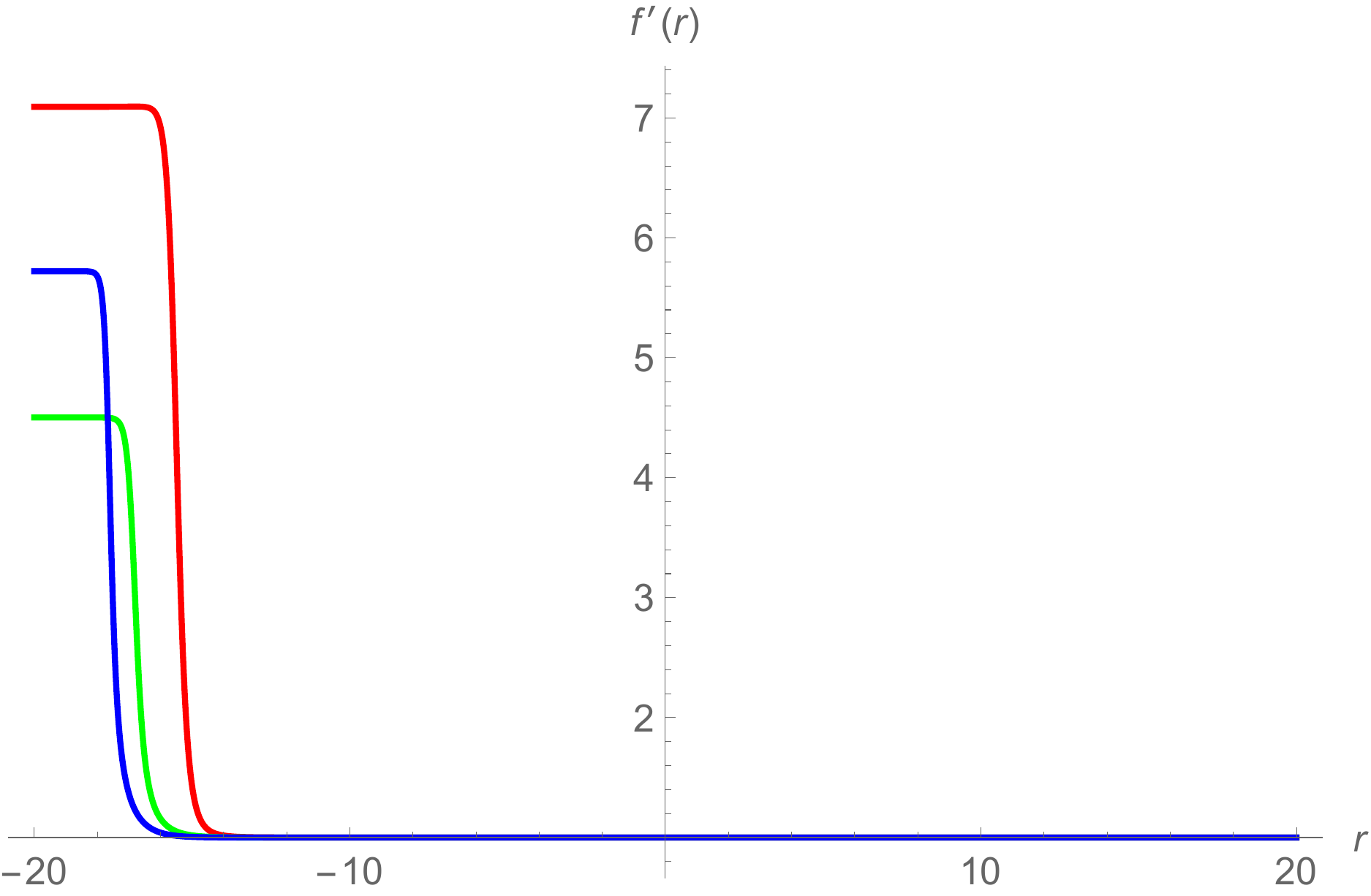}
  \caption{$f'(r)$ solution}
   \end{subfigure} 
  \caption{Supersymmetric $AdS_4$ black holes with $AdS_2\times S^2$ horizon for $g_2=g_1=1$, $p_3=2$, $\kappa=1$ and $p_{12}=3 (\textrm{red}), 6 (\textrm{green}), 9 (\textrm{blue})$.}
  \label{Fig4}
\end{figure}

In this case, the solutions carry magnetic charges of $A^{12-}$, and the resulting BPS equations are given by
\begin{eqnarray}
f'&=&\frac{1}{4}e^{-2h-\frac{\phi}{2}-\phi_2}\left[e^{2h}(g_1+g_1e^{2\phi_2}+2g_2e^{\phi+\phi_2})-2\kappa e_6e^{\phi_2}\right. \nonumber \\
& &\left.+\kappa e^\phi(p_{12}-p_3)-\kappa(p_{12}+p_3)e^{\phi+\phi_2}\right],\\
h'&=&\frac{1}{4}e^{-2h-\frac{\phi}{2}-\phi_2}\left[e^{2h}(g_1+g_1e^{2\phi_2}+2g_2e^{\phi+\phi_2})+2\kappa e_6e^{\phi_2}\right. \nonumber \\
& &\left.-\kappa e^\phi(p_{12}-p_3)+\kappa(p_{12}+p_3)e^{\phi+\phi_2}\right],\\
\phi'&=&\frac{1}{2}e^{-2h-\frac{\phi}{2}-\phi_2}\left[e^{2h}g_1(1+e^{2\phi_2})-2g_2e^{2h+\phi+\phi_2}+2\kappa e_6e^{\phi_2} \right. \nonumber \\
& &\left.+\kappa (p_{12}-p_3)e^\phi-\kappa (p_{12}+p_3)e^{2\phi_4+\phi}\right],\\
\phi_2'&=&-\frac{1}{2}e^{-\frac{\phi}{2}-\phi_2}\left[g_1(e^{2\phi_2}-1)+\kappa e^{-2h+\phi}[p_{12}-p_3+(p_{12}+p_3)e^{2\phi_2}]\right].
\end{eqnarray}
From these equations, we find a family of $AdS_2\times \Sigma^2$ fixed points given by
\begin{eqnarray}
\phi&=&\ln\left[\frac{2e_6e^{2\phi_2}(e^{2\phi_2}-1)}{(1+e^{2\phi_2})[p_{12}-p_3+(p_{12}+p_3)e^{2\phi_2}]}\right],\nonumber \\
h&=&\frac{\phi_2}{2}-\frac{1}{2}\ln\left[-\frac{g_1(1+e^{2\phi_2})}{2\kappa e_6}\right],\nonumber \\
\phi_2&=&\frac{1}{2}\ln\left[\frac{2e_6g_2-g_1p_3+\sqrt{4e_6^2g_2^2+g_1^2p_{12}^2-4e_6p_3g_1g_2}}{g_1(p_{12}+p_3)}\right].
\end{eqnarray}
Similar to the previous case, both $AdS_2\times S^2$ and $AdS_2\times H^2$ geometries are possible depending on the values of various parameters. Examples of flow solutions from the $AdS_4$ vacuum to $AdS_2\times H^2$ fixed points with
\begin{equation}
g_2=g_1=1,\qquad p_3=\frac{1}{4},\qquad \kappa=-1
\end{equation}
and $p_{12}=1,2,3$ are given in figure \ref{Fig3}. For flow solutions to $AdS_2\times S^2$ fixed points, we give some representative solutions for $p_{12}=3,6,9$ and
\begin{equation}
g_2=g_1=1,\qquad p_3=2,\qquad \kappa=1
\end{equation}
in figure \ref{Fig4}.

\subsection{Solutions with $SO(2)_{\textrm{diag}}\times SO(2)_{\textrm{diag}}$ symmetry}
\begin{figure}[H]
  \centering
  \begin{subfigure}[b]{0.45\linewidth}
    \includegraphics[width=\linewidth]{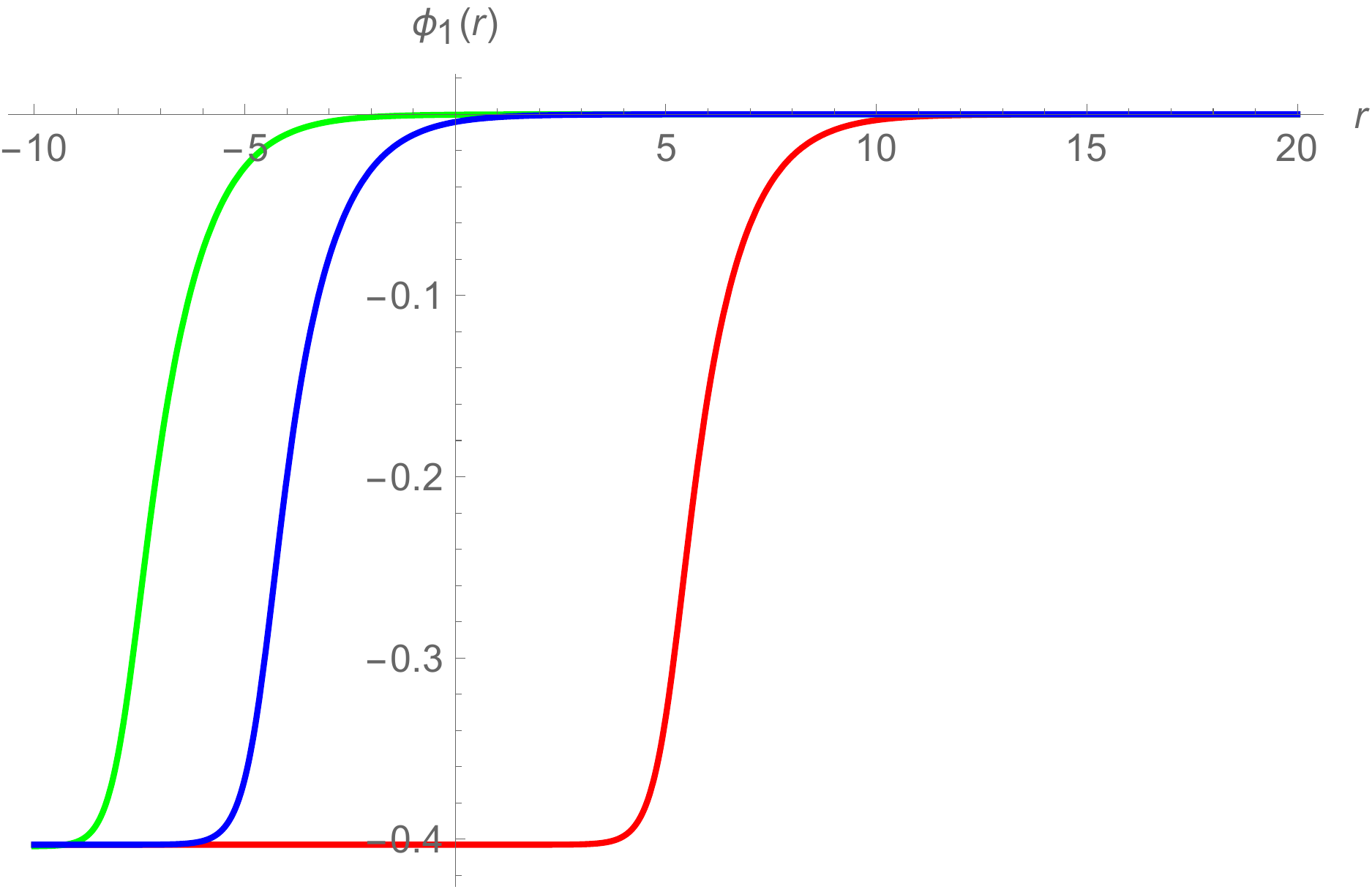}
  \caption{$\phi_1(r)$ solution}
  \end{subfigure}
  \begin{subfigure}[b]{0.45\linewidth}
    \includegraphics[width=\linewidth]{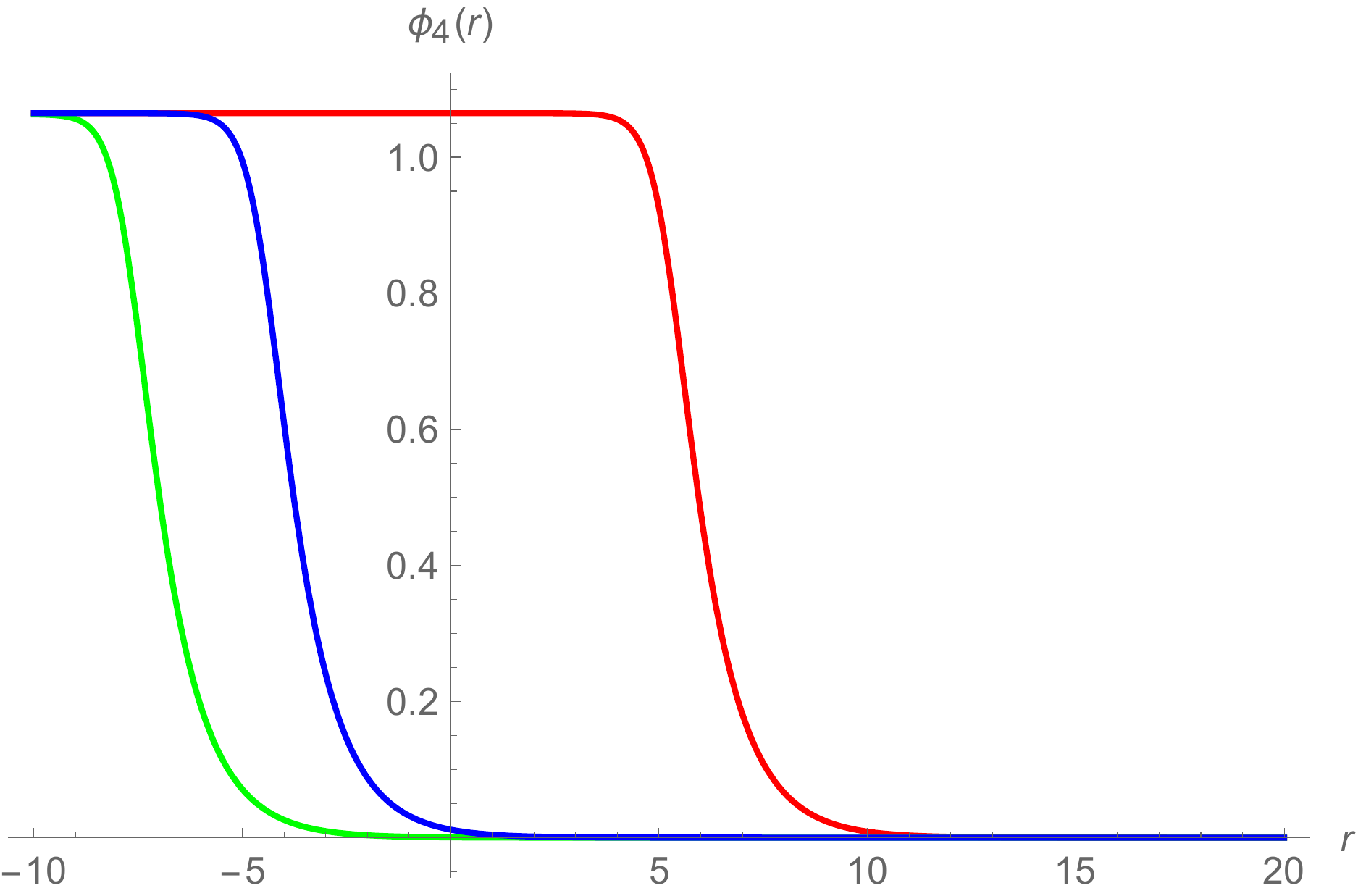}
  \caption{$\phi_4(r)$ solution}
  \end{subfigure}\\
   \begin{subfigure}[b]{0.45\linewidth}
    \includegraphics[width=\linewidth]{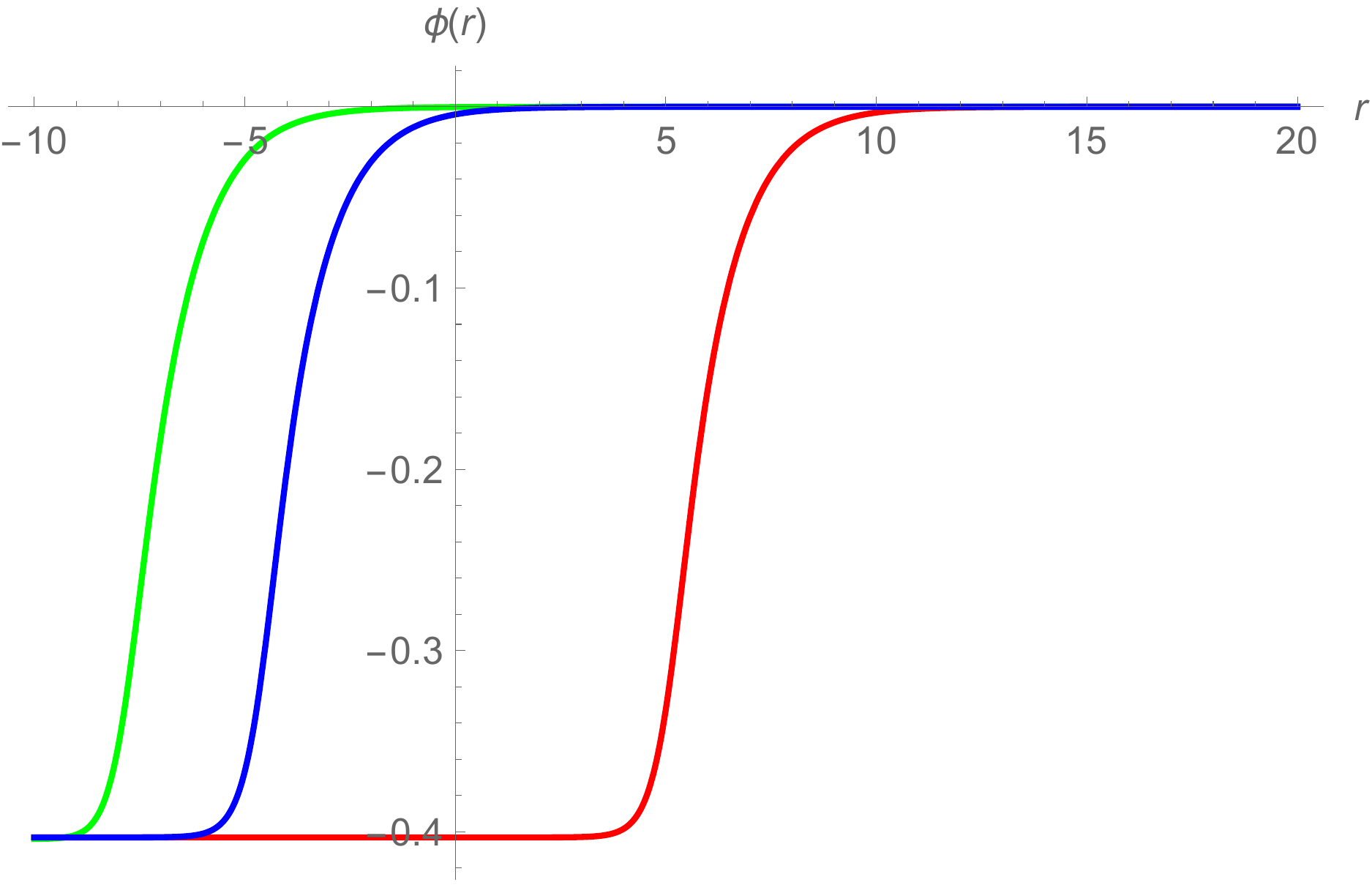}
  \caption{$\phi(r)$ solution}
   \end{subfigure} 
 \begin{subfigure}[b]{0.45\linewidth}
    \includegraphics[width=\linewidth]{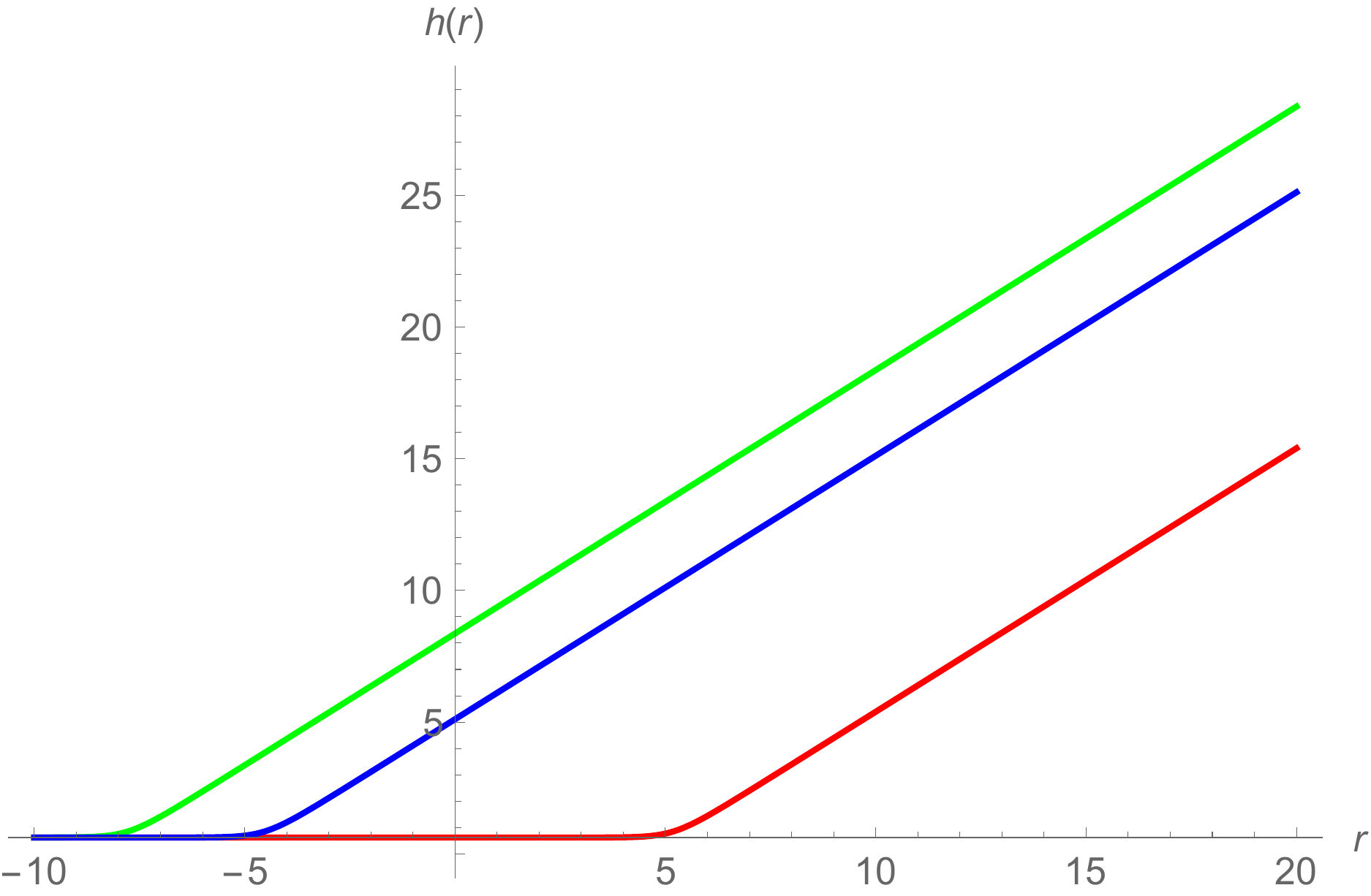}
  \caption{h(r) solution}
   \end{subfigure} \\
 \begin{subfigure}[b]{0.45\linewidth}
    \includegraphics[width=\linewidth]{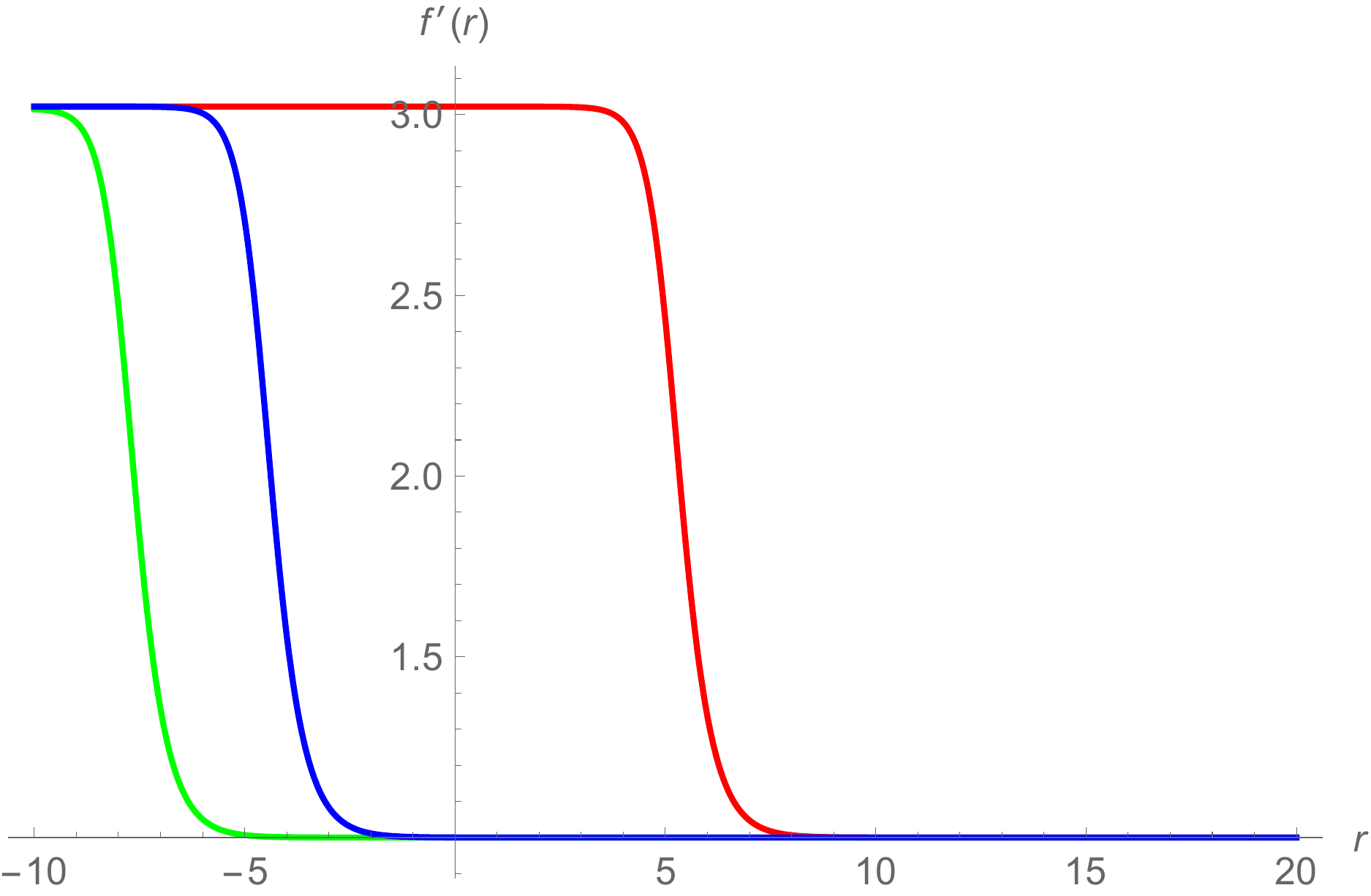}
  \caption{$f'(r)$ solution}
   \end{subfigure}
    \caption{Supersymmetric $AdS_4$ black holes with $AdS_2\times H^2$ horizon $(i)$ for $g_2=g_1=1$, $\tilde{g}_1=2g_1$, $\tilde{g}_2=3g_2$ and $p_3=-3 (\textrm{red}),-3.00000025 (\textrm{blue}),-3.005 (\textrm{green})$.}
  \label{Fig5}
\end{figure}

\begin{figure}[H]
  \centering
  \begin{subfigure}[b]{0.45\linewidth}
    \includegraphics[width=\linewidth]{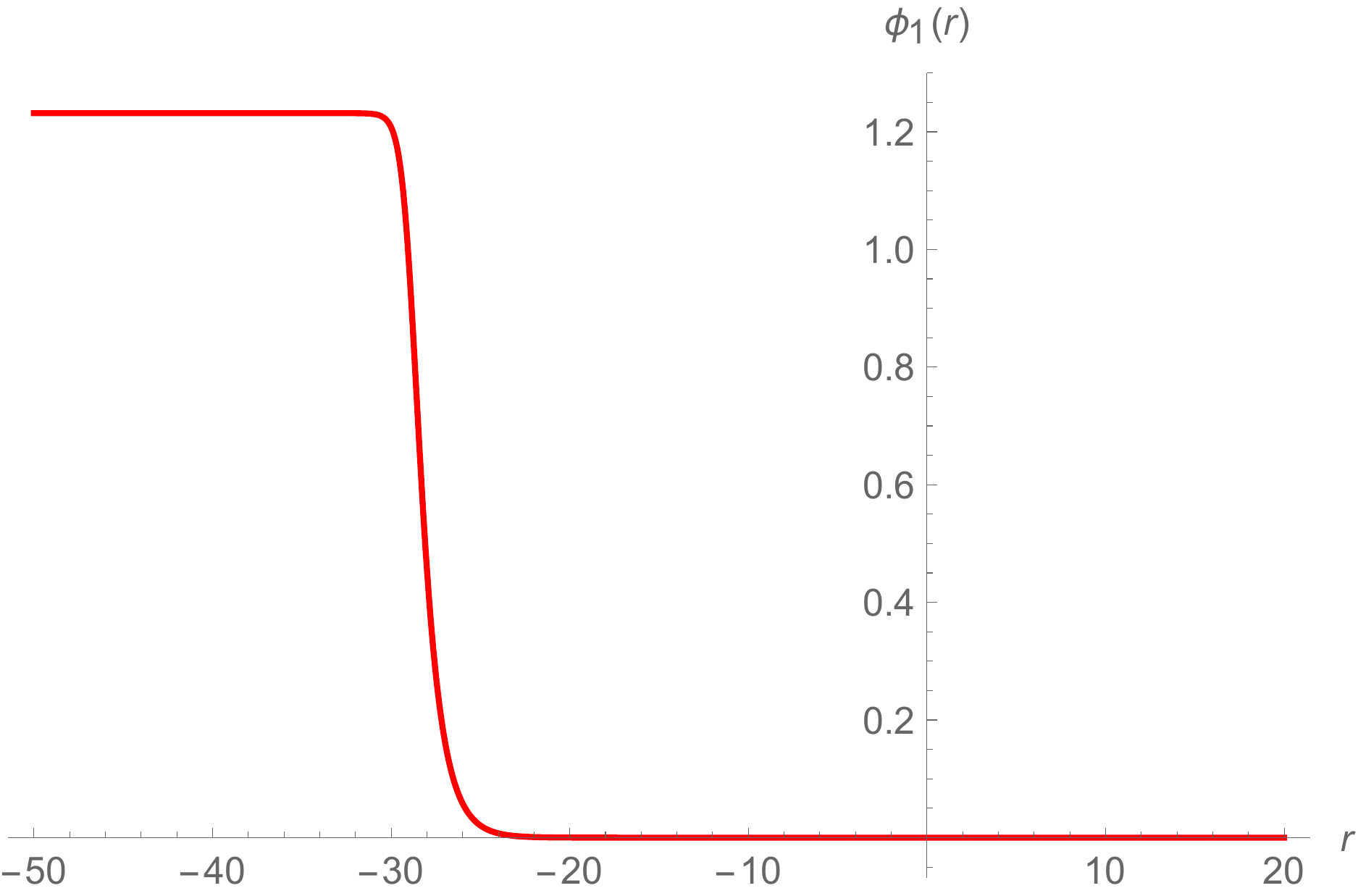}
  \caption{$\phi_1(r)$ solution}
  \end{subfigure}
  \begin{subfigure}[b]{0.45\linewidth}
    \includegraphics[width=\linewidth]{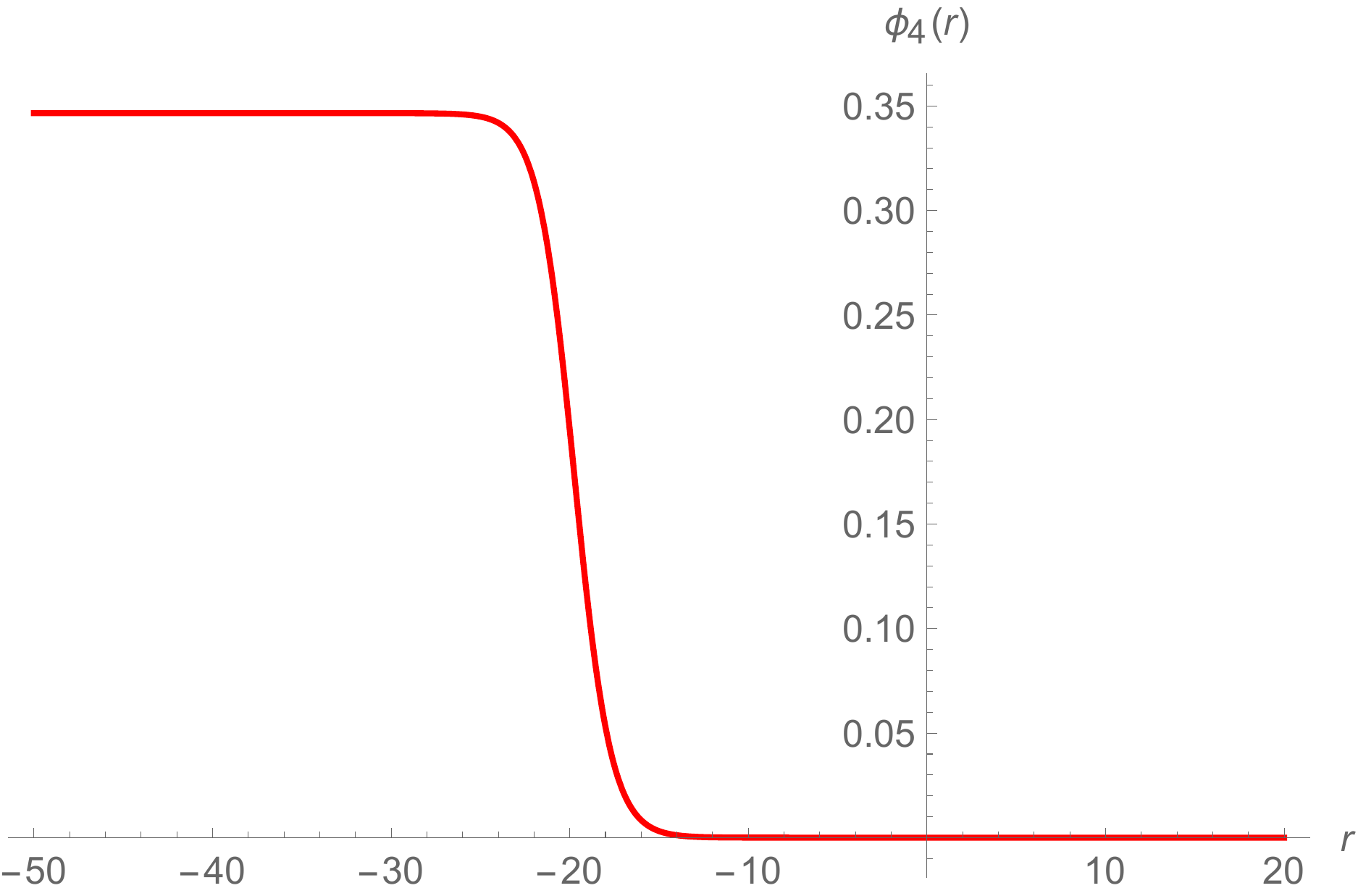}
  \caption{$\phi_4(r)$ solution}
  \end{subfigure}\\
   \begin{subfigure}[b]{0.45\linewidth}
    \includegraphics[width=\linewidth]{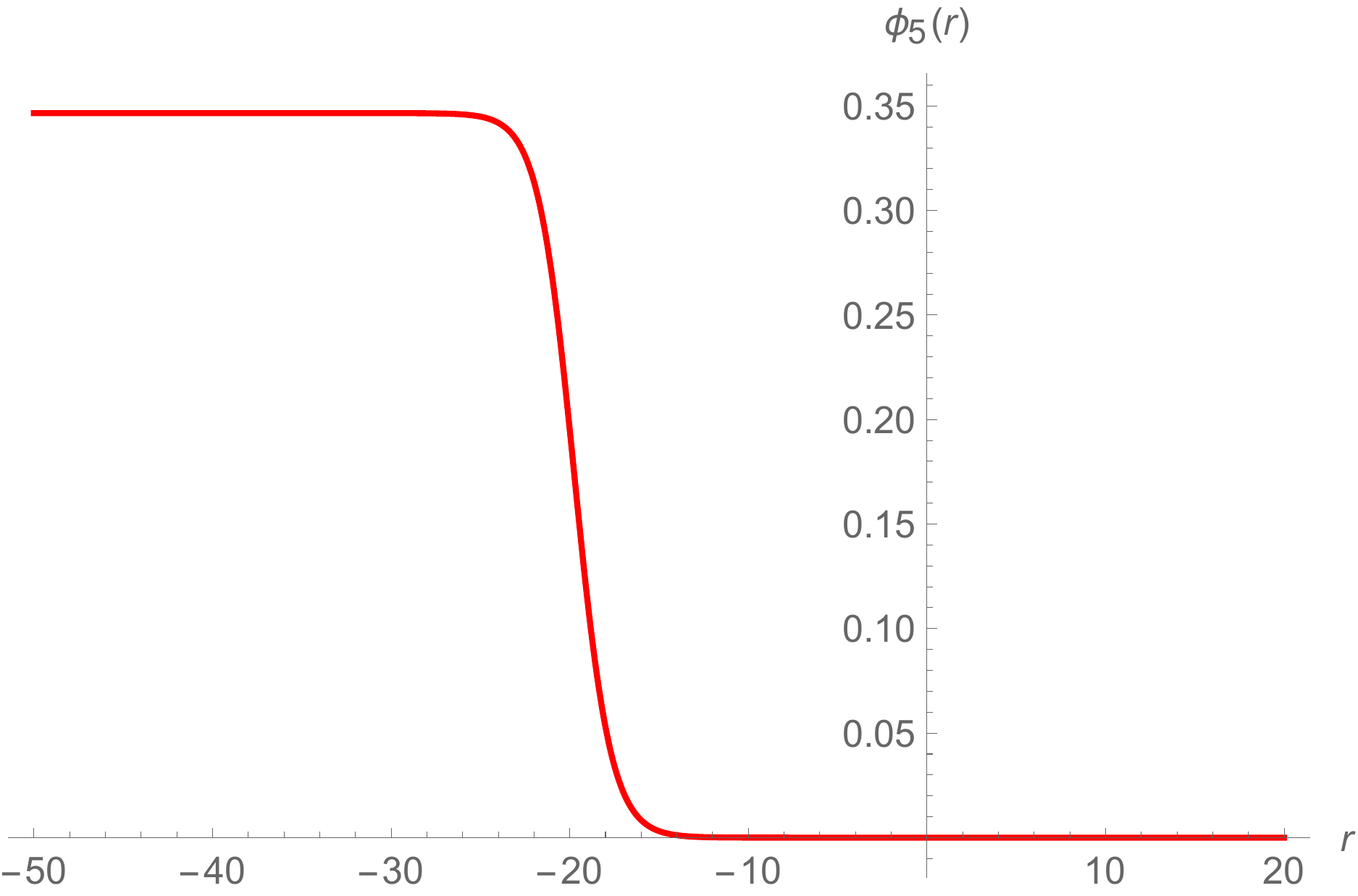}
  \caption{$\phi_5(r)$ solution}
   \end{subfigure} 
 \begin{subfigure}[b]{0.45\linewidth}
    \includegraphics[width=\linewidth]{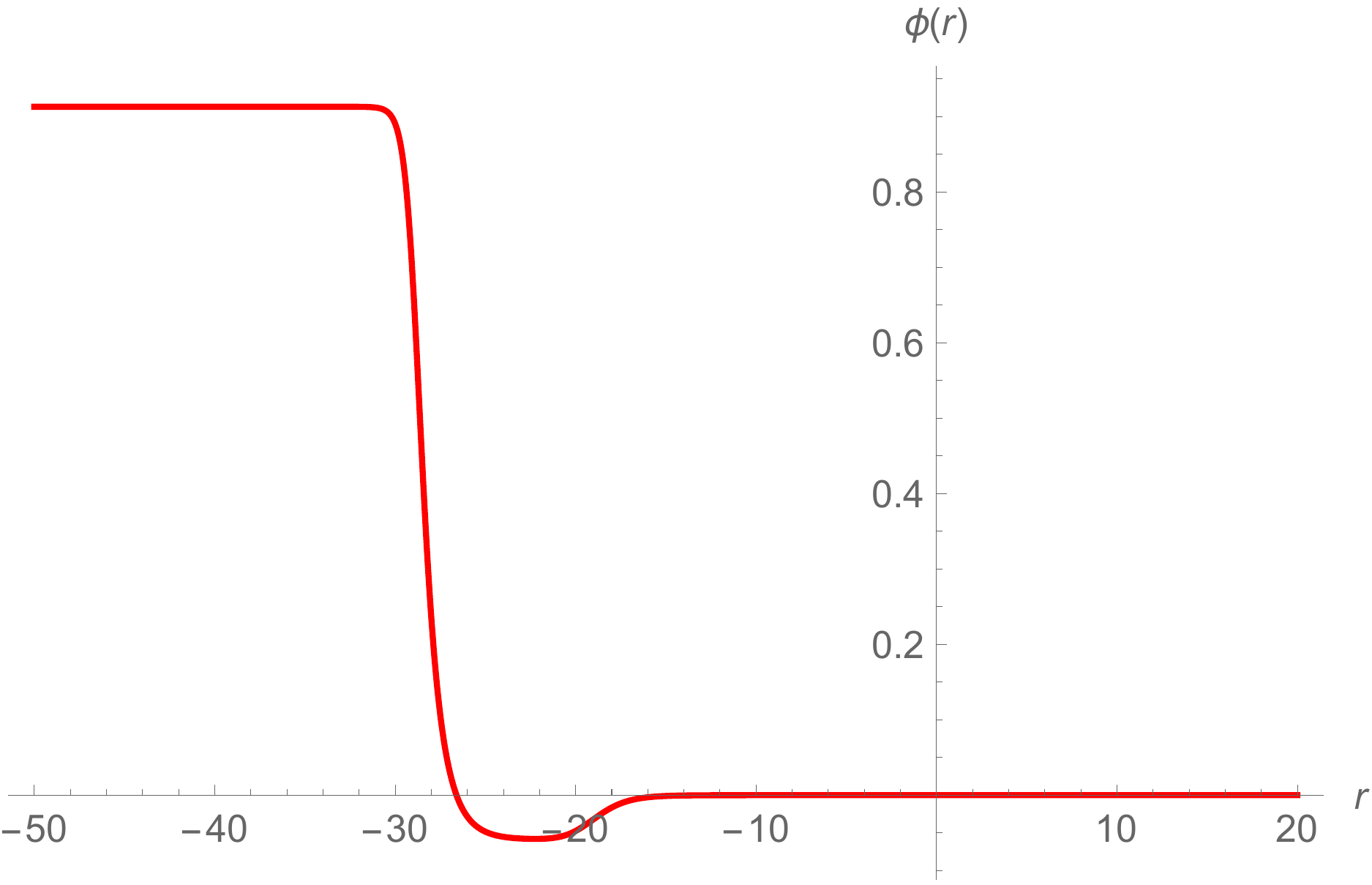}
  \caption{$\phi(r)$ solution}
   \end{subfigure} \\
   \begin{subfigure}[b]{0.45\linewidth}
    \includegraphics[width=\linewidth]{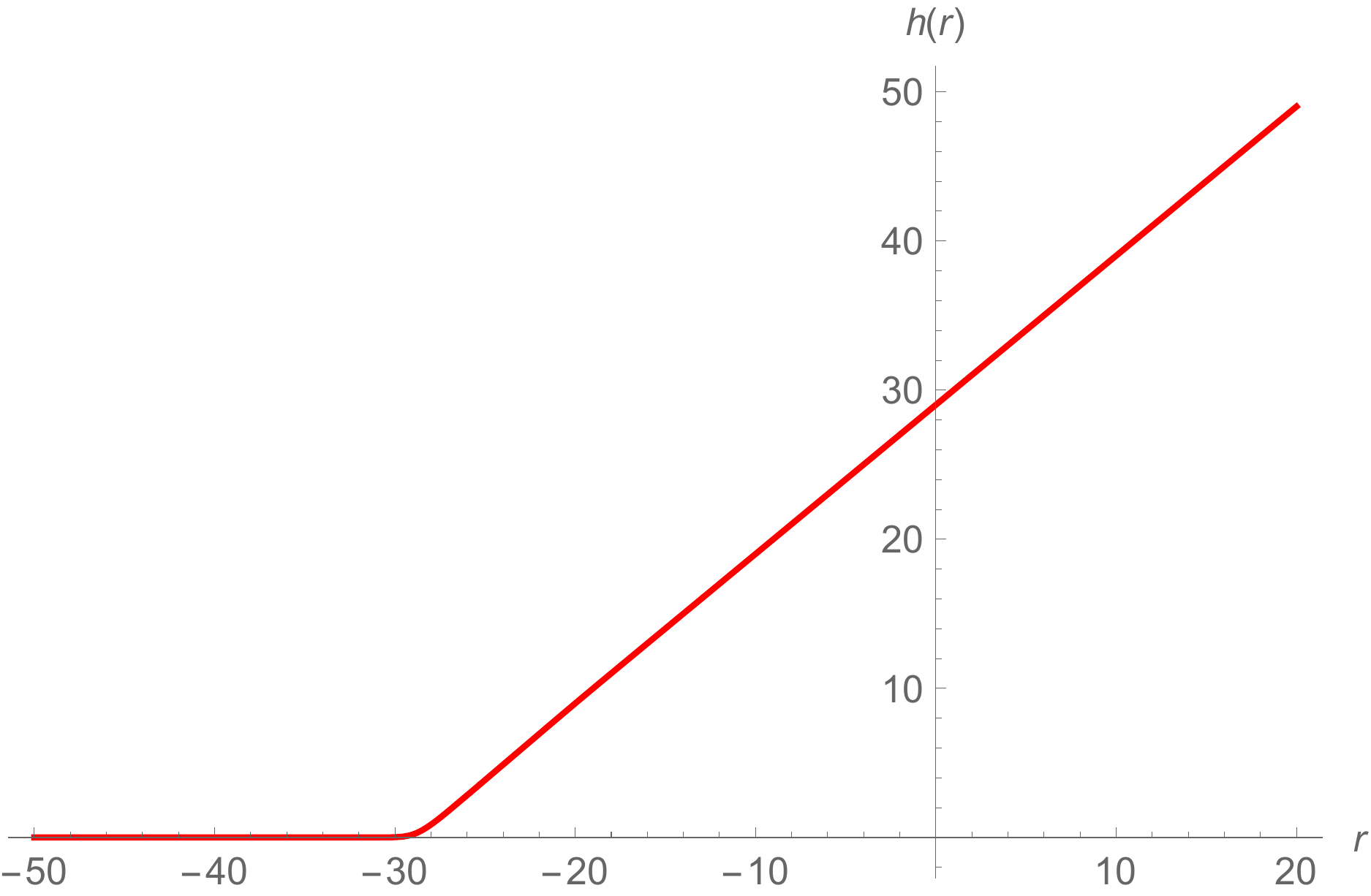}
  \caption{$h(r)$ solution}
   \end{subfigure}
 \begin{subfigure}[b]{0.45\linewidth}
    \includegraphics[width=\linewidth]{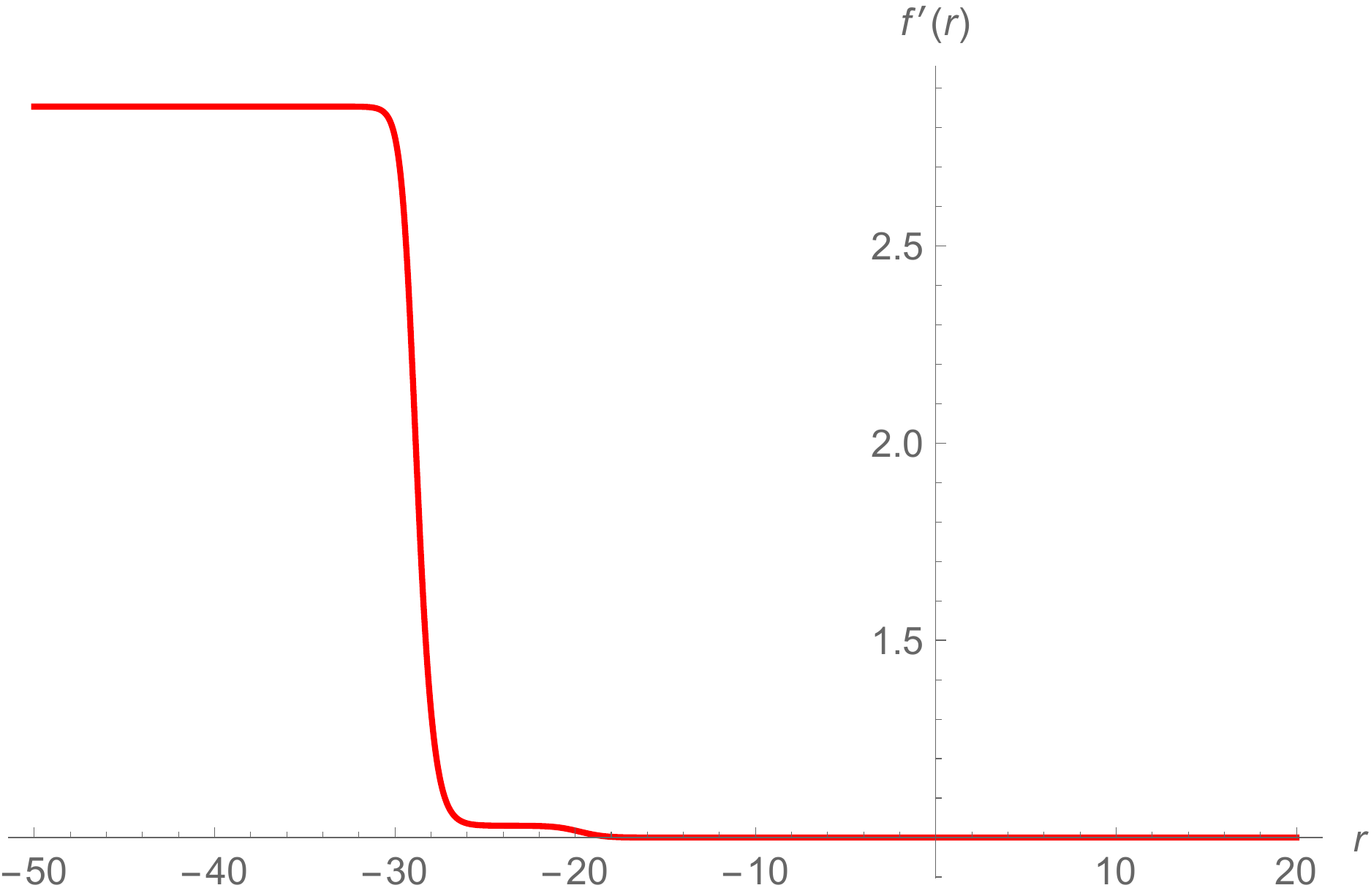}
  \caption{$f'(r)$ solution}
   \end{subfigure}
    \caption{A supersymmetric $AdS_4$ black hole with $AdS_2\times H^2$ horizon $(ii)$ for $g_2=g_1=1$, $\tilde{g}_1=2g_1$, $\tilde{g}_2=3g_2$ and $p_3=-3$.}
  \label{Fig6}
\end{figure}

\begin{figure}[H]
  \centering
  \begin{subfigure}[b]{0.45\linewidth}
    \includegraphics[width=\linewidth]{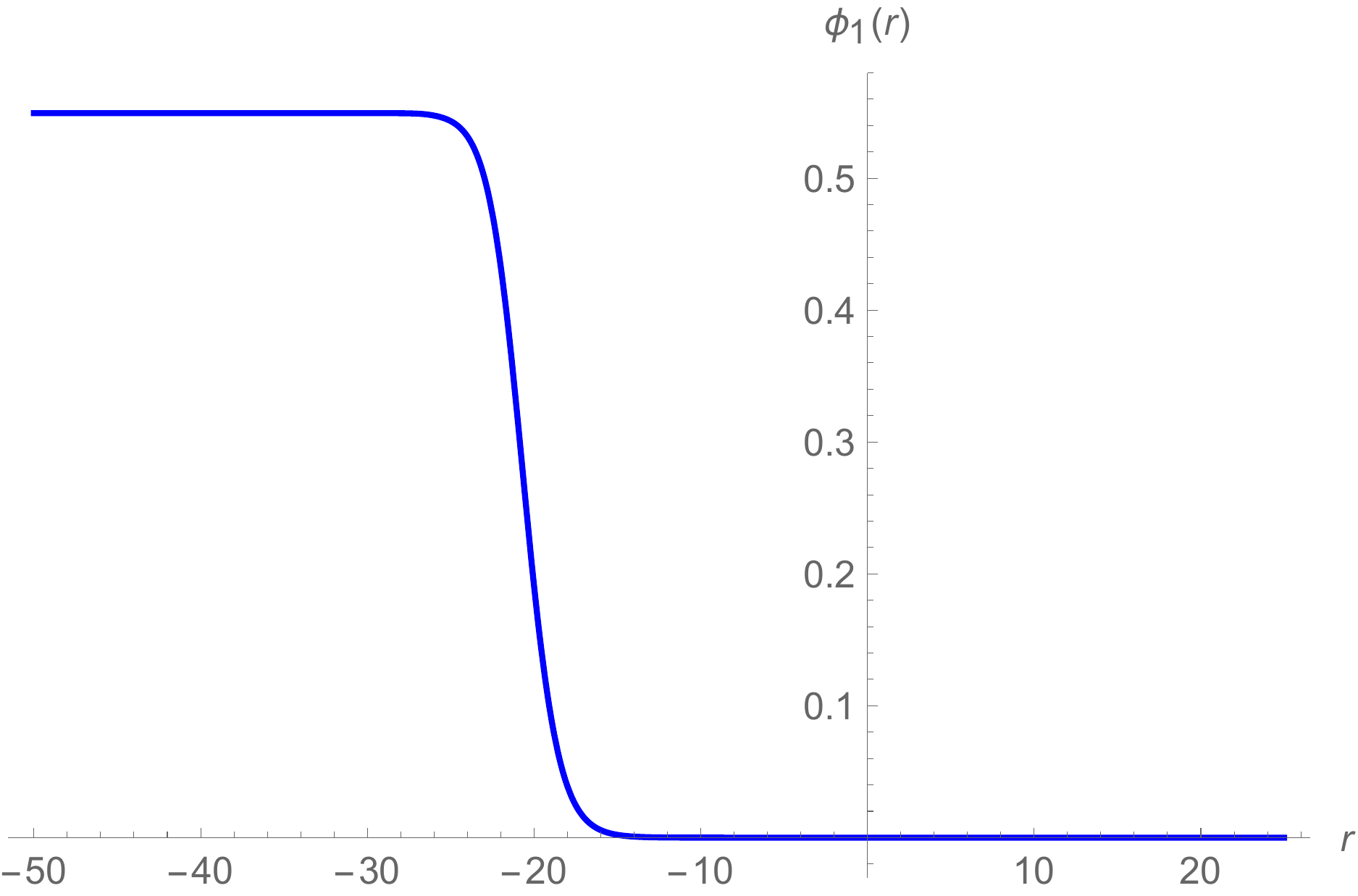}
  \caption{$\phi_1(r)$ solution}
  \end{subfigure}
  \begin{subfigure}[b]{0.45\linewidth}
    \includegraphics[width=\linewidth]{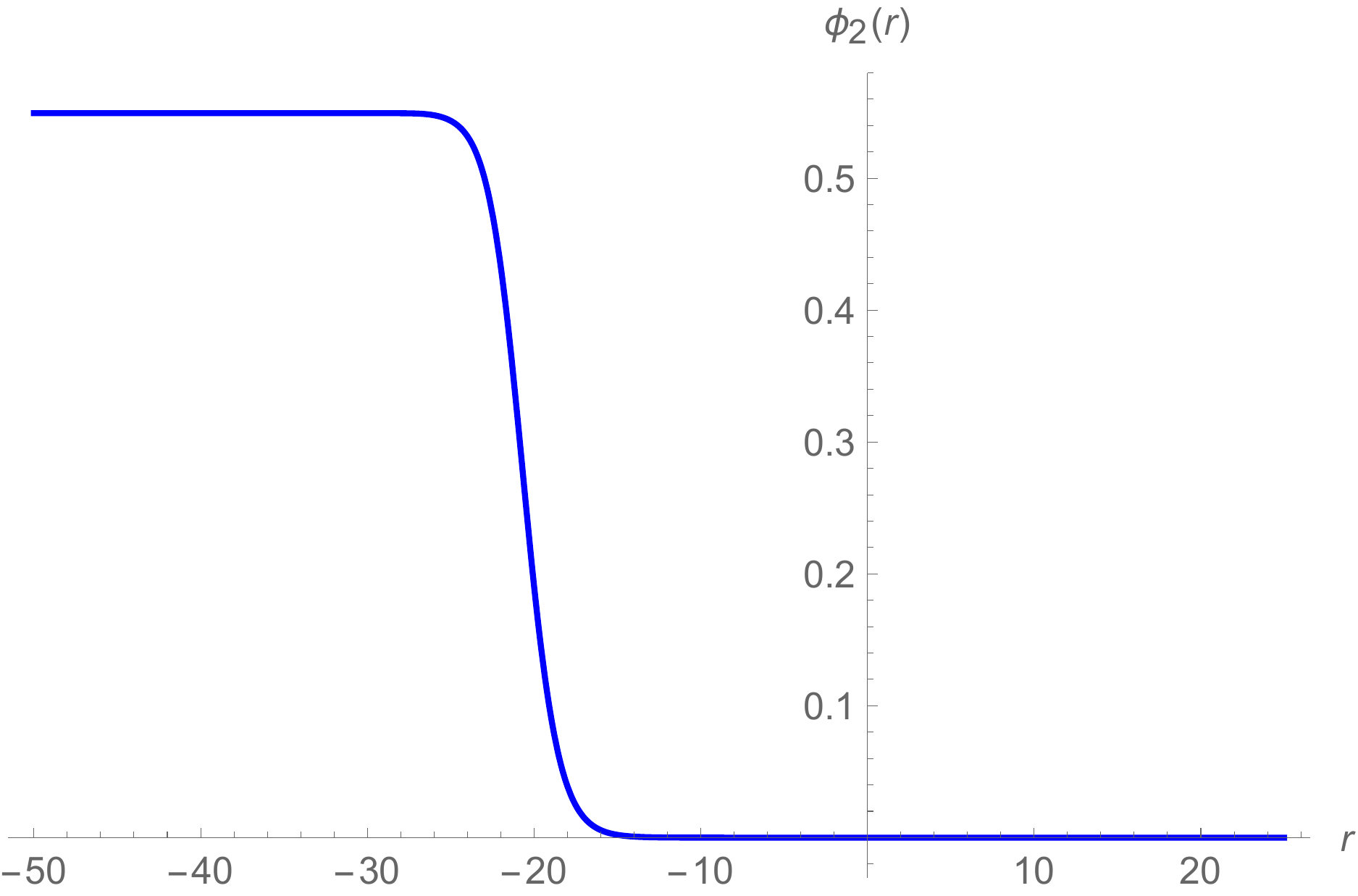}
  \caption{$\phi_2(r)$ solution}
  \end{subfigure}\\
   \begin{subfigure}[b]{0.45\linewidth}
    \includegraphics[width=\linewidth]{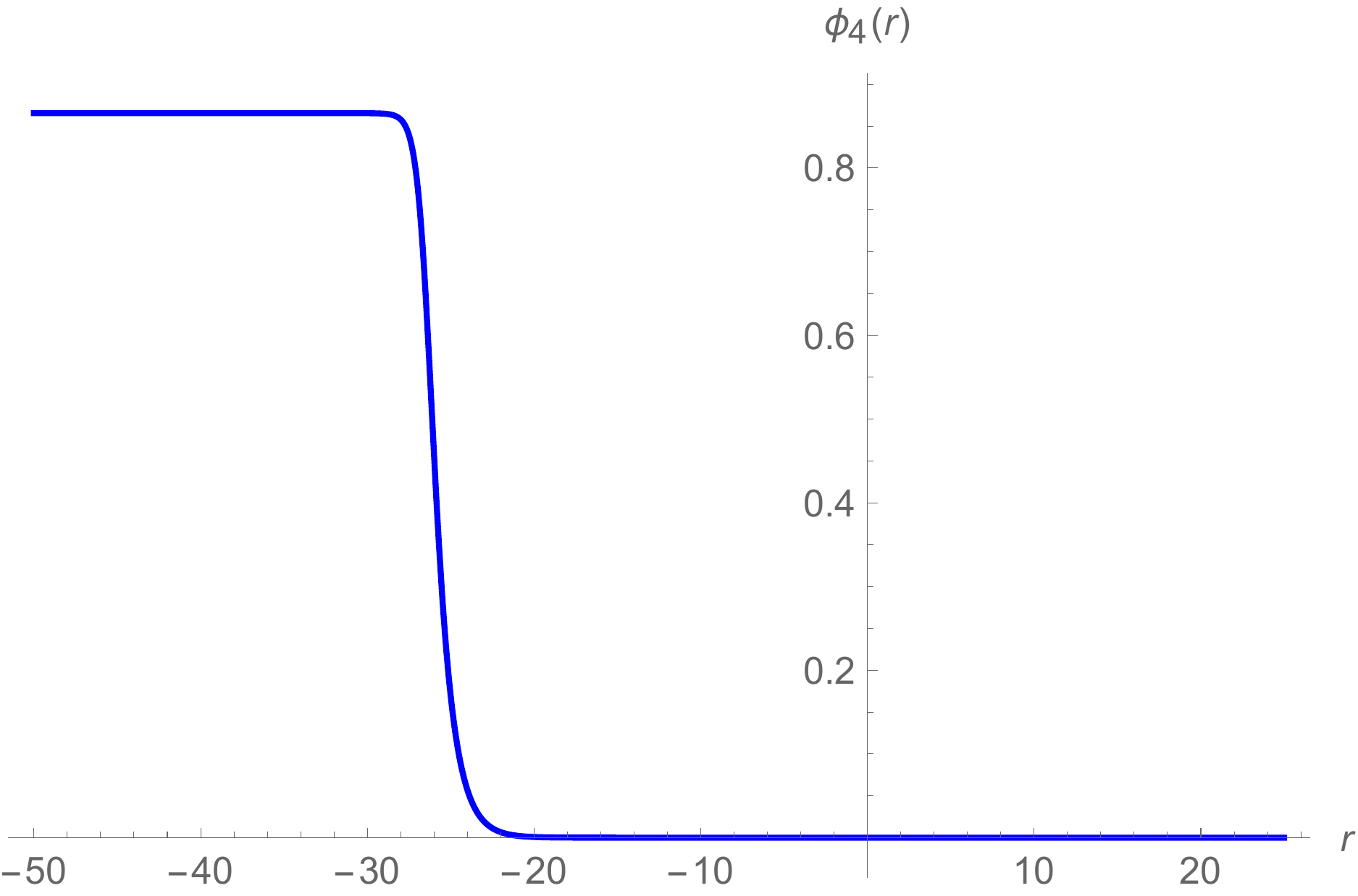}
  \caption{$\phi_4(r)$ solution}
   \end{subfigure} 
 \begin{subfigure}[b]{0.45\linewidth}
    \includegraphics[width=\linewidth]{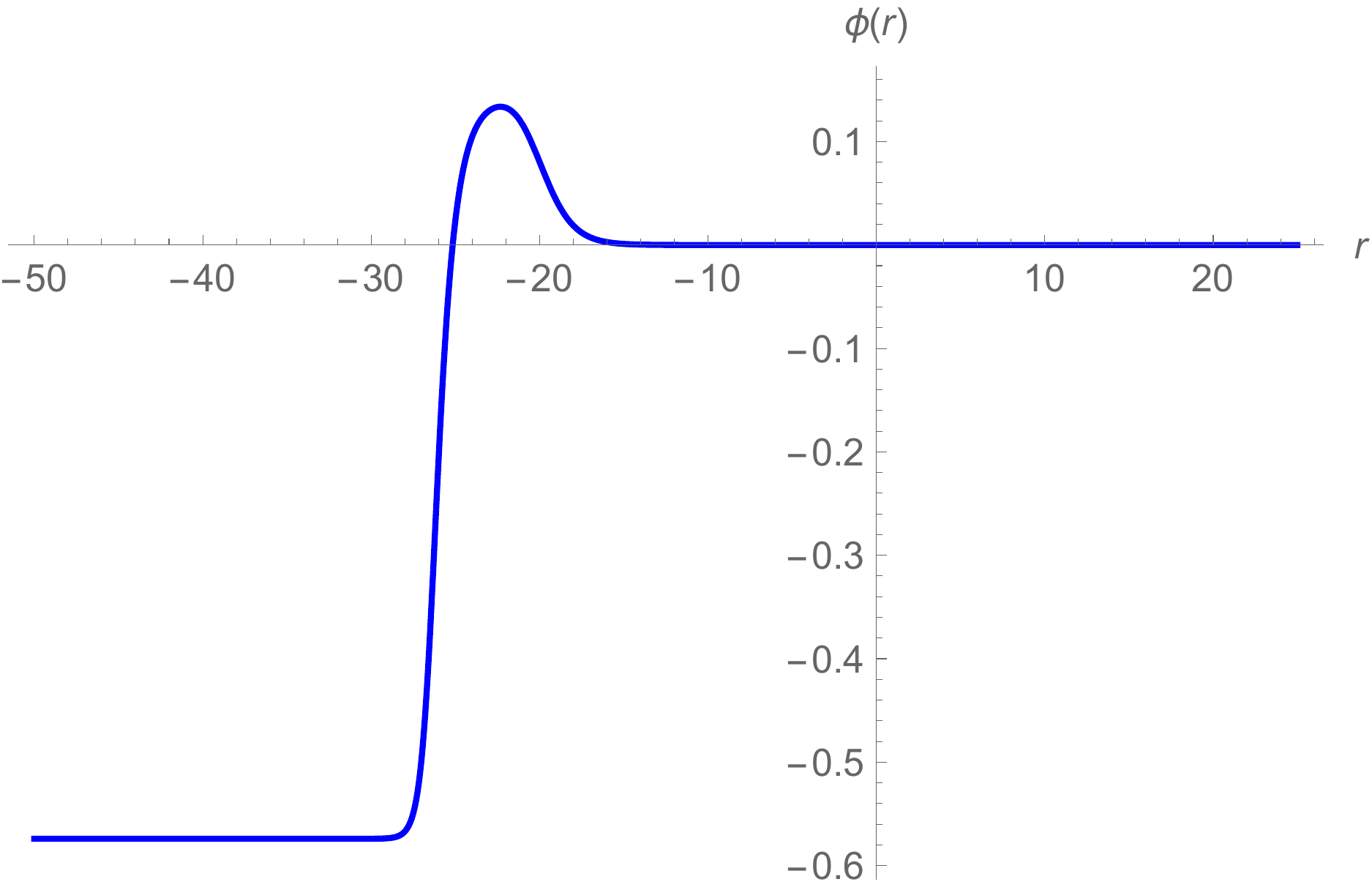}
  \caption{$\phi(r)$ solution}
   \end{subfigure} \\
   \begin{subfigure}[b]{0.45\linewidth}
    \includegraphics[width=\linewidth]{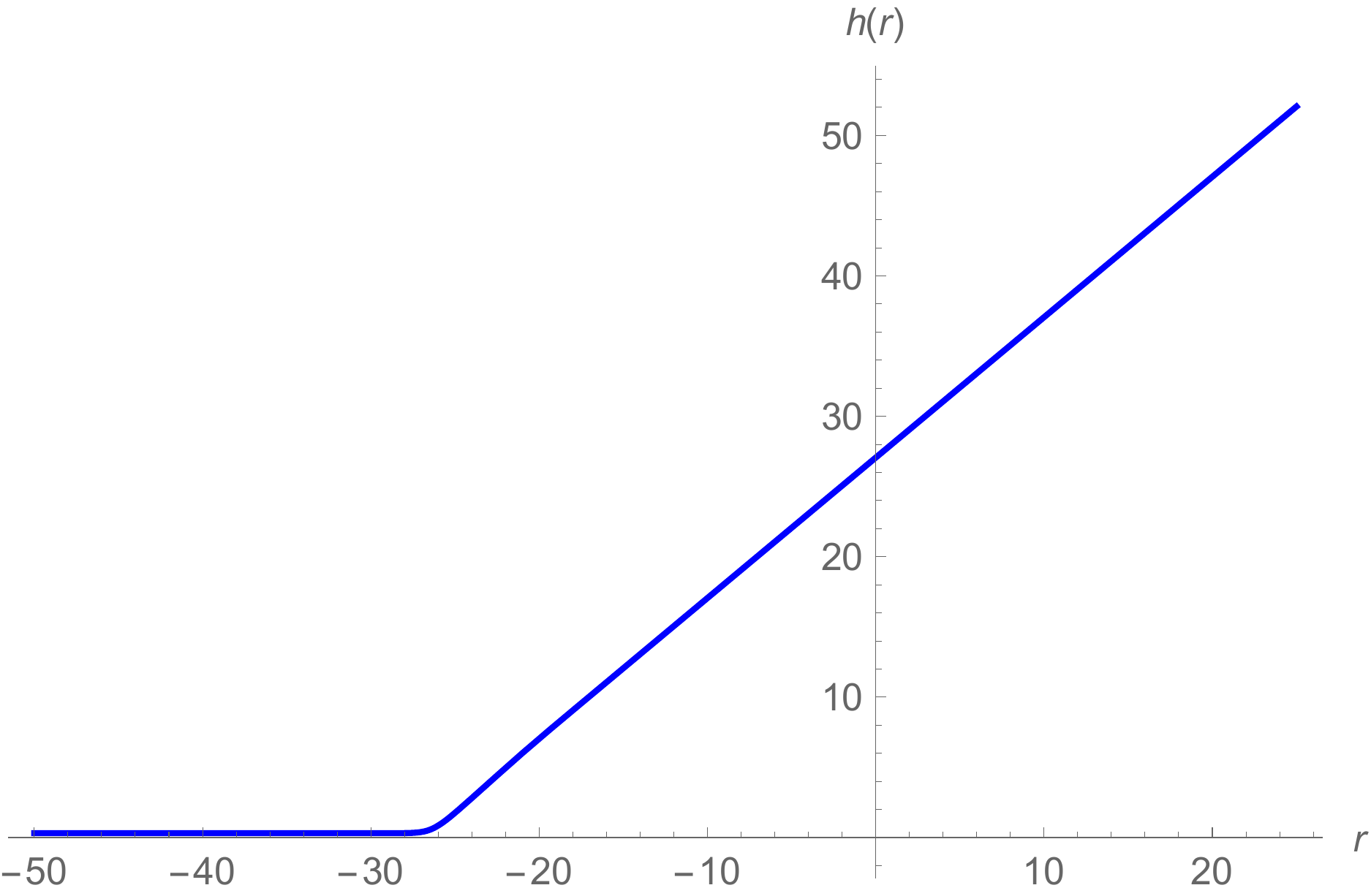}
  \caption{$h(r)$ solution}
   \end{subfigure}
 \begin{subfigure}[b]{0.45\linewidth}
    \includegraphics[width=\linewidth]{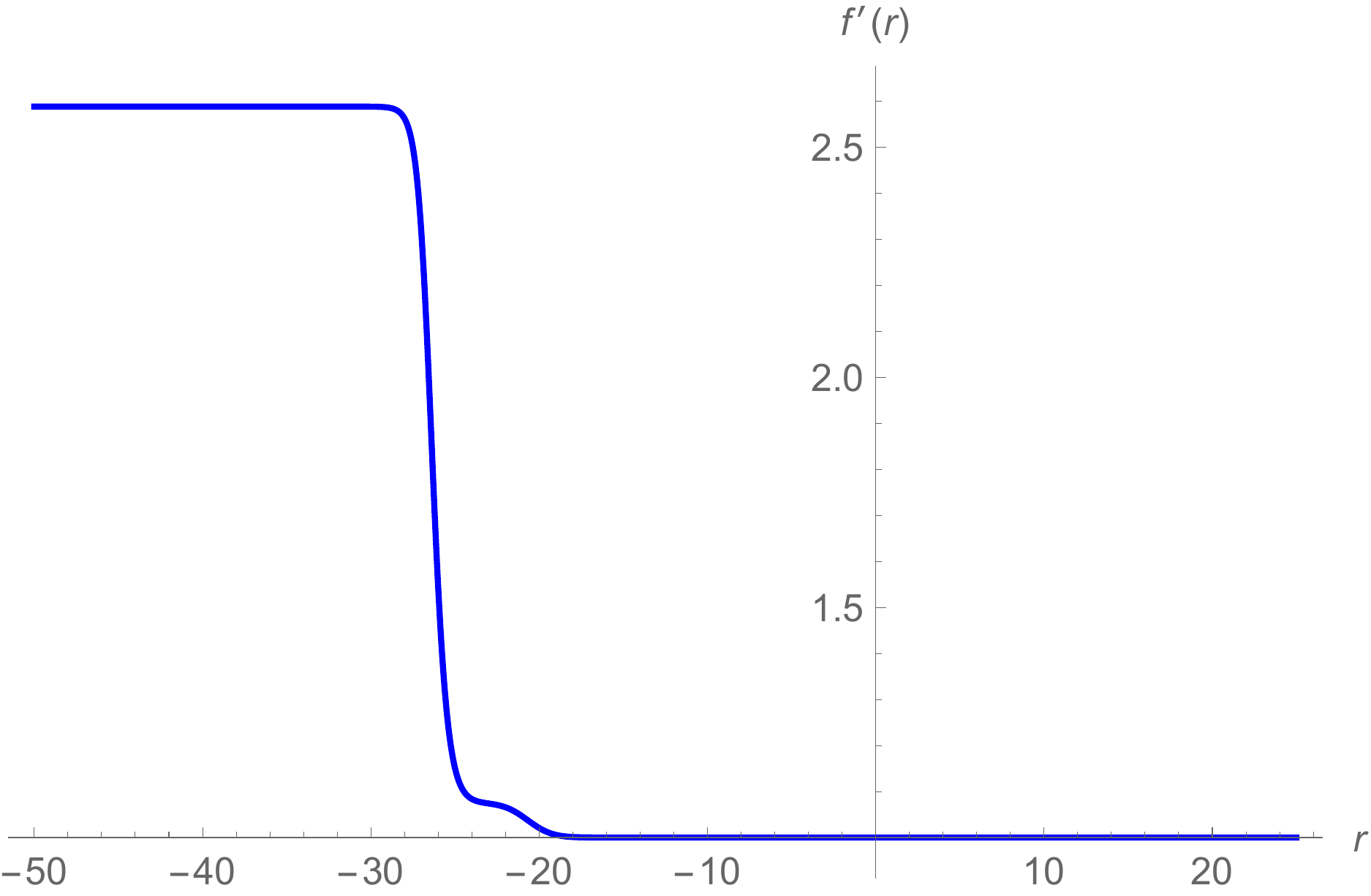}
  \caption{$f'(r)$ solution}
   \end{subfigure}
    \caption{A supersymmetric $AdS_4$ black hole with $AdS_2\times S^2$ horizon $(iii)$ for $g_2=g_1=1$, $\tilde{g}_1=2g_1$, $\tilde{g}_2=3g_2$ and $p_3=3$.}
  \label{Fig7}
\end{figure}

In this section, we repeat the same analysis for a smaller residual symmetry $SO(2)_{\textrm{diag}}\times SO(2)_{\textrm{diag}}$. As we will see, a new feature is the appearance of a number of non-trivial supersymmetric $AdS_4$ vacua. All of these vacua are not new but have recently been found in \cite{4D_N4_flows} to which we refer for more details. Since the analysis of $SO(2)_{\textrm{diag}}\times SO(2)_{\textrm{diag}}$ singlet scalars has not previously appeared, we will give more detail than the $SO(2)\times SO(2)\times SO(2)\times SO(2)$ sector considered in the previous section.  
\\ 
\indent We begin with the scalars from $SO(6,6)/SO(6)\times SO(6)$ coset which contains six singlets corresponding to the following non-compact generators
\begin{eqnarray}
Y_{11},\qquad Y_{11}+Y_{22},\qquad Y_{12}-Y_{21},\qquad Y_{66},\qquad Y_{44}+Y_{55},\qquad Y_{45}-Y_{54}\, .
\end{eqnarray}
The coset representative can be then written as
\begin{equation}
\mc{V}=e^{\phi_1 Y_{11}}e^{\phi_2 (Y_{11}+Y_{22})}e^{\phi_3 (Y_{12}-Y_{21})}e^{\phi_4 Y_{66}}e^{\phi_5 (Y_{44}+Y_{55})}e^{\phi_6 (Y_{45}-Y_{54})}\, .\label{SO2diag_2_coset}
\end{equation}
With this coset representative, scalar kinetic terms are given by
\begin{eqnarray}
e^{-1}\mc{L}_{\textrm{kin}}&=&-\frac{1}{4}({\phi'}^2-e^{-2\phi}{\chi'}^2)-\frac{1}{2}[{\phi_1'}^2+(1+\cosh4\phi_3){\phi_2'}^2+2{\phi_3'}^2]\nonumber \\
& &-\frac{1}{2}[{\phi_4'}^2+(1+\cosh4\phi_6){\phi_5'}^2+2{\phi_6'}^2].
\end{eqnarray}
The tensor $A_1^{ij}$ is proportional to the identity matrix of which the four-fold degenerate eigenvalue gives the superpotential only for $\chi=0$. Since the complete expressions are much more complicated and will not play any important role in subsequent analysis, we will only give the potential and superpotential for the case of $\chi=0$. These are given respectively by 
\begin{eqnarray}
V&=&\frac{1}{8}\left[g_1^2\cosh^2\phi_2[\cosh2(\phi_1-\phi_2)+\cosh2(\phi_1+\phi_2)-2\cosh2\phi_1-4]\times \right. \nonumber \\
& &\cosh^2\phi_2(\cosh\phi-\sin\phi)+g_2^2\cosh^2\phi_5(\cosh\phi+\sinh\phi)\times\nonumber \\
& &[\cosh2(\phi_4-\phi_5)+\cosh2(\phi_4+\phi_5)-2\cosh2\phi_4-4]\nonumber\\
& &+\tilde{g}_1^2[\cosh2(\phi_1-\phi_2)+\cosh2(\phi_1+\phi_2)+2\cosh2\phi_1-4]\times \nonumber \\
& &(\cosh\phi-\sinh\phi)\sinh^2\phi_2+16g_2\tilde{g}_1\cosh\phi_4\cosh^2\phi_5 \sinh\phi_1\times \nonumber \\
& &\sinh^2\phi_2-2g_1\tilde{g}_1(\cosh\phi-\sinh\phi)\sinh2\phi_1\sinh^2\phi_2 \nonumber \\
& &+\tilde{g}_2^2[\cosh2(\phi_4-\phi_5)+\cosh2(\phi_4+\phi_5)+2\cosh2\phi_4-4]\sinh^2\phi_5\times \nonumber \\
& &(\cosh\phi+\sinh\phi)+16g_1\tilde{g}_2\cosh\phi_1\cosh^2\phi_2\sinh\phi_4\sinh^2\phi_5\nonumber \\
& &-16g_1g_2\cosh\phi_1\cosh^2\phi_2\cosh\phi_4\cosh^2\phi_5 \nonumber \\
& &+16\tilde{g}_1\tilde{g}_2\sinh\phi_1\sinh^2\phi_2\sinh\phi_4\sinh^2\phi_5\nonumber \\
& &\left. -2g_2\tilde{g}_2e^\phi\sinh2\phi_4\sinh^22\phi_5\right]
\end{eqnarray}
and 
\begin{eqnarray}
W&=&\frac{1}{4}e^{-\frac{\phi}{2}}\left[g_1\cosh\phi_1(1+\cosh2\phi_2\cosh2\phi_3)+\tilde{g}_1\sinh\phi_1(1-\cosh2\phi_2\cosh2\phi_3) \right.\nonumber \\
& &\left.g_2e^\phi\cosh\phi_4(1+\cosh2\phi_5\cosh2\phi_6)+\tilde{g}_2e^\phi\sinh\phi_4(1-\cosh2\phi_5\cosh2\phi_6) \right].\nonumber \\
& &
\end{eqnarray}
\indent It is straightforward to verify that the superpotential admits the following four supersymmetric $AdS_4$ vacua
\begin{eqnarray}
\textrm{I}&:&\qquad \phi_\alpha=0,\quad \alpha=1,2,\ldots, 6,\quad \phi=\ln\left[\frac{g_1}{g_2}\right],\quad V_0=-3g_1g_2,\quad \\
\textrm{II}&:&\qquad \phi_\alpha=0,\quad \alpha=1,2,3,6,\qquad \phi_4=\pm \phi_5=\frac{1}{2}\ln\left[\frac{\tilde{g}_2+g_2}{\tilde{g}_2-g_2}\right],\nonumber \\
& &\qquad \phi=\frac{1}{2}\ln\left[\frac{g_1^2(\tilde{g}_2^2-g_2^2)}{g_2^2\tilde{g}_2^2}\right],\qquad V_0=-\frac{3g_1g_2\tilde{g}_2}{\sqrt{\tilde{g}_2^2-g_2^2}},\\
\textrm{III}&:&\qquad \phi_\alpha=0,\quad \alpha=3,4,5,6,\qquad \phi_1=\pm \phi_2=\frac{1}{2}\ln\left[\frac{\tilde{g}_1+g_1}{\tilde{g}_1-g_1}\right],\nonumber \\
& &\qquad \phi=-\frac{1}{2}\ln\left[\frac{g_2^2(\tilde{g}_1^2-g_1^2)}{g_1^2\tilde{g}_1^2}\right],\qquad V_0=-\frac{3g_1g_2\tilde{g}_1}{\sqrt{\tilde{g}_1^2-g_2^1}},
\end{eqnarray}
\begin{eqnarray}
\textrm{IV}&:&\qquad \phi_3=\phi_6=0,\qquad \phi_1=\pm \phi_2=\frac{1}{2}\ln\left[\frac{\tilde{g}_1+g_1}{\tilde{g}_1-g_1}\right],\nonumber \\
& &\qquad \phi_4=\pm \phi_5=\frac{1}{2}\ln\left[\frac{\tilde{g}_2+g_2}{\tilde{g}_2-g_2}\right],\qquad
 \phi=\ln\left[\frac{g_1\tilde{g}_1}{g_2\tilde{g}_2}\sqrt{\frac{\tilde{g}_2^2-g_2^2}{\tilde{g}_1^2-g_1^2}}\right],\nonumber \\
 & &\qquad V_0=-\frac{3g_1g_2\tilde{g}_1\tilde{g}_2}{\sqrt{(\tilde{g}_2^2-g_2^2)(\tilde{g}_1^2-g_1^2)}}\, .
\end{eqnarray}
All of these vacua have already been found in \cite{4D_N4_flows}, but we repeat them here for later convenience. We also note the unbroken gauge symmetries for these solutions which are given respectively by $SO(4)\times SO(4)$, $SO(4)\times SO(3)$, $SO(3)\times SO(4)$ and $SO(3)\times SO(3)$.  
\\
\indent To find supersymmetric $AdS_4$ black hole solutions, we now turn to the analysis of Yang-Mills equations. To implement the $SO(2)_{\textrm{diag}}\times SO(2)_{\textrm{diag}}$ symmetry, we impose the following conditions on the gauge fields 
\begin{equation}
g_1A^{3+}=-\tilde{g}_1A^{9+}\qquad \textrm{and}\qquad g_2A^{6-}=-\tilde{g}_2A^{12-}
\end{equation}
which lead to the same composite connection given in \eqref{N4_composite}. Therefore, the twist conditions and relevant projectors are the same.
\\
\indent Unlike the $SO(2)\times SO(2)\times SO(2)\times SO(2)$ case, the YM currents are non-vanishing in this case. From equation \eqref{YM1}, we find 
\begin{eqnarray}
 D\mc{H}^{3-}&=&\frac{1}{2}g_1(\cosh2\phi_2\sinh4\phi_3\phi_2'-2\sinh2\phi_2\phi_3')*dr,\\
 D\mc{H}^{9-}&=&\frac{1}{2}\tilde{g}_1(\cosh2\phi_2\sinh4\phi_3\phi_2'-2\sinh2\phi_2\phi_3')*dr
\end{eqnarray}
which, from the ansatz of the gauge fields, imply that $b_3$ and $b_9$ are constant and
\begin{equation}
\phi_2=0\qquad \textrm{or}\qquad \phi_3=0\, .
\end{equation}
Similarly, equation \eqref{YM2} gives
\begin{eqnarray}
 D\mc{H}^{6+}&=&-\frac{1}{2}g_2(\cosh2\phi_5\sinh4\phi_6\phi_5'-2\sinh2\phi_5\phi_6')*dr,\\
 D\mc{H}^{12+}&=&-\frac{1}{2}\tilde{g}_2(\cosh2\phi_5\sinh4\phi_6\phi_5'-2\sinh2\phi_5\phi_6')*dr
\end{eqnarray}
which lead to constant $b_6$ and $b_{12}$ together with
\begin{equation}
\phi_5=0\qquad \textrm{or}\qquad \phi_6=0\, .
\end{equation}
\indent We also note that the radial component of the composite connection is given by
\begin{equation}
{Q_{ri}}^j=-\cosh\phi_3\sinh\phi_3\phi_2'{(i\sigma_2\otimes \sigma_1)_i}^j-\cosh\phi_6\sinh\phi_6\phi_5'{(\sigma_1\otimes i\sigma_2)_i}^j
\end{equation}
which identically vanishes whenever $\phi_2=0$ or $\phi_3=0$ and $\phi_5=0$ or $\phi_6=0$. In order to find solutions interpolating between supersymmetric $AdS_4$ vacua identified above, we will choose a definite choice
\begin{equation}
\phi_3=\phi_6=0\, .
\end{equation}
\indent We then consider equation \eqref{YM3}. Equations for $\mc{H}^{3-}$ and $\mc{H}^{9-}$ give
\begin{eqnarray}
\tilde{A}_t^{3'}&=&\frac{\kappa p_3}{\tilde{g}_1}e^{\phi+f-2h}(g_1\sinh2\phi_1-\tilde{g}_1\cosh2\phi_1),\\
\tilde{A}_t^{9'}&=&\frac{\kappa p_3}{\tilde{g}_1}e^{\phi+f-2h}(\tilde{g}_1\sinh2\phi_1-g_1\cosh2\phi_1)
\end{eqnarray}
together with
\begin{equation}
A_t^{3'}=-\frac{\tilde{g}_1}{g_1}A_t^{9'}=\frac{g_1e^{f-\phi-2h}(\kappa e_3-g_1b_3)}{\tilde{g}_1\cosh2\phi_1-g_1\sinh2\phi_1}=\frac{\tilde{g}_1e^{f-\phi-2h}(\kappa e_9-g_2b_9)}{g_1\cosh2\phi_1-\tilde{g}_1\sinh2\phi_1}\, .
\end{equation}
For $\tilde{g}_1\neq g_1$ which is needed for the existence of non-trivial $AdS_4$ vacua, the last equation implies 
\begin{equation}
e_3=e_9=b_3=b_9=0
\end{equation}
which in turn gives 
\begin{equation}
A_t^{3'}=A_t^{9'}=0\, .
\end{equation}
Similarly, equations for $\mc{H}^{6-}$ and $\mc{H}^{12-}$ give
\begin{equation}
p_{12}=p_{6}=b_6=b_{12}=0\qquad \textrm{and}\qquad \tilde{A}^{6'}_t=\tilde{A}^{12'}_t=0
\end{equation}
together with
\begin{eqnarray}
A_t^{6'}&=&\frac{\kappa e_6}{\tilde{g}_2}e^{f-2h-\phi}(\tilde{g}_2\cosh2\phi_4-g_2\sinh2\phi_4),\\
A_t^{12'}&=&\frac{\kappa e_6}{\tilde{g}_2}e^{f-2h-\phi}(g_2\cosh2\phi_4-\tilde{g}_2\sinh2\phi_4).
\end{eqnarray}
\indent With $\chi=\phi_3=\phi_6=0$, we find that both $\mc{W}$ and $\mc{Z}$ are real and given by
\begin{eqnarray}
\mc{W}&=&\frac{1}{2}e^{-\frac{\phi}{2}}(g_1\cosh\phi_1\cosh^2\phi_2-\tilde{g}_1\sinh\phi_1\sinh^2\phi_2)\nonumber \\
& &+\frac{1}{2}e^{\frac{\phi}{2}}(g_2\cosh\phi_4\cosh^2\phi_5-\tilde{g}_2\sinh\phi_4\sinh^2\phi_5),\\
\mc{Z}&=&-\frac{\kappa}{2\tilde{g}_1\tilde{g_2}}e^{-2h-\frac{\phi}{2}}\left[e^\phi p_3\tilde{g}_2(\tilde{g}_1\cosh\phi_1-g_1\sinh\phi_1)\right. \nonumber \\
& &\left.+e_6\tilde{g}_1(g_2\sinh\phi_4-\tilde{g}_2\cosh\phi_4)\right].
\end{eqnarray}
It can be readily verified that critical points I, II, III, and IV are critical points of $\mc{W}$ as expected for supersymmetric vacua. 
\\
\indent As in the previous case, there are two possible topological twists, $N=4$ and $N=2$ twists. The $N=4$ twists do not give rise to any $AdS_2\times \Sigma^2$ fixed points, so we will only give the results on $N=2$ twists. Since both $\mc{W}$ and $\mc{Z}$ are real, we find the phase $e^{i\Lambda}=\pm 1$, and the BPS equations are given by
\begin{eqnarray}
f'&=&|\mc{W}-\mc{Z}|\nonumber \\
&=&\frac{1}{2}e^{-\frac{\phi}{2}}\left[g_1\cosh\phi_1\cosh^2\phi_2-\tilde{g}_1\sinh\phi_1\sinh^2\phi_2\right. \nonumber \\ 
& &\left.+g_2e^\phi\cosh\phi_4\cosh^2\phi_5-\tilde{g}_2
e^\phi\sinh\phi_4\sinh^2\phi_5\right]\nonumber \\
& &-\kappa\frac{e^{-\frac{\phi}{2}-2h}}{\tilde{g}_1\tilde{g}_2}\left[e_6\tilde{g}_1(\tilde{g}_2\cosh\phi_4-g_2\sinh\phi_4)-\right. \nonumber \\ 
& &\left.e^\phi\tilde{g}_2p_3(\tilde{g}_1\cosh\phi_1-g_1\sinh\phi_1)\right],\\
 h'&=&|\mc{W}+\mc{Z}|\nonumber \\
 &=&\frac{1}{2}e^{-\frac{\phi}{2}}\left[g_1\cosh\phi_1\cosh^2\phi_2-\tilde{g}_1\sinh\phi_1\sinh^2\phi_2\right. \nonumber \\ 
& &\left.+g_2e^\phi\cosh\phi_4\cosh^2\phi_5-\tilde{g}_2
e^\phi\sinh\phi_4\sinh^2\phi_5\right]\nonumber \\
& &+\kappa\frac{e^{-\frac{\phi}{2}-2h}}{\tilde{g}_1\tilde{g}_2}\left[e_6\tilde{g}_1(\tilde{g}_2\cosh\phi_4-g_2\sinh\phi_4)\right. \nonumber \\ 
& &\left.-e^\phi\tilde{g}_2p_3(\tilde{g}_1\cosh\phi_1-g_1\sinh\phi_1)\right],\\
\phi'&=&-4\frac{\pd |\mc{W}+\mc{Z}|}{\pd \phi}\nonumber \\
&=&
e^{-\frac{\phi}{2}}\left[g_1\cosh\phi_1\cosh^2\phi_2-\tilde{g}_1\sinh\phi_1\sinh^2\phi_2\right. \nonumber \\ 
& &\left.+e^\phi(\tilde{g}_2\sinh\phi_4\sinh^2\phi_5-g_2\cosh\phi_4\cosh^2\phi_5)\right]\nonumber \\
& &+\frac{\kappa}{\tilde{g}_1\tilde{g}_2}e^{-2h-\frac{\phi}{2}}\left[e^\phi\tilde{g}_2p_3(\tilde{g}_1\cosh\phi_1-g_1\sinh\phi_1)\right. \nonumber \\ 
& &\left.+e_6\tilde{g}_1(\tilde{g}_2\cosh\phi_4-g_2\sinh\phi_4)\right],\\
\phi_1'&=&-2\frac{\pd |\mc{W}+\mc{Z}|}{\pd \phi_1}\nonumber \\
&=&e^{-\frac{\phi}{2}}(\tilde{g}_1\cosh\phi_1\sinh^2\phi_2-g_1\cosh^2\phi_2\sinh\phi_1)\nonumber \\ 
& &+\frac{\kappa p_3}{\tilde{g}_1}e^{-2h+\frac{\phi}{2}}(\tilde{g}_1\sinh\phi_1-g_1\cosh\phi_1),\\
\phi_2'&=&-\frac{\pd |\mc{W}+\mc{Z}|}{\pd \phi_2}\nonumber \\
&=&e^{-\frac{\phi}{2}}\cosh\phi_2\sinh\phi_2(\tilde{g}_1\sinh\phi_1-g_1\cosh\phi_1),\\
\phi_4'&=&-2\frac{\pd |\mc{W}+\mc{Z}|}{\pd \phi_4}\nonumber \\
&=&e^{\frac{\phi}{2}}(\tilde{g}_2\cosh\phi_4\sinh^2\phi_5-g_2\cosh^2\phi_5\sinh\phi_4)\nonumber \\
& &-\frac{\kappa e_6}{\tilde{g}_2}e^{-2h-\frac{\phi}{2}}(\tilde{g}_2\sinh\phi_4-g_2\cosh\phi_4),
\end{eqnarray}
\begin{eqnarray}
\phi_5'&=&-\frac{\pd |\mc{W}+\mc{Z}|}{\pd \phi_5}\nonumber \\
&=&e^{\frac{\phi}{2}}\cosh\phi_5\sinh\phi_5(\tilde{g}_2\sinh\phi_4-g_2\cosh\phi_4).
\end{eqnarray}
\indent From $\phi_2'$ and $\phi_5'$ equations, we immediately see that there are four possibilities for $AdS_2\times \Sigma^2$ fixed points to exist:
\begin{eqnarray}
i&:&\qquad \phi_2=\phi_5=0,\nonumber\\
ii&:&\qquad \phi_2=0 \quad \textrm{and}\quad \phi_4=\frac{1}{2}\ln\left[\frac{\tilde{g}_2+g_2}{\tilde{g}_2-g_2}\right],\nonumber \\
iii&:&\qquad \phi_5=0\quad \textrm{and}\quad  \phi_1=\frac{1}{2}\ln\left[\frac{\tilde{g}_1+g_1}{\tilde{g}_1-g_1}\right] \nonumber \\   iv&:&\qquad \phi_1=\frac{1}{2}\ln\left[\frac{\tilde{g}_1+g_1}{\tilde{g}_1-g_1}\right] \quad \textrm{and}\quad \phi_4=\frac{1}{2}\ln\left[\frac{\tilde{g}_2+g_2}{\tilde{g}_2-g_2}\right].
 \end{eqnarray}
 These coincide with the values of scalars at supersymmetric $AdS_4$ vacua I, II, III and IV. However, the last possibility does not lead to any $AdS_2\times \Sigma^2$ fixed points. We then consider only the remaining three cases:
 \begin{itemize}
 \item $i$: In this case, we set $\phi_2=\phi_5=0$ and find an $AdS_2\times \Sigma^2$ fixed point given by
 \begin{eqnarray}
 h&=&\frac{1}{2}\phi+\frac{1}{2}\ln\left[\frac{\kappa p_3(\tilde{g}_1-g_1\coth\phi_1)}{g_1\tilde{g}_1}\right],\nonumber \\
 \phi&=&\frac{1}{2}\ln\left[\frac{e_6g_1\tilde{g}_1(g_2\coth\phi_4-\tilde{g}_2)}{p_3g_2\tilde{g}_2(\tilde{g}_1-g_1\coth\phi_1)}\right],\nonumber \\
\phi_1&=&\frac{1}{2}\ln\left[\frac{g_1(g_2\cosh2\phi_4-\tilde{g}_2\sinh2\phi_4)}{g_2(\tilde{g}_1-g_1)} \right.\nonumber \\
& &\left.+\frac{\sqrt{g_2^2(\tilde{g}_1^2-g_1^2)+g_1^2(g_2\cosh2\phi_4-\tilde{g}_2\sinh2\phi_4)^2}}{g_2(\tilde{g}_1-g_1)}\right],\nonumber \\
 \phi_4&=&\frac{1}{2}\ln\left[\frac{e_6^2g_2^4\tilde{g}_1^2\tilde{g}_2+2e_6g_1g_2^3\tilde{g}_1^2\tilde{g}_2p_3+g_1^4\tilde{g}_2^3p_3^2+g_2\sqrt{X}}{(\tilde{g}_2-g_2)(g_1^4\tilde{g}_2^2p_3^2-e_6^2g_2^4\tilde{g}_1^2)}\right]
 \end{eqnarray}
for 
\begin{eqnarray}
X&=&e_6^4g_2^8\tilde{g}_1^4+4e_6^3g_1g_2^5\tilde{g}_1^4\tilde{g}_2^2p_3+2e_6^2g_1^2g_2^2\tilde{g}_1^2\tilde{g}_2^2
[2g_2^2\tilde{g}_1^2-g_1^2(g_2^2-2\tilde{g}_2^2)]p_3^2\nonumber \\
& & +4e_6g_1^5g_2\tilde{g}_1^2\tilde{g}_2^4p_3^3+g_1^8\tilde{g}_2^4p_3^4\, .
\end{eqnarray}
  \item $ii$: In this case, we have $\phi_2=0$ and 
  \begin{eqnarray}
  \phi_4&=&\phi_5=\frac{1}{2}\ln\left[\frac{\tilde{g}_2+g_2}{\tilde{g}_2-g_2}\right],\nonumber \\
  h&=&\frac{1}{2}\ln\left[\frac{\kappa p_3e^\phi(\tilde{g}_1-g_1\coth\phi_1)}{g_1\tilde{g}_1}\right],\nonumber \\
  \phi&=&\ln\left[\frac{\sqrt{\tilde{g}_2^2-g_2^2}\left[2g_1p_3(g_1\cosh2\phi_1-\tilde{g}_1\sinh2\phi_1)+\sqrt{2g_1p_3Y}\right]}{4g_2\tilde{g}_2p_3(g_1\cosh\phi_1-\tilde{g}_1\sinh\phi_1)}\right],\nonumber \\
  \phi_1&=&\frac{1}{2}\ln\left[\frac{2e_6g_2\tilde{g}_1+g_1\tilde{g}_1p_3+\sqrt{4e_6^2g_2^2\tilde{g}_1^2+4e_6g_1g_2\tilde{g}_1^2p_3+g^4_1p_3^2}}{g_1p_3(g_1-\tilde{g}_1)}\right]
  \end{eqnarray}
  with 
  \begin{eqnarray}
  Y&=&g_1p_3(\tilde{g}_1^2+g_1^2)\cosh4\phi_1-4g_1\tilde{g}_1\sinh2\phi_1(e_6g_2+g_1p_3\cosh2\phi_1)\nonumber \\
  & &+g_1^3p_3-\tilde{g}_1^2(4e_6g_2+g_1p_3)+4e_6g_2\tilde{g}_1^2\cosh2\phi_1\, .
  \end{eqnarray}
\item $iii$: For this final possibility, we have $\phi_5=0$ and
\begin{eqnarray}
\phi_1&=&\phi_2=\frac{1}{2}\ln\left[\frac{\tilde{g}_1+g_1}{\tilde{g}_1-g_1}\right],\nonumber \\
h&=&\frac{1}{2}\ln\left[\frac{\kappa e_6 e^{-\phi}[g_2(1+e^{2\phi_4})+\tilde{g}_2(1-e^{2\phi_4})]}{g_2\tilde{g}_2(e^{2\phi_4}-1)}\right],\nonumber \\
\phi&=&\ln\left[\sqrt{e_6g_2(e_6g_2^3+2g_1\tilde{g}_2^2p_3-2g_1\tilde{g}_2^2p_3\cosh2\phi_4
+2g_1g_2\tilde{g}_2p_3\sinh2\phi_4)}\right.\nonumber \\
& &\left.\phantom{\sqrt{g_2^3}} +e_6g_2^2\right]+\ln\left[\frac{\tilde{g}_1e^{\phi_4}(\coth\phi_4-1)}{2g_2\tilde{g}_2p_3\sqrt{\tilde{g}_1^2-g_1^2}}\right],\nonumber \\
\phi_4&=&\frac{1}{2}\ln\left[\frac{\tilde{g}_2(e_6g_2+2g_1p_3)+\sqrt{e_6^2g_2^4+4e_6g_1g_2\tilde{g}_2^2p_3+4g_1^2\tilde{g}_2^2p_3^2}}{e_6g_2(g_2-\tilde{g}_2)}\right].
\end{eqnarray}
 \end{itemize}
In each case, we have not explicitly given the expressions for $L_{AdS_2}$ due to their complexity. These can be obtained from $f'$ equation by using the values of the other fields at the fixed points. We have verified that all the above three cases indeed lead to valid $AdS_2\times \Sigma^2$ fixed points in each case. This will also be clearly seen later in numerical analyses.  
\\
\indent For critical point $i$, we obtain only $AdS_2\times H^2$ solutions with $\kappa =-1$. Examples of solutions interpolating between the supersymmetric $AdS_4$ critical point I and these $AdS_2\times H^2$ geometries are shown in figure \ref{Fig5} for $g_2=g_1=1$, $\tilde{g}_1=2g_1$, $\tilde{g}_2=3g_2$ and $p_3=-3,-3.00000025,-3.005$. The reason for choosing values of $p_3$ very close to each other is for the convenience in the presentation. The numerical plots for solutions in which the values of $p_3$ are widely separated are very far from each other.  
\\
\indent For critical point $ii$, we have found only $AdS_2\times H^2$ solutions as in critical point $i$. An example of the solutions interpolating between supersymmetric $AdS_4$ critical points I and II and an $AdS_2\times H^2$ geometry with $g_2=g_1=1$, $\tilde{g}_1=2g_1$, $\tilde{g}_2=3g_2$ and $p_3=-3$ is shown in figure \ref{Fig6}. We have set $\phi_2=0$ along the entire solution. We also note that the solution indeed exhibits an intermediate $AdS_4$ critical point II with the value $\phi=-0.05889$ given by the chosen values of various parameters in this solution. 
\\
\indent Unlike the previous two cases, in critical point $iii$, we only find $AdS_2\times S^2$ solutions. An example of flow solutions is shown in figure \ref{Fig7} with $g_2=g_1=1$, $\tilde{g}_1=2g_1$, $\tilde{g}_2=3g_2$ and $p_3=3$. Along the entire flow, we have set $\phi_5=0$. As in the flow solution to $AdS_2\times H^2$ critical point $ii$, the solution exhibits an intermediate $AdS_4$ critical point III with $\phi=0.143841$, so the solution interpolates between $AdS_4$ critical points I and II and $AdS_2\times S^2$ geometry in the IR. The solutions in this case and the flow to critical point $ii$ are similar to solutions describing RG flows across dimensions in half-maximal gauged supergravities in five, six and seven dimensions \cite{5Dtwist,5D_N4_flow,6D_twist,7D_twist}. Moreover, there also exist solutions that flow directly from $AdS_4$ critical point I to these $AdS_2\times S^2$ and $AdS_2\times H^2$ fixed points. We will not give these solutions here since they are similar to the solutions in $SO(2)\times SO(2)\times SO(2)\times SO(2)$ case without non-trivial $AdS_4$ vacua. 
\\
\indent We end this section by noting that there do not exist any $AdS_2\times \Sigma^2$ fixed points for case $iv$ discussed above. Therefore, there are no flow solutions from the supersymmetric $AdS_4$ vacuum IV to $AdS_2\times \Sigma^2$ geometries in the IR. This is in line with the $N=3$ gauged supergravity studied in the previous section in which no $AdS_2\times \Sigma^2$ fixed points exist for RG flows involving the non-trivial $N=3$ $AdS_4$ critical point with $SO(3)$ symmetry. On the other hand, as we have seen above, $AdS_2\times \Sigma^2$ critical points $ii$ and $iii$ do exist and are connected to non-trivial $AdS_4$ critical points II and III. However, the latter do not have an analogue in the case of $N=3$ gauged supergravity.  

\section{Conclusions and discussions}\label{conclusion}
We have studied a number of supersymmetric black hole solutions in asymptotically $AdS_4$ space from matter-coupled $N=3$ and $N=4$ gauged supergravities. In $N=3$ theory, we have found an $AdS_2\times H^2$ solution with $SO(2)\times SO(2)$ symmetry. We have also given a complete solution interpolating between $SO(3)\times SO(3)$ symmetric $AdS_4$ vacuum and this $AdS_2\times H^2$ geometry with a non-vanishing scalar. The resulting solution has a very similar structure to those given in $N=5,6$ gauged supergravities. The solution with vanishing scalars is a solution of pure $N=3$ gauged supergravity and can be embedded in massive type IIA theory using the result of \cite{Varela_pureN3}. We have also shown that there are no $AdS_4$ black hole solutions with $SO(2)_{\textrm{diag}}$ symmetry. Therefore, in $N=3$ gauged supergravity under consideration here, it is clear that there are no other solutions. 
\\
\indent Although we have considered only a particular case of three vector multiplets, it has been shown in \cite{Jan_maximal_AdS} that the $SO(3)_R\subset SU(3)_R$ symmetry must be gauged in order for the gaugings to admit a supersymmetric $AdS_4$ vacuum. This is also an essential part in performing topological twists since the gravitini and Killing spinors are charged exclusively under this symmetry or a diagonal subgroup with parts of the symmetry of vector multiplets. Therefore, even with extra vector multiplets and possibly larger gauge groups, the structure of the topological twists should be the same and eventually leads to a similar conclusion.  
\\
\indent In pure $N=4$ gauged supergravity, we have recovered an $AdS_2\times H^2$ solution studied in \cite{flow_acrossD_bobev}. However, we have included a non-vanishing axion and given the interpolating solutions between this geometry and the supersymmetric $AdS_4$ vacuum. For matter-coupled $N=4$ gauged supergravity, we have found a number of $AdS_2\times S^2$ and $AdS_2\times H^2$ solutions with $SO(2)\times SO(2)\times SO(3)\times SO(2)$ symmetry. We have also given various examples of numerical solutions interpolating between these geometries and the $AdS_4$ vacuum with $SO(4)\times SO(4)$ symmetry. The BPS equations are very complicated, and we are not able to completely carry out the analysis. However, we have given a number of possible $AdS_4$ black hole solutions with both spherical and hyperbolic horizons. We note that unlike $N=5$ and $N=6$ gauged supergravities, there exist matter multiplets in $N=4$ theory, and the two $SO(2)$ factors involving in the twists are not necessarily equal though related, see the twist condition in \eqref{twist_N2_N4}. This gives a weaker constraint on the charges and leaves more freedom to find $AdS_2\times \Sigma^2$ solutions. This is also supported by the fact that, when restricted to the case of pure $N=4$ gauged supergravity, the charges of $A^{3+}$ and $A^{6-}$ must be equal, and only one $AdS_2\times H^2$ solution which is an analogue of similar solutions in $N=5,6$ theories exists.  
\\
\indent We have also found $AdS_2\times S^2$ and $AdS_2\times H^2$ solutions with $SO(2)_{\textrm{diag}}\times SO(2)_{\textrm{diag}}$ symmetry. Similar to the $N=3$ theory, in this case, we have performed a complete analysis and classified all possible supersymmetric $AdS_2\times \Sigma^2$ solutions with the aforementioned residual symmetry at least for the case of six vector multiplets. In this case, apart from the trivial $AdS_4$ critical point with the full $SO(4)\times SO(4)$ symmetry, there exist additional three supersymmetric $AdS_4$ vacua with $SO(4)\times SO(3)$, $SO(3)\times SO(4)$ and $SO(3)\times SO(3)$ symmetries. Except for the last critical point, we have found black hole solutions interpolating between these vacua and $AdS_2\times S^2$ and $AdS_2\times H^2$ geometries. We hope all these solutions could be useful in black hole physics and  holographic studies of twisted compactifications of $N=3$ and $N=4$ SCFTs in three dimensions on a Riemann surface.   
\\
\indent It is interesting to look for more general solutions in the $SO(2)\times SO(2)\times SO(2)\times SO(2)$ case in particular solutions carrying both electric and magnetic charges of the same gauge fields. In this paper, we have given only some representative examples of the possible solutions which carry either electric or magnetic charges of a given gauge field. Another direction is to find an embedding of the solutions given here in string/M-theory. Solutions in pure $N=3$ and $N=4$ gauged supergravities can be embedded in ten and eleven dimensions using consistent truncations given respectively in \cite{Varela_pureN3,N010_truncation_Cassani} and \cite{Pope_4DpureN4}. It would be useful to find similar embedding for the solutions in matter-coupled gauged supergravities. It could also be of particular interest to study the dual three-dimensional $N=3,4$ SCFTs with topological twists and compute microscopic entropy of the black holes. Finally, it would be interesting to study similar solutions in other gauged supergravities such as $\omega$-deformed $N=8$ gauged supergravity and $N=4$ truncation of massive type IIA on $S^6$ given in \cite{omega_N8} and \cite{N4_4D_from_mIIA}, respectively.  
\vspace{0.5cm}\\
{\large{\textbf{Acknowledgement}}} \\
This work is supported by The Thailand Research Fund (TRF) under grant RSA6280022.



\begin{thebibliography}{99}
\bibitem{stronginger_vafa} A. Strominger and C. Vafa, ``Microscopic Origin of the Bekenstein-Hawking Entropy'', Phys. Lett. \textbf{B379} (1996) 99-104, arXiv: hep-th/9601029.
\bibitem{maldacena} J. M. Maldacena, ``The large $N$ limit of
superconformal field theories and supergravity'', Adv. Theor. Math.
Phys. \textbf{2} (1998) 231-252, arXiv: hep-th/9711200.
\bibitem{Gubser_AdS_CFT} S. S. Gubser, I. R. Klebanov and A. M. Polyakov, ``Gauge Theory Correlators from Non-Critical String Theory'', Phys. Lett. \textbf{B428} (1998) 105-114, arXiv: hep-th/9802109.
\bibitem{Witten_AdS_CFT} E. Witten, ``Anti De Sitter Space and holography'', Adv. Theor. Math. Phys. \textbf{2} (1998) 253-291, arXiv: hep-th/9802150.
\bibitem{Zaffaroni_BH1} F. Benini, K. Hristov and A. Zaffaroni, ``Black hole microstates in $AdS_4$ from supersymmetric localization'', JHEP 05 (2016) \textbf{054}, arXiv: 1511.04085.
\bibitem{Zaffaroni_BH2} F. Benini, K. Hristov, and A. Zaffaroni, ``Exact microstate counting for dyonic black holes in $AdS_4$'', Phys. Lett. \textbf{B05} (2017) 076, arXiv: 1608.07294.
\bibitem{Zaffaroni_BH3} S. M. Hosseini and A. Zaffaroni, ``Large N matrix models for 3d $N = 2$ theories: twisted index, free energy and black holes'', JHEP 08 (2016) \textbf{064}, arXiv: 1604.03122.
\bibitem{twisted_index1} F. Benini and A. Zaffaroni, ``A topologically twisted index for three-dimensional supersymmetric theories'', JHEP 07 (2015) \textbf{127}, arXiv: 1504.03698.
\bibitem{twisted_index2} S. M. Hosseini and N. Mekareeya, ``Large N topologically twisted index: necklace quivers, dualities, and Sasaki-Einstein spaces'', JHEP 08 (2016) \textbf{089}, arXiv: 1604.03397.
\bibitem{twisted_index3} F. Benini and A. Zaffaroni, ``Supersymmetric partition functions on Riemann surfaces``, Proc. Symp. Pure Math. \textbf{96} (2017) 13-46, arXiv: 1605.06120.
\bibitem{twisted_index4} C. Closset and H. Kim, ``Comments on twisted indices in 3d supersymmetric gauge theories'', JHEP 08 (2016) \textbf{059}, arXiv: 1605.06531.
\bibitem{twisted_index5} A. Cabo-Bizet, V. I. Giraldo-Rivera, and L. A. Pando Zayas, ``Microstate Counting of $AdS_4$ Hyperbolic Black Hole Entropy via the Topologically Twisted Index'', JHEP 08 (2017) \textbf{023}, arXiv: 1701.07893.
\bibitem{AdS4_BH1} S. L. Cacciatori and D. Klemm, ``Supersymmetric AdS(4) black holes and attractors'', JHEP 01 (2010) \textbf{085}, arXiv: 0911.4926.
\bibitem{AdS4_BH2} G. Dall’Agata and A. Gnecchi, ``Flow equations and attractors for black holes in $N = 2$ $U(1)$ gauged supergravity, JHEP 03 (2011) \textbf{037}, arXiv: 1012.3756.
\bibitem{AdS4_BH3} K. Hristov and S. Vandoren, ``Static supersymmetric black holes in $AdS_4$ with spherical symmetry'', JHEP 04 (2011) \textbf{047}, arXiv: 1012.4314.
\bibitem{AdS4_BH4} N. Halmagyi, ``BPS Black Hole Horizons in $N=2$ Gauged Supergravity'', JHEP 02 (2014) \textbf{051}, arXiv: 1308.1439.
\bibitem{AdS4_BH5} N. Halmagyi, M. Petrini, and A. Zaffaroni, ``BPS black holes in $AdS_4$ from M-theory'', JHEP 08 (2013) \textbf{124}, arXiv: 1305.0730.
\bibitem{Guarino_AdS2_1} A. Guarino and J. Tarrio, ``BPS black holes from massive IIA on $S^6$'', JHEP 09 (2017) \textbf{141}, arXiv: 1703.10833.
\bibitem{Guarino_AdS2_2} A. Guarino, ``BPS black hole horizons from massive IIA'', JHEP 08 (2017) \textbf{100}, arXiv: 1706.01823.
\bibitem{Kim_AdS2} J. P. Gauntlett, N. Kim, S. Pakis and D. Waldram, ``Membranes Wrapped on Holomorphic Curves'', Phys. Rev. \textbf{D65} (2002) 026003, arXiv: hep-th/0105250.
\bibitem{N3_SU2_SU3} P. Karndumri, ``Holographic RG flows in $N=3$ Chern-Simons-Matter theory from $N=3$ 4D gauged supergravity'', Phys. Rev. \textbf{D94} (2016) 045006, arXiv: 1601.05703.
\bibitem{Trisasakian_AdS2} P. Karndumri, ``Supersymmetric $AdS_2\times \Sigma_2$ solutions from tri-sasakian truncation'', Eur. Phys. J. C (2017) \textbf{77}, 689, arXiv: 1707.09633.
\bibitem{N5_flow} P. Karndumri and C. Maneerat, ``Supersymmetric solutions from $N=5$ gauged supergravity'', Phys. Rev. \textbf{D101} (2020) 126015, arXiv: 2003.05889.
\bibitem{N6_flow} P. Karndumri and J. Seeyangnok, ``Supersymmetric solutions from $N=6$ gauged supergravity'', Phys. Rev. \textbf{D103} (2021) 066023, arXiv: 2012.10978.
\bibitem{AdS4_N4_Jan} J. Louis and H. Triendl, ``Maximally supersymmetric $AdS_4$ vacua in $N=4$ supergravity'', JHEP 10 (2014) \textbf{007}, arXiv:1406.3363.
\bibitem{N3_Ferrara} L. Castellani, A. Ceresole, S. Ferrara, R. D'Auria, P. Fre and E. Maina, ``The complete $N=3$ matter coupled supergravity'', Nucl. Phys. \textbf{B268} (1986) 317-348.
\bibitem{N3_Ferrara2} L. Castellani, A. Ceresole, R. D'Auria, S. Ferrara, P. Fre and E. Maina, ``$\sigma$-model, duality transformations and scalar potentials in extended supergravities'', Phys. Lett. \textbf{B161} (1985) 91-95.
\bibitem{Castellani_book} L. Castellani, R. D' Auria and P. Fre, ``Supergravity and Superstring theory: a geometric perspective'', World Scientific, Singapore 1990.
\bibitem{pureN3_1} D. Z. Freedman, ``$SO(3)$ Invariant Extended Supergravity'', Phys. Rev. Lett. \textbf{38} (1977) 105.
\bibitem{pureN3_2} P. Fre, ``Extended Supergravity on the Supergroup Manifold $N=3$ and $N=2$ Theories'', Nucl. Phys. \textbf{B186} (1981) 44-60.
\bibitem{Varela_pureN3} O. Varela, ``Minimal $D=4$ truncations of type IIA'', JHEP 11 (2019) \textbf{009}, arXiv: 1908.00535.
\bibitem{N010_truncation_Cassani} D. Cassani and P. Koerber, ``Tri-Sasakian consistent
reduction'', JHEP 01 (2012) \textbf{086}, arXiv: 1110.5327.
\bibitem{N3_4D_gauging} P. Karndumri and K. Upathambhakul, ``Gaugings of four-dimensional $N=3$ supergravity and AdS$_4$/CFT$_3$ holography'', Phys. Rev. \textbf{D93} (2016) 125017 arXiv: 1602.02254.
\bibitem{N4_gauged_SUGRA} J. Schon and M. Weidner, ``Gauged $N=4$ supergravities'', JHEP 05 (2006) \textbf{034}, arXiv: hep-th/0602024.
\bibitem{Eric_N4_4D} E. Bergshoeff, I. G. Koh and E. Sezgin, ``Coupling of Yang-Mills to $N=4$, $d=4$ supergravity'', Phys. Lett. \textbf{B155} (1985) 71-75.
\bibitem{dS_Roest} D. Roest and J. Rosseel, ``De Sitter in Extended Supergravity'', Phys. Lett. \textbf{B685} (2010) 201-207, arXiv: 0912.4440.
\bibitem{N4_Janus} P. Karndumri, ``Holographic RG flows and Janus solutions from matter-coupled $N=4$ gauged supergravity'', Eur. Phys. J. \textbf{C81} (2021) 520, arXiv: 2102.05532.
\bibitem{Klemm_symplectic} D. Klemm, N. Petri and M. Rabbiosi, ``Symplectically invariant flow equations for $N=2$, $D=4$ gauged supergravity with hypermultiplets'', JHEP 04 (2016) \textbf{008}, arXiv: 1602.01334.
\bibitem{5Dtwist} H. L. Dao and P. Karndumri, ``Supersymmetric $AdS_5$ black holes and strings from 5D $N=4$ gauged supergravity'', Eur. Phys. J. \textbf{C79} (2019) 247, arXiv: 1812.10122.
\bibitem{H_C_twist} D. Gaiotto, ``Twisted compactifications of $3D$ $N=4$ theories and conformal blocks'', JHEP 02 (2019) \textbf{061}, arXiv: 1611.01528.
\bibitem{flow_acrossD_bobev} N. Bobev and P. M. Crichigno, ``Universal RG Flows Across Dimensions and
Holography'', JHEP 12 (2017) \textbf{065}, arXiv: 1708.05052.
\bibitem{4D_N4_flows} P. Karndumri and K. Upathambhakul, "Holographic RG flows in N = 4 SCFTs from half-maximal gauged supergravity", Eur. Phys. J. \textbf{C78} (2018) 626, arXiv:hep-th/1806.01819.
\bibitem{5D_N4_flow} H. L. Dao and P. Karndumri, ``Holographic RG flows and $AdS_5$ black strings from 5D half-maximal gauged supergravity'', Eur. Phys. J. \textbf{C79} (2019) 137, arXiv: 1811.01608.
\bibitem{6D_twist} P. Karndumri, ``Twisted compactification of $N = 2$ 5D SCFTs to three and two dimensions from $F(4)$ gauged supergravity'', JHEP 09 (2015) \textbf{034}, arXiv: 1507.01515.
\bibitem{7D_twist} P. Karndumri, ``RG flows from $(1,0)$ 6D SCFTs to $N = 1$ SCFTs in four and three dimensions'', JHEP 06 (2015) \textbf{027}, arXiv: 1503.04997.
\bibitem{Jan_maximal_AdS} S. Lust, P. Ruter and J. Louis, ``Maximally Supersymmetric AdS Solutions and their Moduli Spaces
'', JHEP 03 (2018) \textbf{019}, arXiv: 1711.06180.
\bibitem{Pope_4DpureN4} M. Cvetic, H. Lu and C. N. Pope, ``Four-dimensional $N=4$ $SO(4)$ Gauged Supergravity from $D=11$'', Nucl. Phys. \textbf{B574} (2000) 761-781, arXiv: hep-th/9910252.
\bibitem{omega_N8} G. Dall’Agata, G. Inverso and M. Trigiante, ``Evidence for a family of $SO(8)$ gauged
supergravity theories'', Phys. Rev. Lett. \textbf{109} (2012) 201301, arXiv:1209.0760.
\bibitem{N4_4D_from_mIIA} A. Guarino, J. Tarrio and O. Varela, ``Halving $ISO(7)$ supergravity'', JHEP 11 (2019) \textbf{143}, arXiv: 1907.11681.
\end{thebibliography}
\end{document}